\documentclass[12pt]{article}
\usepackage[latin1]{inputenc}
\usepackage{amsfonts}
\usepackage[dvips]{graphics}
\usepackage{graphicx}
\usepackage{cite}
\usepackage{stmaryrd}
\usepackage{amssymb}
\usepackage{amsmath}
\usepackage{fancyhdr}
\usepackage[body={6.0in, 8.2in},left=1.25in,right=1.25in,bottom=0.65in]{geometry}
\usepackage{setspace}                    % Line spacing
\usepackage{hyperref}

\let\ssection=\section
\renewcommand{\section}{\setcounter{equation}{0}\ssection}

\newcommand{\be}{\begin{equation}}
\newcommand{\ee}{\end{equation}}
\newcommand{\er}{\ref}

\newcommand{\ca}{$C^*$-algebra} 
 \newcommand{\rep}{representation}
\newcommand{\irrep}{irreducible representation}
\newcommand{\Hs}{Hilbert space}

 \newcommand{\ovl}{\overline}
 \newcommand{\til}{\tilde}
\newcommand{\raw}{\rightarrow}

\newcommand{\ot}{\otimes}

\newcommand\mathC{\mkern1mu\raise2.2pt\hbox{$\scriptscriptstyle|$}
		{\mkern-7mu\rm C}}				%%% The complex  numbers
		
\newcommand{\mathR}{{\rm I\! R}}         %%% The real numbers
 \newcommand{\cci}{C^{\infty}_c}
\newcommand{\half}{\mbox{\footnotesize $\frac{1}{2}$}}
\newcommand{\third}{\mbox{\footnotesize $\frac{1}{3}$}}

\newcommand{\lho}{\lim_{\hbar\rightarrow 0}}   

\newcommand{\qh}{q_{\hbar}}

 \newcommand{\varep}{\varepsilon}

\newcommand{\rh}{\rho} \newcommand{\sg}{\sigma}
  
 \newcommand{\phv}{\varphi}
 \newcommand{\ps}{\psi} \newcommand{\Ps}{\Psi}
\newcommand{\om}{\omega} 

\newcommand{\GS}{\mathfrak{S}}

\newcommand{\CA}{{\mathcal A}} \newcommand{\CB}{{\mathcal B}}

\newcommand{\CG}{{\mathcal G}} \renewcommand{\H}{{\mathcal H}}

 \newcommand{\CS}{{\mathcal S}}
 
\newcommand{\CQ}{{\mathcal Q}}

\newcommand{\N}{{\mathbb N}} \newcommand{\R}{{\mathbb R}}
 \newcommand{\Z}{{\mathbb Z}}
\renewcommand{\qh}{\CQ_{\hbar}}

\def\Dslash{\setbox0=\hbox{$D$}D\hskip-\wd0\hbox to\wd0{\hss\sl/\/\hss}}

\renewcommand{\lho}{\lim_{\hbar\raw 0}}
\newcommand{\lni}{\lim_{N\raw\infty}}
\newcommand{\up}{\uparrow}
\newcommand{\down}{\downarrow}

\begin{document}
\begin{center}
{\large\bf Less is Different: Emergence and Reduction Reconciled}
\end{center}

\vspace{0.8 truecm}
\begin{center}
            J.~Butterfield 
			\\[10pt]
Trinity College, Cambridge University, Cambridge CB2 1TQ; email: jb56@cam.ac.uk
\end{center}

%\begin{center} Tuesday 19 October 2010 \end{center}

%\begin{center}  Forthcoming in {\em Foundations of Physics} \end{center}
\vspace{0.6in}

\begin{abstract}
This is a companion to another paper. Together they rebut two widespread philosophical %%@
doctrines about emergence. The first, and main, doctrine is that emergence is incompatible %%@
with reduction. The second is that emergence is supervenience; or more exactly, supervenience %%@
without reduction.

 In the other paper, I develop these rebuttals in general terms, emphasising the second %%@
rebuttal. Here I discuss the situation in physics, emphasising the first rebuttal. I focus on %%@
limiting relations between theories and illustrate my claims with four examples, each of them %%@
a model or a framework for modelling, from well-established mathematics or physics.

I take emergence as behaviour that is novel and robust relative to some comparison class. I  %%@
take reduction as, essentially, deduction. The main idea of my first rebuttal will be to %%@
perform the deduction after taking a limit of some parameter. Thus my first main claim will %%@
be that in my four examples (and many others), we can deduce a novel and robust behaviour, by %%@
taking the limit $N \raw \infty$ of a parameter $N$.

But on the other hand, this does not show that that the $N = \infty$ limit is ``physically %%@
real'', as some authors have alleged. For my second main claim is that in these same %%@
examples, there is a weaker, yet still vivid, novel and robust behaviour that occurs {\em %%@
before} we get to the limit, i.e. for finite $N$. And it is this weaker behaviour which is %%@
physically real.

My examples are: the method of arbitrary functions (in probability theory); fractals (in %%@
geometry); superselection for infinite systems  (in quantum theory); and phase transitions %%@
for infinite systems (in statistical mechanics).

\end{abstract}
\newpage
\tableofcontents

\newpage
\section{Introduction}\label{Intr}
\subsection{A limited peace}\label{peace}
`More is different!', proclaimed Philip Anderson in a famous paper (1972) advocating the %%@
autonomy of what are often called `special' or `higher-level' sciences or theories. A catchy %%@
slogan, indeed. But his reductionist opponents, such as Weinberg (1987), could have matched %%@
it, by invoking Mies van der Rohe's pithy defence of functionalist architecture: `Less is %%@
more'. Hence my title. For my main point will be that although emergence is usually opposed %%@
to reduction, many examples exhibit both. So my title, `Less is different', is meant as an %%@
irenic combination of the two parties' slogans. I will spell out this reconciliation in two %%@
claims, illustrated by four examples. The two claims, mnemonically labelled (1:Deduce) and %%@
(2:Before), are defined in Section \ref{prosp}; and each example is a model or a framework %%@
for modelling, from well-established mathematics or physics.

My irenic title is also ironic. For it deliberately echoes the sceptical refrain that there %%@
is nothing new
 under the Sun. Though I will not name names, most would agree that there is a good deal of %%@
heat, and rather less light, in the debate about emergence vs. reduction. Here's hoping that %%@
you will not recite that same refrain after reading this paper! Of course the heat and dark %%@
is in part due to different authors giving `emergence' and `reduction' different  meanings. %%@
Thus I do not claim to be the only author to celebrate these words' compatibility. Among %%@
other celebrants, albeit using different meanings, are Simon (1996, pp. 249-251) and Wimsatt %%@
(1997, pp. 99-100 and references therein).\footnote{Other playful variations on Anderson's %%@
slogan occur in Kadanoff's  splendid historico-philosophical introductions to phase %%@
transitions (2009, 2010, 2010a): which I will advert to in Section \ref{phasetr}. Cat (1998) %%@
is a scholarly review of the Anderson-Weinberg debate; Bouatta and Butterfield (2011) also %%@
contains a discussion.\label{leocredit}}

However, this is a companion to another paper (Butterfield 2010). So although the papers can %%@
be read independently, I should begin by describing their common aims and how they share out %%@
the work between them. In brief, both papers construe the contested terms, `emergence' and %%@
`reduction',  as follows; (the other paper gives more details, and a defence of these %%@
construals; cf. its Sections 1.1, 2.1 and 3.1.1.). 

I take emergence as behaviour that is novel and robust relative to some comparison class. In %%@
particular, my examples will be typical of many, by using two widespread conceptions of what %%@
the comparison class is, as follows. (1): {\em Composites}: The system  is a composite; and %%@
its properties and behaviour are novel and robust compared to those of its component systems, %%@
especially its microscopic or even atomic components. (2): {\em Limits}: The system  is a %%@
limit of a sequence of systems, typically as some parameter (in the theory of the systems) %%@
goes to infinity (or some other crucial value, often zero); and its properties and behaviour %%@
are novel and robust compared to those of  systems described with a finite (respectively: %%@
non-zero) parameter. (Section \ref{ssqv} will explain how these ideas, (1) and (2), are %%@
better put in terms of quantities and their values, rather than systems.)

I take reduction as, essentially, deduction; though usually aided by appropriate definitions %%@
or bridge-principles linking the two theories' vocabularies. This will be close to endorsing %%@
the traditional account of Nagel (1961), despite various objections levelled against it. The %%@
picture is that the claims of some worse or less detailed (often earlier) theory can be %%@
deduced within a better or more detailed (often later) theory, once we adjoin to the latter %%@
appropriate definitions of the proprietary terms of the former. I also adopt a mnemonic %%@
notation, writing $T_b$ for the better, bottom or basic theory, and $T_t$ for the tainted, %%@
top or tangible theory; (where `tangible' connotes restriction to the observable, i.e. less %%@
detail). So the picture is, with $D$ standing for the definitions: $T_b \& D \Rightarrow %%@
T_t$. In logicians' jargon: $T_t$ is a {\em definitional extension} of $T_b$.

In both papers, especially the other one, I consider a notion much discussed in the %%@
philosophy (but not physics) literature: supervenience (also known as `determination' or %%@
`implicit definability'). This is a less contested term. It is taken by all to be a relation %%@
between families of properties: the extensions of all the properties in one family relative %%@
to a given domain of objects determine the extension of each property in the other family. %%@
Besides, under wide conditions, this is a weakening of the usual notion of the second family %%@
being definable from the first, which is called `explicit definability'.  Roughly speaking, %%@
this weakening allows a definition of a property $P$ in the second family, in terms of the %%@
first family, to be infinitely long, rather than finite.  

Since the definitions used in a Nagelian reduction are finite, supervenience is widely taken %%@
to be a weakening of Nagelian reduction. Besides, various philosophers have considered the %%@
infinity of ``ways to be $P$'' given by an infinitely long definition to be a good way of %%@
making precise the heterogeneity or multiplicity of realization that philosophers have often %%@
associated with emergence. Thus arose the doctrine that emergence is ``mere supervenience'', %%@
i.e. supervenience without all the definitions being finite, as in a Nagelian  reduction. 

With these construals of the terms, the papers aim to rebut two widespread doctrines about %%@
emergence: the doctrine just mentioned, that emergence is mere supervenience, found in the %%@
philosophy literature; and the more widespread doctrine, found also in the physics literature %%@
(including the Anderson-Weinberg debate), that emergence is incompatible with reduction.

In the other paper, I develop these rebuttals in general terms; including a discussion of %%@
some other possible construals of the contested terms. I also emphasise supervenience, and %%@
thereby the first rebuttal. Thus I give (i) examples of mere supervenience which are not %%@
emergence and (ii) examples of emergence  which are not mere supervenience nor reduction. 

But in this paper, I will discuss the situation in physics and down-play supervenience, thus %%@
emphasising the second rebuttal. That is, I will argue that  emergence is {\em compatible} %%@
with reduction, since physics gives examples combining both. The main idea will be to perform %%@
the reduction, i.e. deduction, after taking a limit of some parameter. Thus my first main %%@
claim, (1:Deduce), will be that in my four examples (and many others), we can deduce a novel %%@
and robust behaviour, by taking the limit $N \raw \infty$ of a parameter $N$.

But on the other hand, this does not show that that the $N = \infty$ limit is ``physically %%@
real'', as some authors have alleged. For my second main claim, (2:Before), is that in these %%@
same examples, there is a weaker, yet still vivid, novel and robust behaviour that occurs %%@
{\em before} we get to the limit, i.e. for finite $N$. And it is this weaker behaviour which %%@
is physically real.

This contrast between strong and weak senses of emergence, and respectively its absence or %%@
presence at finite $N$, will be the main common theme across my four examples. It will also %%@
illuminate another current topic within philosophy of physics, about the significance of  %%@
`singular' limits in a physical theory. In fact, some authors  propose to characterize %%@
emergence in terms of `singular' limits.\footnote{I have used scare-quotes since writers %%@
often use the term loosely---too loosely, as I explain, and complain, in Section \ref{ssqv}. %%@
But for easier reading, I will henceforth drop the scare-quotes.} I deny this proposal. %%@
Although my two claims, and my four examples combining emergence and reduction, involve %%@
taking a limit, the limit is singular in only two of the four examples (the second and %%@
fourth, viz. fractals and phase transitions). So emergence is not always a matter of a %%@
singular limit---just as it is not always a matter of mere supervenience. 

This negative verdict leaves open many questions, in particular: is emergence always a matter %%@
of a limit, whether singular or not? And even though a singular limit is not necessary for %%@
emergence, is it sufficient? In fact I think the answers to these questions are again `No'. %%@
But I will not attempt to give a detailed characterization of emergence, whereby to prove %%@
these last two `No's.\footnote{Incidentally, for the last `No', that a singular limit is not %%@
sufficient for emergence: I agree with Wayne's argument for this (against Rueger (2000, p. %%@
308; 2006, pp. 344-345)). Wayne uses Rueger's own example, of the van der Pol oscillator %%@
(Wayne 2009, Sections 3-5). My second example, fractals, will give another counterexample %%@
(Section \ref{dimwithwithout}): the topological dimension of a sequence of sets $C_N$ is %%@
discontinuous in the limit, i.e. ${\rm{lim}}_{N \raw 0}$ dim $C_N \neq$ dim ${\rm{lim}}_{N %%@
\raw 0} C_N$, but there is no emergence. For persuasive, more general, critiques of %%@
associating emergence or irreducibility with ``singular asymptotics'', cf. Belot (2005, %%@
especially Section 5) and Hooker (2004, pp. 446-458).\label{Wayne}} The literature contains %%@
several such characterizations, with various merits. But as I explain in the other paper %%@
(especially Sections 1.1, 2.1),  I doubt that there is---and that there needs to be---a %%@
single best meaning of `emergence'; and similarly for `reduction'. Anyway, I can develop my %%@
claims and examples while adopting my construals---of `emergence' as novel and robust %%@
behaviour, and of reduction as deduction {\em a la} Nagel.

Before I give a prospectus (Section \ref{prosp}), I should make two final comments about %%@
these construals, and about my choice of examples. First: I submit that my construals of %%@
`emergence' and `reduction' are strong enough to make it worth exhibiting examples that %%@
combine them. Also, they seem to be in tension with each other: since logic teaches us that %%@
valid deduction gives no new ``content'', how can one ever deduce novel behaviour? This %%@
tension is also shown by the fact that many authors who take emergence to involve novel %%@
behaviour thereby take it to also involve irreducibility. The answer to the `how?'  question, %%@
i.e. my reconciliation, will lie in using limits: one performs the deduction after taking a %%@
limit of some parameter. So one main moral will be that in such a limit there can be novelty, %%@
compared with what obtains away from the limit.

Second: there is the issue of how I choose my examples. Here you may suspect what might be %%@
called the `case-study gambit': trying to  support a general conclusion by describing  %%@
examples that have the required features, though in fact the examples are {\em not} typical, %%@
so that the attempt fails, i.e. the general conclusion, that all or most examples have the %%@
features, does not follow. But to this charge also I plead innocent, for the simple reason %%@
that I will not urge so general a conclusion, in the way that a reductionist opponent of %%@
Anderson might. (For example, I think Weinberg's objective reductionism (1987, p. 349-353) %%@
implies that (with my meanings of the terms) all known examples of emergence are also %%@
examples of reduction.) On the other hand, I do aspire to {\em some} generality! It will be %%@
clear that my claims, in particular my two main ones, (1:Deduce) and (2:Before), are %%@
illustrated by many examples beyond the four I have chosen. So I submit that the claims %%@
reflect the amazing power of Nagelian reduction.

\subsection{Prospectus}\label{prosp}
Thus my main aim is to reconcile emergence with reduction, by arguing   
for two main claims, illustrated by four examples. Each example is a model, or a framework %%@
for modelling, from well-established mathematics or physics; and each involves an integer %%@
parameter $N = 1, 2, ...$ and its limit $N \raw \infty$. In three of the examples, $N$ is, %%@
roughly speaking, the number of physical degrees of freedom of the system; in the second %%@
example, it is the number of iterations of a definitional process. 

In all the examples, $N$ is, physically speaking, finite. But we can consider the limit: both %%@
what happens on the way to the limit, and what happens at it. (In Section \ref{ssqv}, I will %%@
be more precise about the meaning of `what happens', in terms of quantities being %%@
well-defined and what their values are.) Doing so yields my two main claims. The first is:
\begin{quote}
 (1:Deduce): Emergence is compatible with reduction. And this is so, with a strong %%@
understanding both of `emergence' (i.e. `novel and robust behaviour') and of `reduction' %%@
(viz. logicians' notion of definitional extension). In short: in the examples, considering $N %%@
\raw \infty$ enables us to {\em deduce} novel and robust behaviour, in strong senses of %%@
`novel' and `robust'.

 Besides, one needs to consider the limit in that: for each example,
choosing a weaker theory using finite $N$ blocks the deduction of this strong sense. And (as %%@
discussed in Section \ref{ssqv}), this weaker theory is appropriate and salient, i.e. liable %%@
to come to mind. Since the theories $T_t$ and $T_b$ are often defined only  vaguely (by %%@
labels like `thermodynamics' and `statistical mechanics'), this swings-and-roundabouts %%@
situation explains away some of the controversy over whether $T_t$ is reducible to $T_b$.
\end{quote}
The second claim is:
\begin{quote}
\indent (2:Before): But on the other hand: emergence, in a weaker yet still vivid sense, %%@
occurs {\em before} we get to the limit. That is: in each example, one can understand `novel %%@
and robust behaviour' weakly enough that it {\em does} occur for finite $N$.
\end{quote}

Of my four examples, I have chosen the first three to be comparatively small, simple and %%@
agreed-upon, so that the philosophical issues stand out more clearly. They are from %%@
probability theory, geometry and quantum theory, respectively. The fourth example is an %%@
enormous topic in physics, with much less agreement. The examples are, in order:\\
\indent 1: The method of arbitrary functions, in probability theory; (Section \ref{maf});\\
\indent 2: Fractals, in geometry; (Section \ref{frac});\\
\indent 3: Superselection for infinite systems, in quantum theory; (Section %%@
\ref{superselec});\\
\indent 4: Phase transitions for infinite systems, in classical statistical mechanics %%@
(Section \ref{phasetr}).

Apart from the contrast between strong and weak senses of emergence shown by (1:Deduce) and %%@
(2:Before), there will be two other philosophical themes in common across the four examples.

 The first is that supervenience is a ``red herring'', i.e. irrelevant. (So this supports the %%@
other paper's rebuttal of the doctrine that emergence is mere supervenience.) For clarity, it %%@
will again be best to give this a mnemonic label, as follows:
\begin{quote}
(3:Herring): Although various supervenience theses are true in the examples (and many %%@
others), the theses yield little or no insight---either into emergence, or more generally, %%@
into ``what is going on'' in the example.
\end{quote}
We can already state the basic reason for this irrelevance. Supervenience allows that for %%@
each property $P$ in the ``higher'' i.e. supervening family of properties, there is, in the %%@
taxonomy given by the lower family, a disjunction of ``ways to be $P$''. But supervenience %%@
gives no ``control'' on this disjunction: not just in the sense that the disjunction might be %%@
infinite, but also that supervenience allows it to be utterly heterogeneous. In particular,  %%@
no kind of limit is taken; and more generally, no connection is made between the variety, or %%@
infinity, of the disjunction and the limit processes, especially $N \raw \infty$, which are %%@
crucial to the example. Thus supervenience is, at least in these examples (and, I submit, %%@
many others), too weak a concept to be enlightening.

The other theme in common across the four examples is that each example becomes, for finite %%@
but very large $N$, {\em unrealistic} in a vivid---one might even say: catastrophic---way.
But this occurs for reasons {\em external} to current debates about emergence, reduction and %%@
the significance of limits of physical theories. It also will {\em not} undermine my %%@
(2:Before). This is because each of my examples illustrates (2:Before) for values of its %%@
parameter $N$ {\em much smaller} than those at which the example becomes unrealistic in the %%@
catastrophic way.  So I will not emphasize this theme. On the other hand, the theme seems to %%@
have been completely neglected in these debates' literature; so it is worth spelling out. I %%@
will do this in Section \ref{unreal}, again giving it a mnemonic label, (4:Unreal). 

After I discuss (4:Unreal), I give in Section \ref{ssqv} a general discussion of physical %%@
systems and their states and quantities, emphasizing the topic of limits: i.e. limits of %%@
systems, states and quantities, as some parameter $N$ (typically the number of degrees of %%@
freedom) goes to infinity. There are two related philosophical questions to be addressed. The %%@
first, mentioned in Section \ref{peace}, is whether emergence can be characterized in terms %%@
of limits, especially singular limits. {\em Contra} some authors, I deny this; (along with %%@
others such as Wayne, Belot and Hooker).

 The second question is whether in some examples, the singular limit is---not just %%@
indispensable for deducing emergence in some strong sense, or for epistemic concerns such as %%@
explanation and understanding---but also `physically real'. These two questions are related %%@
in various ways: most obviously, by a Yes to the second implying that emergence according to %%@
the first would be physically real.

I shall also deny this: again,  {\em contra} some authors. In more detail: my first claim, %%@
(1:Deduce), will illustrate how limits can be indispensable---viz. to deducing some novel and %%@
robust behaviour, where the behaviour in question is taken in a strong sense. And on the %%@
other hand, my second claim, (2:Before), will bring out how the $N = \infty$ limit is not %%@
physically real. That is: only the weaker sense of emergence that occurs at finite $N$ is %%@
physically real.  

So let me sum up my claims. Emergence is not in all cases failure of reduction, even in the %%@
strong sense of reduction given by deduction (cf. (1:Deduce)). (Here, the deduction's need to %%@
invoke auxiliary definitions of the reduced theory's terms is made precise by logicians' %%@
notion of {\em definitional extension}; for details, cf. Section 3.1 of the companion paper).  %%@
Nor does emergence in all cases occur only in the limit of the relevant parameter (cf. %%@
(2:Before)). Nor is emergence in all cases a matter of this limit being ``singular'' in some %%@
sense: my first and third examples will have non-singular limits (cf. also Section %%@
\ref{ssqv}). Nor is emergence in all cases supervenience; nor is it in all cases failure of %%@
supervenience; (cf. Section 5 of the companion paper, and for the latter denial, (1:Deduce)). %%@
In short: we have before us a varied landscape---emergence is {\em independent} of these %%@
other notions.

\section{Becoming unrealistic {\em on the way to} the limit}\label{unreal}
As we will see, my examples (and many other models, such as continuous models of fluids and %%@
solids) are examples of: formulating a formalism by taking an admittedly unrealistic limit of %%@
a parameter's value. But they are also examples of: formulating a formalism by taking a limit %%@
of a description which is admitted to be unrealistic {\em on the way} to the limit. This is %%@
my fourth labelled claim:
\begin{quote}
\indent (4:Unreal): Each of the four examples becomes unrealistic {\em before} one gets to %%@
the $N = \infty$ limit---regardless of any technical issues about that limit, and regardless %%@
of any philosophical controversies about emergence.
\end{quote}

One reason I need to discuss this claim is to show how it is consistent with (2:Before): the %%@
main point will be (as I mentioned) that (2:Before) applies to much smaller values of $N$. %%@
But phase transitions will also yield a remarkable illustration of ``oscillations'' between %%@
(2:Before) and (4:Unreal). In Section \ref{crossover}, we will see how a system can be %%@
manipulated so as to first illustrate (2:Before), i.e. an emergent behaviour at finite $N$, %%@
then lose this behaviour, i.e. illustrate (4:Unreal), and then enter a regime illustrating %%@
some other emergent behaviour (or revert to the first behaviour): a phenomenon called `{\em %%@
cross-over}'.   

There are also two other reasons why it is worth stating this claim, i.e. reasons unrelated %%@
to my own position in debates about emergence. First, almost all discussions of emergence, or %%@
more generally of limiting relations between  theories, in the physics and philosophy %%@
literatures, fail to notice this point. Agreed, some {\em maestros} notice it---though I do %%@
not mean to argue from authority! Thus Feynman: `When you follow any of our physics too far, %%@
you find that you always get into some kind of trouble' (1964, Lecture 28.1).\footnote{I %%@
should mention another meaning of `intermediate between small and infinite values of a %%@
parameter' that {\em is} noticed by the physics literature, under the label `intermediate %%@
asymptotics': namely, a system's behaviour `for times, and distances from boundaries, large %%@
enough for the influence of the fine details of the initial and/or boundary conditions to %%@
disappear, but small enough that the system is far from the ultimate equilibrium state' %%@
(Barenblatt 1996, p. xiii; cf. also p. 19). This meaning is obviously very different from %%@
this Section's `intermediate $N$' regime. But it is worth mentioning, not just because of its %%@
intrinsic importance, but also because:
(i) it is related to renormalization, which I will touch on in Section \ref{crossover} (cf. %%@
also Goldenfeld et al. (1989)); and (ii) some philosophers (to their credit) have discussed %%@
it---though surely Batterman goes much too far when he writes `I think, as should be obvious %%@
by now, that any investigation that remotely addresses a question related to understanding %%@
universal behavior [i.e. in philosophers' terms: multiple realizability] will involve %%@
intermediate asymptotics as understood by Barenblatt' (2002, p. 46).\label{intermed}}

Second, there is a common kind of reason for the un-realisticness (``break-down'') of the %%@
examples. Besides, this kind of reason inevitably besets {\em many} other examples of taking %%@
limits of models as a parameter $N$, encoding physical degrees of freedom or some analogous %%@
concept, goes to $\infty$.
So this commonality is worth registering, especially in discussions of emergence, or more %%@
generally of limits of models as a parameter $N \rightarrow \infty$.  

In short, the commonality is: as $N$ becomes very large, the example runs up against {\em %%@
either} the micro-structure of space and its contents (for short: atomism), {\em or} the %%@
macro-structure of space and its contents (for short: cosmology). Thus my first two examples %%@
will run up against atomism: that is, very large $N$ will correspond to atomic or sub-atomic %%@
lengths, making what the example says utterly unrealistic. And my third and fourth examples %%@
will run up against cosmology: very large $N$ will correspond to cosmic lengths (and so %%@
gravity, and indeed spacetime curvature), again making what the example says utterly %%@
unrealistic. I stress that these break-downs are not internal to the model, but in relation %%@
to the actual world. To take my third and fourth examples: if there were no gravity nor %%@
spacetime curvature, and if space had the structure of $\mathR^3$, these examples, which %%@
postulate a chain of $N$ spins or a gas of $N$ molecules, in $\mathR^3$ without gravity, %%@
would indeed remain realistic as $N$ grows without bound.

I say `in short', because in some examples Feynman's `some kind of trouble' is not just %%@
either atomism or cosmology. The situation can be more varied. I will not enter into details, %%@
let alone try to classify the kinds of trouble. But to illustrate: my first example, the %%@
method of arbitrary functions, 
will include a model of a roulette wheel whose angular velocity tends to infinity; so the %%@
trouble will be, not atomism, but the fact that the model is Newtonian not relativistic! And %%@
more importantly: in my fourth example, phase transitions, some models run up against both %%@
atomism and cosmology. For in some models, the thermodynamic limit is not just the idea that %%@
keeping the density constant, the number $N$ of molecules (and so the volume) tends to %%@
infinity: there are also conditions on the limiting behaviour of short-range forces. 

However, (4:Unreal) plays a different role in my discussion from my other three labelled %%@
claims: so I will not emphasize it as much as the others. There are two differences. First: %%@
with one exception, discussions of these examples and many others---including discussions %%@
about emergence, and  the examples' $N = \infty$ limit---do not, so far as I know, mention %%@
this un-realisticness for very large $N$. (The exception is my second example, viz. %%@
fractals.) Second: in each of my four examples, this un-realisticness for very large $N$ is %%@
{\em not} relevant to the ways that: \\
\indent \indent (1): a strong sense of emergent (i.e. novel and robust) behaviour can be %%@
deduced at the limit (cf. (1:Deduce)); and\\
\indent \indent (2): a weaker, yet still vivid, sense of emergent behaviour occurs on the way %%@
to the limit (cf. (2:Before)); and \\
\indent \indent (3): supervenience is a red herring, giving little or no insight into the %%@
example (cf. (3:Herring)).  

So my discussion of emergence, in particular my main positive aim---the reconciliation got by %%@
combining my first two claims (1:Deduce) and (2:Before)---can proceed without discussing %%@
(4:Unreal). I stress again that each of the four examples illustrates (2:Before) for values %%@
of its parameter $N$ much smaller than those at which the example becomes unrealistic. (And %%@
this point  applies in many other examples of taking limits of models as a degrees-of-freedom %%@
parameter goes to $\infty$.) So to keep the discussion of my examples as simple as possible, %%@
I will not explicitly refer there to (4:Unreal)---{\em except} at (i) the end of the second %%@
example, fractals, for which, as I said, the literature has noticed the point; and at (ii) %%@
the end of the fourth example, where the phenomenon of {\em cross-over} subtly combines %%@
(4:Unreal) with (2:Before).

\section{Systems, states, quantities, values---and their limits}\label{ssqv}
In Section \ref{prosp}, I promised that my two claims, (1:Deduce) and (2:Before), would %%@
clarify---I dare not say resolve!---the question whether in some examples of `infinite' %%@
and-or `singular' limits, the limit is not just epistemically indispensable but also %%@
`physically real'. More specifically, I said I would agree about the indispensability, thanks %%@
to (1:Deduce), but deny the reality, thanks to (2:Before). But even before I show those %%@
claims in my examples, I can defend my general position; and in particular, justify my %%@
denying the physical reality of the limits. That is the job of this Section.

This is a job worth doing for two reasons. First, some discussions of emergence, and more %%@
generally, of limiting relations between theories are sloppy in their use of mathematical %%@
jargon about limits being `singular' vs. `regular/well-behaved/continuous' etc. And as I %%@
mentioned in Section \ref{peace},  some authors even identify emergence with what happens at %%@
a `singular' limit (Batterman (2002, pp. 6, 120, 127, 135), (2006, pp. 902-903), (2009, pp. %%@
23-24); Rueger (2000, p. 308), (2006, pp. 344-345)). At least for my sense of emergence as %%@
novel and robust behaviour, this is  wrong. In two of my four examples, there is nothing %%@
`singular' about the limit. And recall that footnote \ref{Wayne} cited other arguments (and %%@
other authors) to the effect that a singular limit is not sufficient for emergence or %%@
irreducibility. 

Second, some of the literature's physical examples and philosophical discussions are %%@
dauntingly complex. To take just one current philosopher: Batterman's examples include: (a) %%@
ray optics as a limit of wave optics; (b): classical mechanics as a limit of quantum %%@
mechanics; (c): hydrodynamics as a limit of molecular models; (d): phase transitions as %%@
described in the thermodynamic limit of statistical mechanics. Each of these is a  large and %%@
complex area of physics, in which recent decades have seen a lot of deep and beautiful %%@
work---some of whose creators have themselves given masterly philosophical discussions (e.g. %%@
Berry 1994, Goldenfeld et al. 1989, Kadanoff 2009, 2010, 2010a). So there is a great deal for %%@
philosophers to address; (and all credit to Batterman and others for doing so). But we run %%@
the risk of being blinded by science, i.e. being misled by arcane {\em technicalia}. So I %%@
propose to discuss just one area, and even that only briefly: phase transitions, which will %%@
be my fourth example. (As I mentioned in Section \ref{prosp}, I chose my first three examples %%@
partly for their merit of being comparatively small and simple, so that the philosophical %%@
issues are clearer.) There is also a mountain of previous philosophical discussion, far too %%@
large to be addressed here. For apart from current authors like Batterman and Rueger, %%@
limiting relations between physical theories (singular or not) have long been a topic for %%@
authors such as Post, Schaffner, Scheibe, Rohrlich and Redhead. So I propose here just to %%@
spell out the general situation, as I see it. That will be enough to indicate how (at least %%@
in my examples!) there is no reason to believe the limit is physically real---a verdict which %%@
my examples will then confirm.

I divide the task in three subsections, \ref{deflatelimits} to \ref{2genl}. Sections %%@
\ref{deflatelimits} and \ref{atlimit} lay out some distinctions. Then Section \ref{2genl} %%@
addresses the philosophical issue of what justifies our using a description with $N = %%@
\infty$. There I argue that even when the relevant limits are singular, a straightforward and %%@
broadly instrumentalist justification, viz. mathematical convenience and empirical %%@
correctness, applies: so that we need not believe the limit is physically real.

\subsection{Emergence with and without infinite systems---and with ordinary %%@
limits}\label{deflatelimits}
We begin by envisaging physical systems, $\sigma$ say, each labelled by its parameter  $N$, %%@
and thus a sequence of ever larger systems $\sigma(N)$. In all that follows (including my %%@
examples) $N \in {\bf N}$ := the set of natural numbers. But nothing in this Section or the %%@
sequel depends on this: we could have $N \in \mathR$. We need to distinguish three questions, %%@
about systems, quantities and values respectively.\\
\indent (1): One can ask whether this sequence has as a limit, in the sense of there being %%@
(as a mathematical entity) a natural well-defined infinite system $\sigma(\infty)$.\\
\indent (2): One can ask whether a sequence of quantities on successive systems, say $f(N): = %%@
f(\sigma(N))$, has a limit, which we might denote by $f(\infty)$. (Of course, the physical %%@
idea of each member of such a sequence will be in common, e.g. energy or momentum: but we %%@
distinguish the members by their being quantities on different (sizes of) system.)\\
\indent (3): Finally, one can ask whether a sequence of real number values of quantities on %%@
successive systems, say $v(f(N)): = v(f(\sigma(N)))$, has a limit.

\indent \indent Of course, question (3) is the most familiar. The notion of limit is the %%@
elementary notion from calculus, ${\rm{lim}}_{N \rightarrow \infty} \; v(f(N))$. Here a %%@
sequence of states, $s_N$ say, on the $\sigma(N)$ is to be implicitly understood, so as to %%@
define values for the quantities $f(N)$; but to simplify notation, I will for the most part %%@
not mention $s_N$, and indeed take states as understood. Recall also from the calculus that %%@
if a real sequence $v_N \in \mathR$ grows without bound, i.e. for any number $M$ the $v_N$ %%@
eventually remain greater than $M$, we write: ${\rm{lim}}_{N \raw \infty} \; v_N = \infty$. %%@
This is of course {\em different} from the idea (in  Section \ref{atlimit} below) of taking %%@
$\infty$ as a possible value of the  parameter or label on $v$, i.e. the idea of a sequence %%@
element $v_{\infty} \in \mathR$, which is after the denumerable sequence $v_N$, $N \in {\bf %%@
N}$.\footnote{My examples will of course need rather more calculus: e.g. we will need to %%@
distinguish between different kinds of convergence.}

 But we can also make sense of the first two questions.  As to (1): in both classical and %%@
quantum physics we can often define the limit of a sequence $\sigma(N)$. Some approaches %%@
individuate a system by its state-space, and then use infinite cartesian or tensor products %%@
(for the classical and quantum cases respectively). Other approaches individuate a system by %%@
its set (in fact: algebra) of quantities, and then define limit algebras. This leads to how %%@
we make sense of (2). The algebra of quantities usually has a mathematical structure (in %%@
particular a topology) that enables one to define the limit of a sequence of quantities (i.e. %%@
not just, as in (1), a limit of a sequence of their values).

Note  that the existence of an infinite system $\sigma(\infty)$ should {\em not} in general  %%@
be identified with the existence of a limit quantity $f(\infty)$, or even several such; nor %%@
with the sequence of values $v(f(N))$ having a limit in the ordinary calculus sense. Indeed, %%@
my first example (the method of arbitrary functions) will illustrate this. There will be no %%@
infinite system $\sigma(\infty)$, but the sequences of values $v(f(N))$ will each have a %%@
limit in the ordinary sense---in fact a finite one, viz. $\frac{1}{2}$. These limits are in %%@
no way `singular'. Yet there will be emergence, i.e. novel and robust behaviour.

There are also cases where there is (as a mathematical entity) an infinite system, and %%@
quantities defined on it whose values are the ordinary (in no way `singular') limits of %%@
values on the finite systems; and where there is emergence. My third example (superselection %%@
in quantum theory) will illustrate this.

And finally there are cases that suit the enthusiastic talk about singular limits! That is: %%@
cases where there is (as a mathematical entity) an infinite system, and quantities defined on %%@
it that take ``new'' values, i.e. values different from the limits of values on the finite %%@
systems. My second and fourth examples (fractals and phase transitions) will illustrate  %%@
this, the emergence being shown by these new values.\footnote{But as argued in footnote %%@
\ref{Wayne}, such discontinuous limits are not sufficient  for emergence.} Section %%@
\ref{atlimit} gives a few more details.

\subsection{The limit of a sequence vs. what is true at that limit}\label{atlimit}
The mathematical idea of this distinction is elementary. Recall that if we adjoin the number %%@
$\infty$ to the natural numbers ${\bf N}$, then we can consider sequences of real numbers %%@
$v_n \in \mathR$, with $n \in {\bf {N}} \cup \{ \infty \}$, i.e. sequences of order-type %%@
$\omega + 1$. For such sequences we can define the ordinary notion of limit, i.e. %%@
${\rm{lim}}_{n \in {\bf N}} \; v_n$; and then of course we recognize that there are  cases in %%@
which ${\rm{lim}} \; v_n := {\rm{lim}}_{n \in {\bf N}} \; v_n$ exists and is {\em not} equal %%@
to  $v_{\infty}$.
For $v_{\infty}$ means the $(\omega + 1)$-th member of the sequence---a quite different idea %%@
from the ordinary limit!

Section \ref{deflatelimits}'s idea of an infinite system $\sigma(\infty)$ allows us to apply %%@
this mathematical idea. We simply interpret adjoining the number $\infty$ to the set of  %%@
finite values of $N$ as considering the infinite system $\sigma(\infty)$, as well as the %%@
finite systems $\sigma(N)$.  I shall spell this out: first (a) for values of quantities, and %%@
then (b) for quantities themselves.

\noindent (a): {\em Values of quantities}: Suppose:
 (i) a sequence $v(f(N))$ of values of a quantity has a limit ${\rm{lim}}_{N \rightarrow %%@
\infty} \; v(f(N))$ as $N$ tends to infinity (as mentioned in Section \ref{deflatelimits}, a %%@
sequence of states $s_N$ is here understood, so that one might write $v(f(N), s_N)$).
And suppose also:\\
\indent (ii) there is also a well-defined infinite system $\sigma(\infty)$ on which:\\
\indent \indent the common physical idea of the various $f(N)$ makes sense and gives a %%@
natural well-defined limit quantity, which we might write as $f(\sigma(\infty))$ (on %%@
$\sigma(\infty)$); and on which \\
\indent \indent there is a natural well-defined limit state, $s$ say.

Then we need to distinguish:\\
\indent (i) the given limit of the values, ${\rm{lim}}_{N \rightarrow \infty} \; v(f(N)) %%@
\equiv {\rm{lim}}_{N \rightarrow \infty} \; v(f(N, s_N))$, from\\
\indent (ii) the value $v(f(\sigma(\infty), s)$ of the natural limit quantity %%@
$f(\sigma(\infty))$ in the  natural limit state, $s$.

\noindent (b): {\em Quantities}: For quantities themselves, rather than values, the point is %%@
in essence the same. The statement is a close parallel of that in (a): indeed, shorter since %%@
we refer only to quantities, not to values of quantities---albeit thereby more abstract. Thus %%@
suppose: (i) a sequence of quantities $f(N)$ has a limit, dubbed $f(\infty)$ in Section %%@
\ref{deflatelimits}. And suppose also: (ii) there is also a well-defined infinite system %%@
$\sigma(\infty)$ on which the common physical idea of the various $f(N)$ makes sense and %%@
gives a natural well-defined limit quantity, which we might write as $f(\sigma(\infty))$ (on %%@
$\sigma(\infty)$).

Then we need to distinguish:\\
\indent (i) the given limit, $f(\infty) := {\rm{lim}}_{N \rightarrow \infty} f(N)$, from\\
\indent (ii) the natural definition of the quantity $f(\sigma(\infty))$ on $\sigma(\infty)$.

\subsection{Justifying $N = \infty$}\label{2genl}
\subsubsection{Distinguishing straightforward from mysterious cases}\label{distmyst}
`Justifying $N = \infty$' is of course a shorthand! For---to sum up Sections %%@
\ref{deflatelimits} and \ref{atlimit}---we have just learnt to distinguish two numbers: %%@
although in some models they are both well-defined and equal, they need not be! Namely:\\
\indent (i): the limit ${\rm{lim}}_{N \rightarrow \infty} \; v(f(N))$ of a sequence of values %%@
(which limit might equal $\pm \infty$);\\
\indent (ii): the value $v(f(\sigma(\infty))$ of the natural limit quantity %%@
$f(\sigma(\infty))$ on the infinite system $\sigma(\infty)$.\\
So if we ask the question what justifies an ``$N = \infty$'' model or description of a %%@
system, for which $N$ is actually finite, we must allow that the answers may be different for %%@
different models. (Here and in the rest of this Subsection, I consider, for simplicity, just %%@
values of quantities as in (a) of Section \ref{atlimit}: not quantities themselves, as in (b) %%@
of Section \ref{atlimit}.)

We of course expect a straightforward justification for the two cases of `non-singular' %%@
limits, i.e. the cases:\\
\indent (a):  (i) is well-defined (though perhaps $= \pm \infty$), but there is no infinite %%@
system so that (ii) is ill-defined; (cf. my first example, the method of arbitrary %%@
functions);\\
\indent (b):  there is an infinite system, and (i) and (ii) are both well-defined and are %%@
equal; (cf. my third example, superselection in quantum theory).

Namely, we expect a justification in terms of convenience and correctness, along the lines:
\begin{quote}
(Straightforward Justification): The use of the infinite limit---i.e. the use of (i) for case %%@
(a), and the use of (i) = (ii) for case (b)---is justified, despite $N$ being actually %%@
finite, by its being mathematically convenient and empirically correct (up to the required %%@
accuracy).
\end{quote}
I shall develop and endorse this Justification in Section \ref{strtfwd}.

On the other hand, for `singular limits', i.e. cases where (i) and (ii) are both well-defined %%@
but are {\em not} equal, and (ii) rather than (i) is empirically correct, matters are surely %%@
not straightforward. Such cases seem mysterious. Faced with such a case, should we give up %%@
the assumption that $N$ is actually finite? But in some examples, e.g. where $N$ is the %%@
number of molecules in a sample of gas (as in my fourth example, phase transitions), this %%@
apparently amounts to giving up the atomic constitution of matter!\footnote{Thanks to John %%@
Norton for stressing this point---as a {\em reductio}, of course.} 

Nevertheless, some advocates of the philosophical importance of singular limits give up, or %%@
at least come very close to giving up, $N$'s finiteness. I take as examples, three quotes %%@
from Batterman (his italics):
\begin{quote} 
`a {\em physically} singular problem ... the ``blow-ups'' or divergences ... are the result %%@
of the singular nature of the physics' (2002, p. 56); `{\em real} systems exhibit {\em %%@
physical} discontinuities ... genuine physical discontinuities---real singularities in the %%@
physical system' (2005, pp. 235-236); `no de-idealizing story is possible even in principle' %%@
(2010, p. 17).
\end{quote} 
Agreed, in other passages, he holds back (thank goodness!):
\begin{quote} 
`in (2005), I do speak rather sloppily of genuine physical singularities. It is best to think %%@
instead in terms of some kind of genuine {\em qualitative} change in the system at a given %%@
scale' (2010, p. 22); `fluids {\em are} composed of a finite number of molecules' (2006, p. %%@
903); `water in real tea kettles consists of a finite number of molecules' (2010, p. 7; this %%@
quotation also occurs, together with its surrounding passage, at 2009, p. 9).
\end{quote} 

Note that this mysteriousness does not depend on (i) being well defined. If the $v(f(N))$ %%@
have no limit, not even $\pm \infty$, nevertheless the actual value is presumably $v(f(N_0))$ %%@
for some actual but unknown $N_0$. So (ii) being empirically correct means that $v(f(N_0)) %%@
\approx v(f(\sigma(\infty)))$ up to the required accuracy. But how can that be?

\subsubsection{Dissolving the mystery}\label{dissolvemyst}
I think the mystery can be dissolved, in two stages. (1): First, I will concede that to deny %%@
that $N$ is finite might be a {\em reasonable} move. But in all the examples I know, in %%@
particular in all of my examples, this move is  wrong. (Here, the important point is that in %%@
my second and fourth examples, fractals and phase transitions, this move is wrong: for as %%@
noted in Section \ref{distmyst}, my first and third examples of emergence have no mysterious %%@
`singular limit'.) So the more important stage will be the second one, (2) below: viz., that %%@
we need to consider quantities {\em other than} $f$. I turn to details.

(1): {\em Denying that $N$ is finite; other degrees of freedom}:--- I admit it can be %%@
reasonable to deny that $N$ is finite. But this means something less radical than denying %%@
atomism! Rather we conclude that the finite-$N$ model has not picked the right, or not {\em %%@
all} the right, degrees of freedom for understanding the system; and that the (model of the) %%@
infinite system has somehow `clued in to' the missing relevant degrees of freedom, as shown %%@
by its empirical correctness.

My fourth example, phase transitions in statistical mechanics, provides a putative example. %%@
Assuming that the correct description of a boiling kettle requires infinitely many degrees of %%@
freedom, it is reasonable to say that, since the kettle contains finitely many atoms, and so %%@
finitely many mechanical degrees of freedom, other degrees of freedom---e.g. of the %%@
electromagnetic field---must somehow be involved. Reasonable: but very programmatic! In fact, %%@
there is good evidence that the electromagnetic field is not involved in phase %%@
transitions---suggesting that the answer to the mystery lies elsewhere ...

(2): {\em Other quantities}:--- The mystery is an artefact of focussing on just one quantity %%@
($f$ in my notation). Once we consider appropriate other quantities (and maybe related %%@
mathematical notions), the mystery dissolves. Thus in my second and fourth examples (fractals %%@
and phase transitions), there are other quantities, for which (despite $f$'s singular limit) %%@
the finite-$N$ model, for large $N$, {\em is} close to the values given by the infinite %%@
model: and is thereby also empirically correct. In fact, these other quantities are `cousins' %%@
of the  quantity $f$ which we first considered. Thus the mystery will be dissolved by my %%@
second claim, (2:Before): namely,  we see a weak yet vivid version of the emergent behaviour %%@
{\em before} we get to the limit. Besides, I would claim---though I cannot defend it in this %%@
paper---that this is so in all of physics' similar cases (in particular, in Batterman's %%@
examples from optics, semiclassical mechanics and hydrodynamics).

Agreed, for me to say `there are other quantities or notions for which the finite-$N$ model %%@
is close to the infinite model' or `we see a weak version of emergence before the limit', is %%@
unsatisfyingly abstract. Indeed, it is dismayingly close to the mysterious {\em explanandum}, %%@
viz. that the infinite model is empirically correct! But I submit that at this very general %%@
level, these formulations are the best one can do. To see vividly how the mystery dissolves, %%@
one has to look at examples; cf. my second and fourth examples. But here is a simple %%@
mathematical example illustrating the issues---and that there really is no mystery! As we %%@
shall see, it is {\em not} just a mathematical toy: it models physical situations, especially %%@
phase transitions. 

Consider the sequence of real functions $g_N: \mathR \raw \mathR$, $N \in {\bf N}$,  defined %%@
by
\begin{eqnarray}
g_N(x) := -1 \;\; {\rm{iff}} \;\;  x \leq \frac{-1}{N} \; ; \\
g_N(x) := N x \;\; {\rm{iff}} \;\; \frac{-1}{N} \leq x \leq \frac{1}{N} \; ;  \\
g_N(x) := +1 \;\; {\rm{iff}} \;\; \frac{1}{N} \leq x  \; .
\end{eqnarray}
Thus $g_N(x)$ is constant and equal to $-1$ (respectively +1) for $x$ less than %%@
$\frac{-1}{N}$ (respectively: greater than $\frac{1}{N}$); and it increases linearly, with %%@
gradient $N$, over the interval $[\frac{-1}{N}, \frac{1}{N}]$, so that for all $N$, $g_N(0) = %%@
0$. Each $g_N$ is continuous; but the sequence has as its limit the  function $g_{\infty}$ %%@
given by
\begin{equation}
g_{\infty}(x) = -1 \;\; {\rm{iff}} \;\;  x < 0 \;\; ; \;\;
g_{\infty}(0) = 0 \;\; ; \;\;
g_{\infty}(x) = 1 \;\; {\rm{iff}} \;\;  0 < x \;\; ;
\end{equation}
which is discontinuous at $0$.\footnote{The convergence is pointwise not uniform: uniform %%@
limits of continuous functions are continuous.} So this limit is `singular' in the sense that %%@
continuity is lost. 

We can make this more formal by introducing a two-valued quantity $f_N$, $N \in {\bf {N}} %%@
\cup \{ \infty \}$  that encodes whether or not $g_N$ is continuous: $f_N := 0$ if $g_N$ is %%@
continuous and $f_N := 1$ if $g_N$ is discontinuous. Then we have: $f_N = 0$ for all finite %%@
$N \in {\bf N}$,  but $f_{\infty} = 1$. So in our (i)/(ii) notation (from Sections %%@
\ref{atlimit} and \ref{distmyst}), we have a case where (i) and (ii) are both well-defined %%@
but are unequal.

But there is no mystery here! There only seems to be a mystery if we look solely at the %%@
two-valued quantity $f_N$, whose values report that the limit is `singular', but which say %%@
nothing about how ``close'', for large $N$, the $g_N$ are to $g_{\infty}$.

Besides, there remains no mystery if we add some physical interpretation to the discussion.
Thus imagine that the values of $g_N$ in a neighbourhood of 0, or the slope of $g_N$ %%@
thereabouts, are part of a model of a system with $N$ degrees of freedom. $N$ varies, and is %%@
in general large; so that one considers the sequence of functions $g_N$.  Now imagine that %%@
for large $N$, it is hard to know the actual value $N_0$ of $N$ and-or hard to calculate the %%@
value of $g_N$, even if you know $x$. (Agreed: my example is so simple that only a dimwit %%@
could find the calculation hard! Such is the price of a simple example ...) In this %%@
situation, it obviously could be both (a) mathematically convenient and (b) empirically %%@
accurate---i.e. close enough to the predictions made by $g_{N_0}(x)$ for the actual $x$---to %%@
work with $g_{\infty}$.

 For as to (a): $g_{\infty}$'s being discontinuous need not make it inconvenient. Better the %%@
discontinuous $g_{\infty}$ that you can get a grip on, than the hard-to-know and-or %%@
hard-to-calculate $g_{N_0}$! And as to (b): as $N$ grows, the range of $x$ for which %%@
$g_{N}(x) \neq g_{\infty}(x)$ becomes arbitrarily small. Besides, for $x = 0$---which might %%@
be a physically significant argument---$g_{\infty}$ is completely accurate: i.e. for all $N$, %%@
$g_{N}(0) = g_{\infty}(0)$. In short: again, no mystery. There only seems to be a mystery if %%@
we look solely at $f_N$, and ignore the details about $g_N$ and $g_{\infty}$.

Finally, I stress that this mathematical example has two other features that make it a good %%@
prototype for my ``singular limit'' examples---i.e. my second and fourth, fractals and phase %%@
transitions; (hence my choice of it!). First: each example will have a two-valued quantity %%@
$f_N$, $N \in {\bf {N}} \cup \{ \infty \}$, with $f_N = 0$ for all $N \in {\bf N}$ and %%@
$f_{\infty} = 1$. In fact, this quantity simply records the presence or absence of the %%@
emergent novel property; with presence encoded by $f = 1$. So the jump in the value of $f$ %%@
corresponds to my claim (1:Deduce).

 Second: in phase transitions (my fourth example, Section \ref{phasetr}) there are physical %%@
quantities for finite models whose gradients grow without bound as $N \raw \infty$, just like %%@
this example's gradients of $g_N$ in a neighbourhood of 0. So the remarks here, about the %%@
unmysterious mathematical convenience and empirical accuracy of $g_{\infty}$, will %%@
apply---word for word!

Let me look ahead a little to Section \ref{phasetr}, especially Section \ref{needTD}.B (if %%@
only to placate {\em afficionados}!). Consider the phase transition of a ferromagnet at %%@
sub-critical temperatures, as described by the Ising model with $N$ sites (in two or more %%@
spatial dimensions). The magnetization behaves, as a function of the applied magnetic field, %%@
very like this example's $g_N$. Thus suppose our variable $x$ represents the value of the %%@
applied field (in a given spatial direction). Then to a good approximation, $g_N$ represents %%@
the average magnetization (in appropriate units). So as the applied field passes from %%@
negative to positive values, the ferromagnet's magnetization flips from -1 (i.e. alignment %%@
with the field in the negative direction) to +1 (alignment in the positive direction). But %%@
for larger $N$, the ferromagnet ``lingers longer'': the larger number of sites gives it more %%@
``inertia'' before the rising value of $x$ succeeds in flipping the magnetization from -1 to %%@
+1. (Here, my qualifying phrase `to a good approximation' refers to the Ising model's %%@
magnetization being a smooth function of the applied field (in fact given, in mean field %%@
theory, by the hyperbolic tangent function tanh), and so without sharp corners at $\pm 1/N$ %%@
like my $g_N$.) Thus the magnetic susceptibility, defined as the derivative of magnetization %%@
with respect to magnetic field, is, in the neighbourhood of 0, larger for larger $N$, and %%@
tends to infinity as $N \rightarrow \infty$: compare the gradients of $g_N$ in my example. %%@
Very similar remarks apply to liquid-gas phase transition, i.e. boiling. Here the quantity %%@
which becomes infinite in the $N \rightarrow \infty$ limit, i.e. the analogue of the magnetic %%@
susceptibility, is the compressibility, defined as the derivative of the density with respect %%@
to the pressure.\footnote{Cf. also Kadanoff (2010, p. 20, Figure 5); Menon and Callender %%@
(2011) is a discussion of phase transitions concordant with mine, here and in Section %%@
\ref{phasetr}. You may well ask: Is my mathematical example also a good prototype for %%@
dissolving the corresponding alleged mystery in physics' other `singular' limits,  e.g. from %%@
optics, semiclassical mechanics and hydrodynamics? My view is: Yes. For a masterly %%@
philosopher's survey of the first two cases, cf. Belot (2005, Sections 3, 4 and %%@
Appendix).\label{hunch}}

To sum up: I have dissolved the mystery about cases in which (i), i.e. the limit of the %%@
finite model, is not equal to (ii), the infinite model, and in which (ii) is empirically %%@
correct, by arguing that there are other quantities ($g$ rather than $f$, in my notation) for %%@
which (i) {\em is} close to (ii) (and so, also, empirically correct). I can therefore turn to %%@
elaborating and endorsing the Straightforward Justification which I announced in Section %%@
\ref{distmyst}: in short, mathematical convenience and empirical correctness. For I now %%@
maintain that it applies to all my four examples.

\subsubsection{Developing the Straightforward Justification}\label{strtfwd}
This Justification consists of two obvious, very general, broadly instrumentalist, reasons %%@
for using a model that adopts the limit $N = \infty$: mathematical convenience, and empirical %%@
adequacy (upto a required accuracy). So it also applies to {\em many} other models that are %%@
almost never cited in philosophical discussions of emergence and reduction. In particular, it %%@
applies to the many classical continuum models of fluids and solids, that are obtained by %%@
taking a limit of a classical atomistic model as the number of atoms $N$ tends to infinity %%@
(in an appropriate way, e.g. keeping mass density constant).

`Mathematical convenience and empirical correctness': merits that are so easy to state! But %%@
as all physicists know, and as echoed in the companion paper's discussion of good variables %%@
and approximation schemes: both can be very hard to attain---indeed, most of a physicist's %%@
work with a model is devoted to attaining them! But {\em if} they are attained by adopting %%@
the limit $N = \infty$, they surely justify using the limit. (At least, they do so, once we %%@
have disposed of any suspicious threat of mystery, such as refuting the atomic constitution %%@
of matter!)

Though the details vary widely among the countless models adopting some $N = \infty$ limit, %%@
this justification  involves two themes that are common to {\em so} many such models that I %%@
should articulate them. The first theme is abstraction from finitary effects. That is: the %%@
mathematical convenience and empirical adequacy of many such models arises, at least in part, %%@
by abstracting from such effects. Consider (a) how transient effects die out as time tends to %%@
infinity; and (b) how edge/boundary effects are absent in an infinitely large %%@
system.\footnote{As to (a), it is worth recalling the witty definition, attributed to %%@
Feynman, of that (invaluable but much-contested!) concept, `equilibrium': `the state the %%@
system gets into after the fast stuff [e.g. relaxation, transients] is finished and the slow %%@
stuff [e.g. Poincar\'{e} recurrence] has not yet started'. For apart from being witty, the %%@
mention of `the slow stuff' echoes Section \ref{unreal}'s warning (4:Unreal). That is: we %%@
should beware that for very large times (not just for very large $N$) physical theories and %%@
models often become unrealistic. And as to both (a) and (b), recall also footnote %%@
\ref{intermed}'s idea of intermediate asymptotics. Thus Feynman's witty definition should be %%@
revised along the lines `the state the system gets into after both the really fast stuff, and %%@
the intermediate stuff, is finished and ...'.\label{Feynmaneqbm}}

The second theme is that the mathematics of infinity is often much more convenient than the %%@
mathematics of the large finite. The paradigm example is of course the convenience of the %%@
calculus: it is usually much easier to manipulate a differentiable real function than some %%@
function on a large discrete subset of $\mathR$ that approximates it.\footnote{But smoothness %%@
is not everything! In some cases, as we saw with Section \ref{dissolvemyst}'s $g_{\infty}$, a %%@
discontinuous function is more convenient than a continuous one.} I shall just spell out two  %%@
advantages which are endemic. We can begin with the simple case where we consider just the %%@
limit of the values, i.e. (i) of Section \ref{atlimit}; so we set aside for the moment the %%@
infinite model, (ii) of Section \ref{atlimit}.

Thus consider a model in which the actual value of the relevant quantity for  realistic, i.e. %%@
large but finite, $N$, say $N = 10^{23}$---the value $v(f(10^{23}))$ in Section %%@
\ref{atlimit}'s notation, taking the state as understood---is negligibly close to the limit %%@
${\rm{lim}}_{N \rightarrow \infty} \; v(f(N))$. And let us assume  that the value will remain %%@
close as $N$ grows: so the values obey $v(f(10^{23})) \approx v(f(10^{46})) \approx %%@
v(f(10^{69}))$ etc. Working with the limit rather than the actual value promises two %%@
advantages. (Here of course we set aside Section \ref{unreal}'s warning (4:Unreal), that for %%@
many models, the values for vastly larger $N$ will eventually be unrealistic.)

The first is that it may be much easier to know, or at least estimate, the limit's value  %%@
than the actual value---not least because of the first theme, the abstraction from finitary %%@
effects. And {\em ex hypothesi}, working with it involves a negligible inaccuracy about the %%@
actual value.

The second advantage is more  theoretical, and will lead back to Section \ref{atlimit}'s %%@
(ii), i.e. the value of a limit quantity on an infinite system. The idea here is that for %%@
most models and quantities $f$, there is, for a fixed $N$, not a single value $v(f(N))$, but %%@
a range of values, to be considered. That is: $v(f(N))$ is a function of some other variable %%@
which has so far been suppressed in my notation. And to make this function easily %%@
manipulated, e.g. continuous or differentiable so that it can be treated with the calculus, %%@
we often need to have each {\em value} of the function be defined as a limit (namely, of %%@
values of another function).

Continuum models of solids and fluids provide paradigm examples of this. For example, %%@
consider the mass density varying along a rod, or within a fluid. For an atomistic model of %%@
the rod or fluid, that postulates $N$ atoms per unit volume, the average mass-density might %%@
be written as a function of both position $\bf x$ within the rod or fluid, and the %%@
side-length $L$ of the volume $L^3$ centred on $\bf x$, over which the mass-density is %%@
computed: $f(N, {\bf x}, L)$. Now the point is that for fixed $N$, this function is liable to %%@
be intractably sensitive to $\bf x$ and $L$. In particular, if atoms are or contain %%@
point-particles the function will jump when $L$ is varied so as to include or exclude one %%@
such particle. That is: it will not be continuous  in $\bf x$ and $L$. But by taking  a %%@
continuum limit $N \raw \infty$, with $L \raw 0$ (and atomic masses going to zero %%@
appropriately, so that quantities like density do not ``blow up''), we can define a %%@
continuous, maybe even differentiable, mass-density function $\rho(x)$ as a function of %%@
position---and then enjoy all the convenience of the calculus.

So much by way of showing in general terms how the use of an infinite limit $N = \infty$ can %%@
be justified---but not mysterious! At this point, the general philosophical argument  of this %%@
paper is complete! The subsequent Sections present my examples. It will be clear that each %%@
example represents a large field of study. So to save space, I will have to be brutally %%@
brief, both about the examples' details and about references.

\section{The method of arbitrary functions}\label{maf}
My first example is the method of arbitrary functions in probability theory. It is a %%@
venerable tradition, initiated by Poincar\'{e} in his {\em Calcul de Probabilities} (1896), %%@
and developed by many authors including Borel, Fr\'{e}chet and Hopf. Recent presentations %%@
include Engel (1992) and Kritzer (2003); and von Plato (1983, 1994, pp. 168-178) summarizes %%@
the history. But until recently it seems to have been largely neglected in the philosophy of %%@
probability, despite its offering an attractive way to reconcile non-trivial probabilities %%@
(i.e. probabilities that are neither 0 nor 1) with determinism at an `underlying' level---and %%@
despite being the topic of Reichenbach's dissertation!\footnote{I say `until recently' for %%@
two reasons. First: Strevens (2003) has revived the main idea; though he is wary of the %%@
philosophical value of theorems about limiting behaviour, which figure prominently in the %%@
tradition and which I will emphasize. For assessments of Strevens, cf. Colyvan (2005) and %%@
Werndl (2010). Second: some recent papers revive the main idea: Sober (2010), Frigg and %%@
Hoefer (2010), Myrvold (2011).\label{strevens}}

The main idea of the method is best introduced by an example, and I will follow Poincar\'{e} %%@
(and most discussions) in choosing a roulette wheel, with alternating arcs of red and black %%@
(Section \ref{poinc}). Thus we will be concerned with the probability that the wheel stops %%@
with a red (respectively, black) arc opposite a pointer. For this example, the main idea will %%@
be that under certain assumptions, this probability tends to 0.5, as the number $N$ of arcs %%@
goes to infinity---whatever the details of the spinning and slowing of the wheel. Section %%@
\ref{poinc} will also discuss how this result can be generalized. Then in Section %%@
\ref{emergeequiprob}, I describe how this equiprobability in the limit $N \raw \infty$ counts %%@
as emergent behaviour in my sense, and how it illustrates my claims, (1:Deduce) etc. 

\subsection{Poincar\'{e}'s legacy}\label{poinc}
\subsubsection{Poincar\'{e}'s roulette wheel}\label{poincrw}
Suppose that a roulette wheel with arcs of red and black is spun many times, eventually %%@
coming to a stop with a red or a black arc opposite a pointer. We suppose that it is spun %%@
using various unknown initial conditions, i.e. initial positions relative to the pointer and %%@
initial angular velocities; and that it is slowed and eventually stopped by some unknown %%@
regime of friction. If this is all we know, we can conclude essentially nothing about the %%@
long-run frequency (or probability, in any sense) of it stopping at Red (i.e. with a red arc %%@
opposite the pointer). For the variety of initial conditions and the regime of friction, %%@
taken together, amount to an unknown profile of biassing. This profile might be expressed as %%@
a function giving, for each arc, the probability of the wheel stopping there. And for all we %%@
have so far assumed, this function might make Red very probable (frequent)---or very %%@
improbable (infrequent). 

But suppose we also assume that: \\
\indent \indent (i): there are {\em very many alternating} arcs of red and black;\\
\indent \indent (ii): whatever the unknown profile of biassing might be, it favours and %%@
disfavours {\em large} segments, i.e. segments each of which contains many red and many black %%@
arcs; \\
\indent \indent (iii): within one of these large segments, the bias is not too ``wiggly'' in %%@
the sense that two adjacent arcs get nearly equal biasses.\\
\indent {\em Then} we can be confident that the long-run frequency of Red (and of Black) is %%@
about 50$\%$. For assumptions (i) to (iii) mean that if the profile is expressed as a %%@
probability function, each of its peaks (corresponding to a favoured segment) contains many %%@
red {\em and} many black arcs---and so do each of its troughs (corresponding to a disfavoured %%@
segment). Thus the contribution of any peak to the overall probability (or frequency) of %%@
stopping at Red will be about equal to the peak's contribution to the probability of stopping %%@
at Black; and similarly for any trough. So summing over all the peaks and troughs, the %%@
honours will be about even between Red and Black: there will be approximate equiprobability. %%@
To sum up: (i) to (iii) imply that the idiosyncrasies of the biassing profile get washed out.

This is a beautiful and compelling idea; (originally due, apparently, to an 1886 book by von %%@
Kries; cf. von Plato 1983, p. 38; 1994, p. 169). Expressing it in general and probabilistic %%@
terms, we expect the following. Let a sample space $(X, \mu)$ be partitioned into two %%@
subsets, say $R$ and $B$, in a very ``intricate'' or ``filamentous'' way.  Then for any %%@
probability density function $f$ that is not too ``wiggly'' (say: whose derivative is %%@
bounded: $\mid f' \mid < M$) the probabilities of $R$ and $B$ are about equal:
\be
\int_R \; f d \mu \; \approx  \; \int_B \; f d \mu \; \approx \frac{1}{2} .
\label{mafeq}
\ee
And we expect: that, for any bound $M$ on the derivative of the density $f$, as the partition %%@
becomes more intricate or filamentous,  the difference from exact equiprobability (and so to %%@
both probabilities equalling $\frac{1}{2}$) will tend to 0.

Indeed, Poincar\'{e} (1912, p. 148ff.) turned this idea into a theorem, for a simple model of %%@
the roulette wheel. So we take $X$ to be the circle $[0, 2\pi]$, and the intricate %%@
partitioning of $X$ to be the division into $N$ equal intervals, labelled alternatingly `red' %%@
and `black'. We assume the distribution of the point $x \in X$ at which the wheel stops (i.e. %%@
which is eventually opposite the pointer) is given by a probability density function $f: [0, %%@
2\pi] \raw \mathR$. We assume that $f$ is differentiable, and its derivative is bounded by %%@
$M$, i.e. $\mid f' \mid < M \in \mathR$. This of course makes precise assumptions (ii) and %%@
(iii) above.\footnote{We might also assume that the support of $f$ intersects all $N$ cells %%@
of the partition. This is one way (among several) to represent the natural requirement that %%@
the wheel is spun fast enough, at least sometimes, to prevent it stopping after just a few %%@
arcs have passed the pointer.} Then Poincar\'{e} showed:
\begin{quote}
For any $M \in \mathR$, for all density functions $f$ with derivative bounded by $M$, $\mid %%@
f' \mid < M$: as $N$ = the number of arcs goes to infinity:\\
$\int_R \; f \; d \mu \; \equiv \; {\rm{prob(Red)}} \; \raw \frac{1}{2} \; \;  ; \; \; %%@
{\rm{and}} \;  \; \int_B \; f \; d \mu \; \equiv \; {\rm{prob(Black)}} \; \raw \frac{1}{2}$.
\end{quote}
To sum up: any biassing profile, no matter how wiggly, i.e. sensitive to the wheel's angular %%@
position (no matter how large $M$), can be washed out, so as to give equiprobability up to an %%@
arbitrary accuracy, by a sufficiently intricate partition, i.e. by a sufficiently large $N$.

\subsubsection{Generalizations: statistical stability}\label{poincgeneralzd}
Subsequently, Poincar\'{e}'s theorem was generalized in two main ways. The first way was %%@
historically earlier and is less connected to later developments, especially of probabilistic %%@
methods in the study of dynamical systems. But it is easier to report since its conception of %%@
the parameter $N$ is very close to Poincar\'{e}'s original: it measures the fineness of the %%@
partition of the sample space. In the second way, on the other hand, one takes a different %%@
limit, usually depending on the details of the dynamical system concerned.

I will now sketch both ways. But as regards illustrating my claims about emergence, I should %%@
stress the following points.\\
\indent (a) The illustrations do not need any of these generalizations; so the reader %%@
uniniterested in probability theory can now {\em skip} to Section \ref{emergeequiprob}.\\
\indent (b) The first  way leads to illustrations of my claims that are exactly parallel to %%@
the original illustration given by Poincar\'{e}'s theorem: a happy circumstance, since it %%@
supports my view that my claims have a  wide validity.\\
\indent (c) The second way also illustrates my claims. But because a different, and even %%@
system-dependent, limit is taken, these illustrations are rather different from the %%@
Poincar\'{e} original. So to save space, I will not pursue the details.\\
\indent (d) Poincar\'{e}'s theorem and its generalizations (in both ways) are very suggestive %%@
for the philosophy of probability. As we will see, they hint that even with an underlying %%@
determinism, taking an appropriate limit can define non-trivial probabilities that are %%@
``objectively correct''. But again, to save space, I must make a self-denying ordinance about %%@
this.

The first way generalized the assumptions of the model of the wheel, and adapted them to %%@
other chance set-ups. At first the conditions on the initial density function $f$ were %%@
weakened, by authors such as Borel and Fr\'{e}chet. In short, Borel assumed merely that $f$ %%@
was continuous; and Fr\'{e}chet merely that it was Riemann-integrable. 

As to other chance set-ups, one paradigm example, which had the merit of extending the method %%@
of arbitrary functions to densities of more than one variable, was Buffon's needle. In this %%@
problem a person throws a needle of length $l$ on to a table on which a pattern of parallel %%@
lines at a distance $d$ ($d > l$) has been ruled. One asks: what is the probability that the %%@
needle lands so as to intersect one of the lines? The elementary treatment assumes that the %%@
point where the centre of the needle lands has a uniform probability density (in the interval %%@
$[0,d]$ for simplicity); and similarly that the angle between the needle and the lines is %%@
uniformly distributed. It then follows by an elementary argument that the probability of %%@
intersection is $2l/d\pi$. 

But it is more realistic to assume that there is some unknown (``arbitrary'') density %%@
function, perhaps peaked near the centre of the table, for the point where the centre of the %%@
needle lands.\footnote{Similarly, one might say, for the angle at which the needle lands. But %%@
I will not pursue how to relax this assumption.} Can we again apply von Kries' and %%@
Poincar\'{e}'s idea that a more and more intricate partition of the sample space (here, the %%@
table) will wash out the influence of the peaks (and troughs) of the unknown density %%@
function? Yes! Borel indicated, and Hostinsk\'{y} showed in detail, that one can recover the %%@
familiar answer, $2l/d\pi$, by taking the limit as the number $N$ of lines on the table goes %%@
to infinity. For this theorem, Hostinsk\'{y} assumed that the partial derivatives of the %%@
density function exist, are continuous and are bounded. And he takes the limit, $N \raw %%@
\infty$, while (i) the table size is constant, so that the lines' separation $d$ goes to %%@
zero, and (ii) the ratio $l/d$ is constant.

At this point, we must concede that the theorems reported so far have an obvious limitation: %%@
the limit, $N \raw \infty$, is unrealistic. The number of arcs on a roulette wheel, and the %%@
number of parallel lines on any table, is in fact fixed. (So this sense of being unrealistic %%@
is more straightforward, and in practice arises for much smaller $N$, than the idea of %%@
running up against the atomic constitution of matter, involved in my (4:Unreal) of Section %%@
\ref{unreal}.) Can we respect this fact, and yet still apply our initial idea that an %%@
intricate partition of the sample space washes out the influence of the peaks and troughs of %%@
an unknown density function?

As I see matters, there are two broad strategies one can adopt. Both are important; and %%@
fortunately, they are compatible. The first strategy is piecemeal, and takes no limits. One %%@
models each chance set-up as realistically as one wishes or is able to; and then calculates, %%@
perhaps numerically, how wiggly (in some sense) the density function could be, while yielding %%@
approximately the probabilities we observe and-or desire---e.g. for the roulette wheel, %%@
equiprobability of Red and Black. This strategy is obviously sensible; and in Section %%@
\ref{equiprobbefore} we will see how it illustrates my claim (2:Before). But for now, I turn %%@
to the other strategy.

This is what I called the `second way' of generalizing Poincar\'{e}'s theorem. In short: to %%@
derive the observed or desired probabilities, a {\em different} limit is taken. This strategy %%@
can also be piecemeal: the details of the chance set-up suggest what limit to take. I shall %%@
briefly report two impressively neat examples of this: Hopf's analysis of the roulette wheel, %%@
and Keller's analysis of coin-tossing. Then I shall report how this second way leads to the %%@
important idea of {\em statistical stability}.

Hopf's idea is that for a roulette wheel with a fixed number $N$ of arcs, the equiprobability %%@
of Red and Black will follow from allowing higher and higher initial angular velocities. Thus %%@
the basic insight is that even with $N$ fixed, a higher initial angular velocity implies that %%@
the width of an interval of velocities that lead to a specific arc stopping opposite the %%@
pointer is {\em smaller}. Or to make the same point at the opposite extreme: with just a few %%@
arcs (say, two!), and initial angular velocities so small that at most one rotation occurs, %%@
even a ham-fisted croupier can fix the wheel, i.e. guarantee stopping at Red, or at Black. In %%@
more detail: Hopf considers the total angle $\theta \in [0, \infty]$ through which some %%@
fiducial point on the wheel's circumference turns before the wheel stops. Higher initial %%@
angular velocities will make $\theta$ larger; and Red or Black is determined by $\theta$ mod %%@
$2\pi$.
The regime of spinning and friction is summarized in an unknown density function $f$ on the %%@
initial angular velocity $\omega$, with bounded support. But higher velocities are considered %%@
by translating $f$ by a constant $C$, i.e. by defining $f^*(\omega) := f(\omega - C)$; and by %%@
letting $C \raw \infty$. Hopf also allows the frictional force (the braking) to depend, not %%@
only on the present angular velocity, but also on the angle so far turned through; that is, %%@
he allows for an unbalanced wheel. Hopf then proves that as $C \raw \infty$, the distribution %%@
of $\theta$ mod $2\pi$ tends to being uniform on $[0,2\pi]$.

Keller gives a broadly similar analysis of coin-tossing (1986; developed by Diaconis et al. %%@
2007). He takes the coin to be a circular lamina which is initially horizontal: it is tossed %%@
in a vertical line with an initial angular velocity $\omega$ and initial vertical velocity %%@
$u$, and falls under gravity onto a horizontal table where it settles with either Heads or %%@
Tails facing upward. Like Poincar\'{e}s or Hopf's wheel, the sample space of initial %%@
conditions is intricately partitioned into subsets that lead eventually, and %%@
deterministically, to Heads or to Tails. But like Buffon's needle, the sample space is %%@
two-dimensional. It is the positive quadrant of the $(\omega, u)$-plane. So the probabilities %%@
of Heads and Tails are given by integrating over the Heads and Tails subsets, respectively, %%@
an unknown density function $f(\omega, u)$, which Keller takes to be continuous. 

Keller shows that the pattern of Heads and Tails subsets is like a ``hyperbolic zebra''. Each %%@
subset is a thin strip lying along one of a series of hyperbolas, i.e. curves like $\omega = %%@
n K/u$ with $n$ a natural number. Besides, Heads strips alternate with Tails strips; and for %%@
higher values of $\omega$ and $u$ (i.e. as we move North-East in the positive quadrant), the %%@
strips become thinner. This means that, in the now-familiar way, the  integral of $f$, for %%@
Heads or for Tails, over these higher values becomes less sensitive to wiggles in $f$. That %%@
is: as the support of $f$ (or even just $f$'s ``preponderant weight'') tends ``North-East'', %%@
Heads and Tails tend towards being equiprobable---whatever the density function.

Agreed, you might object that these analyses of Hopf and Keller, though neat, are again %%@
unrealistic. No roulette wheel is spun, and no coin is tossed, arbitrarily fast! But the %%@
reply is clear. It has two parts. Analyses like Hopf's and Keller's can give information %%@
about the speed of convergence towards their limit; and this can reassure us that realistic %%@
initial conditions lead to the desired probabilities (here: equiprobability), up to a good %%@
accuracy, for a wide class of density functions. Here of course we return to two previous %%@
themes:\\
\indent (i) in general terms, the two merits of Section \ref{strtfwd}'s Straightforward %%@
Justification of taking a limit: mathematical convenience and empirical success; and\\
\indent (ii) specifically, the value of modelling without taking a limit, i.e. the first %%@
strategy above, and my claim (2:Before). Recall my remark above that the two strategies are %%@
compatible.

Finally, Hopf's and Keller's analyses prompt the idea of {\em statistical stability}, which %%@
has been very important for the probabilistic study of dynamical systems. I will not go in to %%@
the measure-theoretic technicalities (about absolute continuity and types of convergence) %%@
that are needed for an exact definition, but just convey the main idea. (This occurs, under %%@
the label `statistical regularity' in Hopf's own analysis of the roulette wheel.) The general %%@
scenario is that we are given: (i) two probability spaces $(X, \mu)$ and $(Y, \nu)$, i.e. %%@
$\mu, \nu$ are probability measures on appropriate fields of subsets of $X, Y$ respectively; %%@
(ii) a family of maps $F_{\lambda}: X \raw Y$, labelled by a parameter $\lambda \in \mathR$ %%@
or perhaps $\in {\bf N}$. Thus in our examples above, $X$ was the space of initial conditions %%@
and $Y$ was the two element space $\{$ Red, Black $\}$ or $\{$ Heads, Tails $\}$; and each %%@
$F_{\lambda}$ is a deterministic map sending an initial condition $x \in X$ to an outcome $y %%@
\in Y$. 

Returning to the general scenario: $\mu_{\lambda} := \mu \circ F^{-1}_{\lambda}$ is a %%@
probability measure on $Y$, and we can ask whether there is a measure on $Y$ to which %%@
$\mu_{\lambda}$ converges as $\lambda \raw \infty$: or even a measure on $Y$ to which %%@
$\mu_{\lambda}$ converges, for all $\mu$ on $X$ in some suitable class. If so, we say the %%@
family $F_{\lambda}$ is statistically stable. In studying complicated, even ``chaotic'', %%@
deterministic systems, this idea has an important special case: namely, $X = Y, \mu = \nu, %%@
\lambda \in {\bf N}$ and the family $F_{\lambda}$ arises just by iterating a map $T: X \raw %%@
X$, i.e. $F_{\lambda} := T^{\lambda}$ represents a discrete-time evolution. In this case, the %%@
limit measure, $\mu^*$ say, characterizes the long-time statistical behaviour of the system. %%@
In particular, it is readily shown to be invariant under the time-evolution. That is, $T$ %%@
induces an evolution $P_T$ on measures (and their densities) in the natural way, and we have: %%@
$P_T(\mu^*) = \mu^*$.

\subsection{The claims illustrated by emergent equiprobability}\label{emergeequiprob}
I turn to describing how the limiting probabilities of Section \ref{poinc} count as emergent %%@
behaviour in my sense, and how they illustrate my claims (1:Deduce), (2:Before) and %%@
(3:Herring) (listed in Section \ref{prosp}). As I announced, I will for simplicity emphasize %%@
the original Poincar\'{e} theorem, stated at the end of Section \ref{poincrw}. But it will be %%@
clear how the claims are also illustrated by the generalizations given in Section %%@
\ref{poincgeneralzd}, including the closing idea of an invariant limit measure $\mu^*$.

The illustrations unfold immediately, once we stipulate that the limiting probabilities are %%@
to be the emergent behaviour. For me, this means behaviour that is novel or surprising, and %%@
robust, relative to a comparison class. As discussed in the companion paper, this class is %%@
liable to be fixed contextually, and even to be vague or subjective---but nevermind, since %%@
there does not need to be an exact meaning of `emergence'. Here I concede that the limiting %%@
probabilities, especially the equiprobability of Red and Black, or Heads and Tails, are not %%@
novel or surprising---though  I submit that it is surprising that one can deduce them from an %%@
arbitrary density function. In any case, they are robust in a vivid sense: the whole point of %%@
the method of arbitrary functions is that they are invariant under a choice of a density %%@
function from a wide class.

\subsubsection{Emergence in the limit: with reduction---and without}\label{emergelimitMAF}
As to (1:Deduce): we have `reduction as deduction'  in as strong a sense as you could %%@
demand---provided we take the limit. Thus for Poincar\'{e}'s theorem, we take $T_t$ to be %%@
just the statement of equiprobability in the limit of infinite $N$, and $T_b$ to be a
model of the wheel, including enough measure theory and calculus to cover both: (i) the %%@
postulation of various possible density functions $f$ on $[0, 2\pi]$; and (ii) consideration %%@
of the infinite limit $N \raw \infty$. And similarly for Section \ref{poinc}'s other %%@
examples.

(1:Deduce) also concerns ``the other side of the coin'': how the emergent behaviour, here %%@
equiprobability,  is not deducible, if we do {\em not} take the limit but instead confine %%@
$T_b$ to finite $N$. This also is illustrated by Section \ref{poinc}. Thus in particular, for %%@
Poincar\'{e}'s roulette wheel: For any finite $N$, no matter how large, equiprobability will %%@
fail, as badly as you may care to require, for a sufficiently ``wiggly''  density function, %%@
i.e. a sufficiently position-sensitive biassing regime. That is, we have:
\begin{quote}
For all $\varepsilon > 0$, for all positive integers $N$, there is $M \in \mathR$ and a %%@
density function $f$ with  $\mid f' \mid < M$ such that: $\;\;
\int_R \; f \; d \mu \; \equiv \; {\rm{prob(Red)}} \; > 1 - \varepsilon.$
\end{quote}
So here is emergence {\em without} reduction to a weaker finitary $T_b$. Since this weaker  %%@
$T_b$ is a salient theory, one can be tempted to speak of irreducibility. Similarly for %%@
Section \ref{poinc}'s other examples.

It is worth displaying the two sides of (1:Deduce)'s ``coin''---equiprobability's %%@
deducibility in the limit, and its non-deducibility before---in terms of a shift of %%@
quantifiers. Thus the ``form'' of Poincar\'{e}'s theorem is:
\begin{quote}
$\forall \varepsilon > 0, \forall M \in \mathR$, $\forall f$ with  $\mid f' \mid < M$,  %%@
$\exists N$ s.t. $\forall N^* > N$:
$\mid \int_{R; N^* \mbox{arcs}} \; f \; d \mu - \frac{1}{2} \mid \; < \; \varepsilon;$
\end{quote}
while ``the other side of the coin'' is:
\begin{quote}
$\forall \varepsilon > 0, \forall N, \exists M \in \mathR$, and  $f$ with  $\mid f' \mid < %%@
M$, s.t.: 
$\mid \int_{R; N \mbox{arcs}} \; f \; d \mu - \frac{1}{2} \mid \;\; > \; \varepsilon.$
\end{quote}
One can easily check that in Section \ref{poinc}'s other examples, including Buffon's needle, %%@
Hopf's roulette wheel and Keller's tossed coin, the two sides of (1:Deduce)'s ``coin'' %%@
involve a similar quantifier-shift.

Finally, I stress the point announced in Sections \ref{peace} and \ref{deflatelimits}: that %%@
the limits we are concerned with are in no way singular---so a singular limit is not %%@
necessary for emergence. Nor is there any infinite system corresponding to $N = \infty$ (i.e. %%@
$\sigma(\infty)$ in Section \ref{deflatelimits}'s notation). For the roulette wheel, that %%@
would mean a division of $[0, 2\pi]$ in to a denumerable number of equal-length segments! And %%@
similarly for the other limits: e.g. Hopf's roulette wheel spun, or Keller's coin tossed, %%@
with an infinite initial angular velocity.\footnote{I said there can be no division of $[0, %%@
2\pi]$ in to a denumerable number of equal-length segments. No sooner said than doubted---as %%@
so often in philosophy. I am grateful to Alan Hajek for pointing me to Edward Nelson's %%@
adaptation of the ideas of non-standard analysis to probability theory; cf. Nelson (1987, %%@
especially Chapters 4 to 7).}

\subsubsection{Emergence before the limit}\label{equiprobbefore}
(2:Before) claims that before the limit, there is emergence in a weaker but still vivid %%@
sense. Here the weaker sense is approximate rather than exact equiprobability, for some %%@
realistic model of the roulette wheel (or other chance set-up). So we already saw in Section %%@
\ref{poincgeneralzd} how the method of arbitrary functions illustrates this claim: namely, in %%@
the discussion of the finite parameter case, both (i) as a first strategy for defending %%@
Poincar\'{e}'s roulette wheel and (ii) as a reply to the parallel objection to Hopf or %%@
Keller, that no wheel is spun, no coin is tossed, arbitrarily fast. For both (i) and (ii), we %%@
calculate, perhaps numerically, how wiggly (in some sense) the density function could be, %%@
while yielding approximately the probabilities we observe and-or desire---e.g. for the %%@
roulette wheel, equiprobability of Red and Black.  

Speaking of desire raises issues of engineering: indeed, of the profitability of casinos.  
We know that casinos manage to get profitably close to equiprobability, with some small %%@
number, $N \approx 50$, of arcs. And we surmise that even if they had a worryingly wiggly %%@
$f$, they could get profitably close to equiprobability by putting $N$ up to say about 200; %%@
or---following Hopf's idea---by spinning the wheel, on average, some two to three times %%@
faster. 
Here we meet the multi-faceted, even interest-relative, even subjective, question: how close %%@
is close enough? `Close enough for all practical purposes': but what exactly are the %%@
practical purposes? How wiggly an $f$ need the casino guard against?

But I submit that this is a question for casino-owners---who can no doubt pay staff well %%@
enough to answer it accurately for them. At our (typically philosophical!) level of %%@
generality, we do not need to try and answer it. For us, it is enough that given a resolution %%@
of this and similar questions, including vaguenesses, we get a notion of approximate %%@
equiprobability, which can indeed be deduced from a $T_b$ with parameters that are not only %%@
finite, but also realistic. In particular, $T_b$ can imply profitable---for the gamblers: %%@
indiscernible---closeness to equiprobability, using some $N \approx 50$ arcs on the roulette %%@
wheel, and an initial velocity of some $10 \pi$ to $30 \pi$ radians per second (5 to 15 %%@
revolutions per second).

\subsubsection{Supervenience is a red herring}\label{redherring1}
I turn to my third, ancillary, claim (3:Herring). Namely: although various supervenience %%@
theses are true, they yield little or no insight into emergence, or more generally, into %%@
``what is going on'' in the example.

This is well illustrated by Poincar\'{e}'s roulette wheel, and Section \ref{poinc}'s other %%@
examples. For any sequence of spins of the wheel, with any number $N$ of arcs, and any regime %%@
governing its initial velocities, the frequency of Red is of course determined by, %%@
supervenient upon, all the microscopic details of the wheel and its many %%@
spinnings.\footnote{In Section \ref{poinc}, we assumed, for the most part implicitly, that %%@
these details were based on classical mechanics. But the same supervenience thesis would hold %%@
if we assumed instead that they were based on quantum theory. At least, this is so if we set %%@
aside the quantum measurement problem, which threatens to deny us any definite macroscopic %%@
events. The companion paper discusses some dangers in the idea of supervenience on the %%@
microscopic details, ``whatever they might be''.} This supervenience thesis holds for a %%@
finite sequence of spins; or an infinite one, with frequency defined as limiting relative %%@
frequency. And there are analogous supervenience theses for probability, rather than %%@
frequency: the probability of Red is determined by the details of the wheel, especially the %%@
choice of probability density function. Similarly of course, for coin-tosses, and the %%@
frequency or probability of Heads.

I submit that these supervenience theses, whether for frequency or probability, shed no light %%@
on the matters at hand. For they make no connection with the basic idea of the method of %%@
arbitrary functions: that intricate partitions of a sample space can wash out the peaks and %%@
troughs of an unknown density function, and secure robust probabilities. This is a good %%@
illustration of my general reasons (in Section \ref{prosp}) for supervenience theses' %%@
irrelevance: that they make 
no connection between their idea of a variety, perhaps even infinity, of ways to have the %%@
higher-level property $P$, and the limit processes on which the example turns. Thus here, $P$ %%@
is the property that a frequency or probability of Red (or of Heads) is $\frac{1}{2}$, or is %%@
the property of two events being equiprobable; and the example's limit processes are the %%@
number of arcs, or the initial velocities, going to infinity, so as to implement the basic %%@
idea of washing out peaks and troughs. Or we can eschew the limit and use only finite %%@
parameters, as in (2:Before). But again, these supervenience theses shed no helpful %%@
light.\footnote{A {\em caveat}. I agree that these supervenience theses are relevant to the %%@
philosophy  of probability, especially for an empiricist. For example: if we maintain that %%@
the empiricist should accept the model's microscopic details, say because they are %%@
``occurrent'', then the supervenience theses for frequencies support the idea that they %%@
should also accept frequencies---as a metaphysical free lunch, as people say. But in (d) at %%@
the start of Section \ref{poincgeneralzd}, I foreswore the philosophy of probability: for %%@
some discussion in the context of the method of arbitrary functions, cf. the papers by Frigg %%@
and Hoefer, and Myrvold cited in footnote \ref{strevens}.}

\section{Fractals}\label{frac}
My second example is fractals, or rather, one small aspect of this large field: namely, the %%@
idea that a set of spatial points, i.e a subset of $\mathR^n$ ($n = 1, 2, ...$), can have a %%@
dimension that is {\em not} an integer. As we shall see, one can define various notions of %%@
dimension; and much of the discussion and results carry over to spaces more general than %%@
Euclidean space $\mathR^n$. However, I will emphasise one notion of dimension, {\em scaling %%@
dimension} (also known as: {\em similarity dimension}), and confine myself to $\mathR^n$. %%@
Even a very short introduction to this topic (Section \ref{scaldim}) will be enough to %%@
illustrate my claims. For my first three claims, details are in Section \ref{emergedim}. The %%@
discussion is similar to that in Section \ref{emergeequiprob}.\footnote{I should mention a %%@
reason for restricting attention to Euclidean space $\mathR^n$. Namely: Euclidean geometry %%@
admits similarity (of triangles and other figures), while non-Euclidean geometries in general %%@
do not; and on our approach, the definition of fractals needs the idea of similar figures. %%@
Section \ref{nature?} will return to this point.\label{onlyeucsimilar}} 

But as I mentioned in Section \ref{unreal}, I propose for fractals to also discuss my fourth %%@
claim (4:Unreal): that for large but finite $N$, the example becomes unrealistic---for %%@
reasons that are usually ignored in discussions of emergence. I do this in Section %%@
\ref{nature?}. This will mean that in Sections \ref{scaldim} and \ref{emergedim}, the pure %%@
mathematics of dimension in Euclidean geometry will be prominent: the empirical world will %%@
come to the foreground only in Section \ref{nature?}. For in this fractals example, large $N$ %%@
corresponds to very small length-scales; so that here, (4:Unreal) amounts to a `No' answer to %%@
the question `Is fractal geometry the geometry of nature?' In other words: (4:Unreal) denies %%@
that  fractal descriptions of physical objects are literally true: a denial which my first %%@
three claims can largely ignore. Section \ref{fractalsumup} will sum up.

\subsection{Self-similarity and dimension as an exponent}\label{scaldim}
The key innovation of fractals is to extend, from familiar geometric objects such as squares %%@
and cubes to much more ``irregular'' sets, two related ideas: (i) self-similarity and (ii) %%@
dimension as an exponent.

Recall that a square with edge $l$ is the union of $l^2$ unit squares; e.g. a square whose %%@
edge is $l = 3$ units long  is the union of $3^2 = 9$ unit squares.
And a cube with edge $l$ is the union of $l^3$ unit cubes; e.g. a cube whose edge is $l = 3$ %%@
units long is the union of $3^3 = 27$ unit cubes. These examples exhibit both the ideas (i) %%@
and (ii), as follows.\\
\indent (i): The square or cube is a union of smaller copies of itself; and the decomposition %%@
involved can be iterated indefinitely---imagine repeatedly shrinking the unit of length $l$ %%@
by some factor.\\
\indent (ii): In the formula for the measure (area, volume) of the object (i.e. the number of %%@
unit building blocks in it), the dimension occurs as an exponent, and takes the same value, %%@
however fine the decomposition i.e. however small we choose the unit of length. 
\begin{equation}
\mbox{number of unit blocks in object with edge } l = l^{\mbox{dimension of object}}.
\label{defdim}
\end{equation}

So the main idea of fractals is that similarly:---\\
\indent (i'): Some ``irregular'' sets of points are unions of smaller copies of themselves; %%@
where, again, the decomposition involved can be iterated indefinitely. Among these sets will %%@
be some famous examples, which were treated as ``pathological'' when first explored some %%@
hundred years ago: in particular, the Cantor `middle thirds' set $C$ which is a subset of the %%@
unit interval $[0,1] \subset \mathR$ (1872), and the Koch snowflake $K$ which is a subset of %%@
the unit square (1906).\\
\indent (ii'): Applying the idea of eq. \ref{defdim} to such sets, we find that they have %%@
non-integral dimensions. For example, the Cantor set has dimension  about 0.63, and the Koch %%@
snowflake has dimension about 1.26.

These ideas are connected to my themes of emergence and reduction, owing to the fact that %%@
these sets are defined by taking a limit $N \raw \infty$ of an iterated process of %%@
definition. Thus in Section \ref{emergedim} I will take non-integral dimension to be the %%@
emergent (i.e. novel and robust behaviour), which is deduced (and so reduced!) in the limit.

I shall now develop ideas (i) and (ii), especially eq. \ref{defdim}, more formally. But how %%@
fractals illustrate my  claims about emergence and reduction does {\em not} depend on these %%@
details, and the reader uninterested in geometry can now {\em skip} to Section %%@
\ref{emergedim}. But I should also stress, on the other hand, that what follows is the merest %%@
glimpse of the modern theory of dimension. I shall rein in the exposition, and say only %%@
enough: (a) to define the scaling dimension, and see how it can be non-integral (Section %%@
\ref{Cantor}), and (b) to sketch how scaling dimension relates to other concepts of dimension %%@
(Section \ref{moddimtheory}).

\subsubsection{Examples: scaling dimension}\label{Cantor}
I begin by defining the Cantor set and Koch snowflake. This will show that they are %%@
self-similar, i.e. unions of smaller copies of themselves; and this will imply that using eq. %%@
\ref{defdim}'s idea of dimension as exponent, both these sets have a non-integer dimension. %%@
Then I give a general definition of scaling dimension.

{\em 5.1.1.A: The Cantor set $C$}:--- This is defined as the intersection of infinitely many %%@
other subsets, which we will call `stages', labelled 0, 1, 2,... The unit interval $[0,1]$ is %%@
stage 0. After stage 0, each later stage is obtained by deleting the open middle third of %%@
each closed interval of its predecessor.
So stage 1 is $[0,1]$, minus its open middle third. That is: stage 1 is $[0,\frac{1}{3}] \cup %%@
[\frac{2}{3}, 1]$. Then stage 2 is defined by deleting the open middle third of each of %%@
$[0,\frac{1}{3}]$ and $[\frac{2}{3}, 1]$. So stage 2 consists of four disjoint closed %%@
intervals: it  is the set $[0,\frac{1}{9}] \cup [\frac{2}{9}, \frac{1}{3}] \cup [\frac{2}{3}, %%@
\frac{7}{9}] \cup [\frac{8}{9}, 1]$. And so on. 
Thus stage $N$ is the union of $2^N$ intervals, each interval being of length %%@
$(\frac{1}{3})^N$. So the total length of stage $N$ is $2^N \times (\frac{1}{3})^N \equiv %%@
(\frac{2}{3})^N$. So as $N$ goes to infinity, the length of stage $N$ goes to 0.

$C$ is defined to be the intersection of all the stages. Thus $C$ contains those real numbers %%@
between 0 and and 1 whose ternary expansion (i.e. using digits 0,1,2) has no digit 1: so $C$ %%@
is uncountable. Agreed, $C$ is hard to visualize! Its topological properties include: it is %%@
closed, it is nowhere-dense (i.e. its closure has an empty interior) and its complement is a %%@
dense subset of $[0,1]$. 

Now we apply to $C$ the idea of eq. \ref{defdim}. Think of $C$ as the unit block of ``Cantor %%@
type''. And observe that $C$ is the union of two shrunken copies of itself, each smaller by a %%@
factor of 3. That is: one shrunken copy is built by applying the infinite `delete and take %%@
intersection' process to $[0, \frac{1}{3}]$, and the other shrunken copy  by applying the  %%@
process to $[\frac{2}{3}, 1]$.\footnote{This observation can be iterated ``downward''. $C$ is %%@
also the union of $2^2$ shrunken copies of itself, each smaller by a factor of $3^2$. And $C$ %%@
is  the union of $2^3$ shrunken copies of itself, each smaller by a factor of $3^3$; and so %%@
on... for each $N$, $C$ is the union of $2^N$ shrunken copies of itself, each smaller by a %%@
factor of $3^N$.}

 This observation can be reproduced at the next scale up. That is: we can define the ``Cantor %%@
type'' object of scale 3, call it $C'$, as the set that results from applying the infinite %%@
`delete and take intersection' process to $[0, 3]$, rather than to $[0,1]$. Then just as our %%@
original $C$ is the union of two shrunken copies of itself, each smaller by a factor of 3, so %%@
also is $C'$. That is: $C'$ is the union of two unit-size Cantor sets. Now we apply the idea %%@
of eq. \ref{defdim}, getting
\be
\mbox{number of unit Cantor sets in Cantor object of scale } 3 \equiv 2 = 3^{\mbox{dimension %%@
of $C$}}.
\label{defdimCantor}
\ee
Now we recall that for any logarithm base $a$, $b = c^{(\log_a b / \log_a c)}$,
so that in the case of interest: $2 = 3^{(\log 2 / \log 3)}$. Here we drop the suffix stating %%@
the base, since the ratio of logarithms is independent of the base. That is: the dimension of %%@
$C$ is $\log 2 / \log 3$: which is about 0.63.

{\em 5.1.1.B: The Koch snowflake $K$}:--- This also has an iterative construction. Roughly %%@
speaking: we erect smaller and smaller equilateral triangles in the middles of the sides of a %%@
polygon, and define $K$ as the limit. Thus stage 0 is an equilateral triangle. Stage $N+1$ is %%@
constructed from stage $N$ by replacing each line segment of stage $N$ by 4 line segments, %%@
each one-third the length of the original. It follows that the perimeter of the polygon grows %%@
without bound: if $P$ is the perimeter of the initial triangle, then stage $N$ consists of $3 %%@
\times 4^N$ segments each of length $P/3^{N+1}$, so that its perimeter is $(4/3)^N P$.

So this is different from the Cantor set in that $K$ is not itself the union of similar %%@
smaller snowflakes. But each ``side'' of $K$ is the union of four smaller similar curves, %%@
each smaller by a factor 3. So applying again the idea of eq. \ref{defdim}, we get:
\be
4 = 3^{\mbox{dimension of }K} \; \; \; \mbox{ so that: } \;\;\;
{\mbox{dimension of }K} = \frac{\log 4}{\log 3} \approx 1.26. 
\ee

{\em 5.1.1.C: Scaling dimension defined}:--- With these examples as motivation, I proceed to %%@
a general definition. The main effort is in defining the preliminary notion of %%@
self-similarity. For in general we need to allow that the smaller copies (of which the %%@
object, i.e. set, we are concerned with is a union) overlap, i.e. have non-empty %%@
intersection. But we require them to overlap ``minimally'' in the sense that their %%@
intersection is of lower dimension---in the usual integer-valued sense!---than the copies %%@
themselves. Examples include: two continuous curves that have a finite set of points in %%@
common; two rectangles that have parts of the boundaries in common.

For the moment, I will take the usual integer-valued notion of dimension for granted; %%@
(Section \ref{moddimtheory} will rehearse a standard definition of it). Then we say that a %%@
set $X \subset \mathR^n$ is a {\em almost-disjoint union} of two sets $Y, Z$ iff $X = Y \cup %%@
Z$ and $Y \cap Z$ has lower dimension than the dimensions of $Y$ and $Z$. One similarly %%@
defines almost-disjoint unions of more than two sets. And one defines $X$ to be self-similar %%@
if it is an almost-disjoint union of shrunken copies of itself. Here `shrunken copies' can be %%@
made precise by using the vector space structure of $\mathR^n$: (i) to scalar-multiply the %%@
vectors in the set $X$ by a common contraction factor, and (ii) to translate the resulting %%@
shrunken copies out of coincidence with one another, so as to give an almost-disjoint union. %%@
Thus we say: $X$ is {\em self-similar} if it is the almost-disjoint union of $m$ copies of %%@
$X$, each contracted by a common factor $k$, and then translated by a (non-common) vector %%@
$v$. Thus in an obvious notation
\be
X \; = \; \cup^m_{i = 1} \; [\frac{1}{k}X + v_i ] \; .
\label{eqscaldim}
\ee 
Then we define the {\em scaling dimension} of $X$ to be: $\log m /\log k$.

\subsubsection{Generalizations: three other concepts of dimension}\label{moddimtheory} 
Our definition of scaling dimension, eq. \ref{eqscaldim}, is limited to exactly self-similar %%@
objects. But the idea that a dimension occurring as an exponent in a power law can be %%@
non-integral can be developed for much more general kinds of object. These include: (i) %%@
allowing the contraction factor for the building-block set $X$ to be anisotropic (called %%@
`self-affinity', instead of `self-similarity'); and (ii) introducing probabilities governing %%@
the contractions and-or translations of $X$, so that one considers an ensemble of random %%@
fractals, almost all of which are not exactly self-similar.

 These developments have both empirical and theoretical aspects: which have of course %%@
influenced one another over the years. In this Subsection, I round off our glimpse of the %%@
modern theory of dimension by  sketching some of these developments: first the empirical, %%@
then the theoretical. There will be a common key idea: to substitute for Section %%@
\ref{Cantor}'s  contractions of a figure, the complementary idea of contracting a grid of %%@
lines (or planes or hyperplanes), or something analogous to such a grid, like a family of %%@
boxes or discs that appropriately cover the figure.

{\em 5.1.2.A: Empirical aspects}:--- Countless empirical studies have found power law %%@
behaviour with a dimension as a non-integral exponent. One famous example is Richardson's %%@
(1961) discussion of measuring the length of a coastline by traversing it from point to %%@
point, as if with a pair of dividers. Richardson envisages indefinitely improving the %%@
resolution, i.e. reducing the divider-distance. For a continuous curve, we would have the %%@
familiar limit: as the resolution length (divider-distance) $\delta \raw 0$, the number of %%@
steps $n(\delta)$ needed to traverse the coastline grows unboundedly in such a way that the %%@
estimate of the length, $n(\delta)\,\delta$, tends to $l$: where $l$ is the usual length of %%@
the curve, given by  calculus as $l = \int \; ds$. We can express this as a power law with %%@
the curve's dimension $D = 1$ as an exponent. Namely, we would have: 
\be
n(\delta) \approx l/{\delta} \equiv l/{\delta^1} \equiv l/{\delta^D} \; {\rm{with}} \; D = 1.
\ee
But applying the dividers method to ever-larger scale maps suggests instead that as $\delta %%@
\raw 0$, the estimated length $n(\delta)\,\delta$ increases without bound, i.e. $n(\delta) %%@
\approx {\rm{constant}} \, \times \, \delta^D$ with $D$ strictly greater than 1. This is of %%@
course like Section \ref{Cantor}'s discussion of the length, and the dimension, of (the side %%@
of) the Koch snowflake: except that intuitively a coastline has bays as well as %%@
promontories---concave portions as well as convex ones. But this can be modelled using random %%@
fractals, as mentioned in (ii) above.

There are many other examples of such empirical power laws: often, as in this example, with %%@
the quantity of interest, $f$ say, proportional to a power of a resolution $\delta$:
$f = {\rm{constant}} \, \times \, \delta^{D}$. In many cases, of course, the exponent %%@
represents, not length, area or volume, but some other physical quantity. But there are also %%@
plenty of cases where the exponent {\em is} a dimension (in our sense, not the more general %%@
sense of `physical dimension'!). Thus Brady and Ball (1984) studied the dendritic growth of %%@
copper electrodeposited on to an initially pointlike cathode. They found that the volume (or %%@
mass) of copper was proportional to $R^D$, where $R$ is the radius and $D$ was about %%@
2.43---in good agreement with computer-simulations.

{\em 5.1.2.B: Theoretical aspects}:---  For my purposes, the main point here is that the %%@
modern theory of dimension recognizes several different concepts, and of course includes many %%@
theorems relating the agreements and differences in the dimensions assigned to various sets. %%@
I shall sketch three such concepts. As I mentioned at the start of this Subsection, they %%@
share a common general idea: viz. successively finer covers of the set in question, or %%@
something analogous like successively finer grids of lines or (hyper)planes.

I start, for the sake of completeness, with the traditional, i.e. topological, integer-valued %%@
notion. My other two notions are the Hausdorff dimension and the box dimension. They are like %%@
scaling dimension, not just in taking non-integral values; but also in the general underlying %%@
reason for this, viz. some quantity showing power law behaviour. Besides, the dimension they %%@
assign to an exactly self-similar set, as in eq. \ref{eqscaldim}, is equal to the scaling %%@
dimension: viz., in the notation of eq. \ref{eqscaldim}, $\log m / \log k$ (Falconer 2003, %%@
pp. xxiv, 129; Hastings and Sugihara 1993, p. 31, 34, 40). So they are generalizations of %%@
scaling dimension, in the clear sense that if a set $X$ has a scaling dimension $D$, then it %%@
also has both Hausdorff and box dimension equal to $D$. But as I mentioned, each of these %%@
notions also applies to a much wider class of sets. Besides, they have in common that the %%@
power law behaviour occurs as a cover of a set, or something analogous like a grid of lines %%@
or planes, becomes finer. But they are inequivalent notions: for some sets, their values %%@
disagree. 

{\em Topological dimension}: This can be defined for a general topological space; but I %%@
restrict myself to compact subsets of $\mathR^n$. There are various ways to motivate the %%@
definition. Among the clearest is to consider the task of covering the unit square with %%@
closed rectangles in such a way that as few rectangles as possible have points in common. %%@
Suppose we cover the square with a lattice of rectangles; (so the square is their %%@
almost-disjoint union). Then a point at a corner of the lattice is in four rectangles. If %%@
instead we stagger the rectangles, giving a brick-wall pattern, then each point at a corner %%@
is in only three rectangles. On the other hand, it seems this arrangement cannot be %%@
improved---except of course by making the rectangles so that we only need two, or even %%@
one(!), to cover the square. Similarly in three dimensions. A few minutes' reflection %%@
suggests that: (a) for covering the unit cube with arbitrarily small closed rectangular %%@
solids, one can arrange that no point in the cube is contained in more than four solids; but %%@
(b) for sufficiently small solids, at least four solids have a common point. Similarly, one %%@
naturally conjectures, for unit hypercubes $[0,1]^n \subset \mathR^n$: (a) $[0,1]^n$ can be %%@
decomposed as an almost-disjoint union of arbitrarily small $n$-rectangles, in such a way %%@
that no more than $n+1$ of them have a common point; but (b) $n+1$ is the least such number, %%@
i.e. in any such decomposition of $[0,1]^n$, there must be a point common to at least $n+1$ %%@
of the $n$-rectangles.

This prompts the following definitions. Let $X \subset \mathR^n$. Let $\cal U$ be a cover of %%@
$X$ by finitely many sets, and let $\delta > 0$. $\cal U$ is called an $\delta$-{\em cover} %%@
if each element of $\cal U$ has diameter less than $\delta$. (The diameter diam $U$ of a set %%@
$U$ is $\sup_{x,y \in U} \mid x - y \mid$.) The {\em order}, ord $\cal U$, of $\cal U$ is the %%@
natural number $m \in {\bf N}$ for which there is a point of $X$ belonging to $m$ elements of %%@
$\cal U$, but no point belonging to $m+1$ elements. Then we say that $X$ has {\em topological %%@
dimension} $m$ iff $m$ is the least integer for which, for any  $\delta > 0$, there is a %%@
finite closed  $\delta$-cover of $X$ of order $m+1$.

This definition is the beginning of a rich theory. In particular, one shows that it gives the %%@
intuitive verdicts about familiar sets of points: finite sets of points, lines, planes and %%@
solids get dimensions 0, 1, 2 and 3 respectively; and so on for $\mathR^n$. One shows that %%@
dimension thus defined is a topological invariant. It is also easy to check that the Cantor %%@
set has topological dimension 0.   

{\em Hausdorff dimension}:
The definition proceeds in two main steps. (1): We first  sum the diameters, raised to a %%@
power $s$, of the elements of a cover of the set $X$, and consider the limit as the supremum %%@
of these diameters goes to 0. As the set $X$ varies, we get for fixed $s$ a function $H^s$, %%@
which is (an outer measure, and thereby) a measure on an appropriate field of sets---which %%@
includes the Borel sets. (2): These measures, parameterized by $s$, have the curious property %%@
that for any given set $X$, the value $H^s(X)$ is zero or infinity for most $s$. It is this %%@
property that yields the definition of the dimension. The details are as follows.

(1): Let $X \subset \mathR^n$ and $s > 0$ and $\delta > 0$. We define
\be
H^s_{\delta}(X) \;\; = \;\; {\rm{inf}} \;  \Sigma^{\infty}_{i = 1} ({\rm{diam}} \, U_i)^s
\ee
where the infimum is over all countable $\delta$-covers $\{ U_i \}$ of $X$. (One can check %%@
that $H^s_{\delta}$ is an outer measure on $\mathR^n$.) Now we let $\delta \raw 0$:
\be
H^s(X) \, := \, {\rm{lim}}_{\delta \raw 0} \, H^s_{\delta}(X) \, = \, {\rm{sup}}_{\delta > 0} %%@
\, H^s_{\delta}(X). 
\ee
This limit exists; but may be infinite because  $H^s_{\delta}$ increases as $\delta$ %%@
decreases. $H^s$ is an outer measure, and so restricts to a measure on the $\sigma$-field of %%@
$H^s$-measurable sets. This includes the Borel sets, and the measure is called {\em Hausdorff %%@
$s$-dimensional measure}. 

(2): For any $X$, $H^s(X)$ is clearly non-increasing as $s$ increases from $0$ to $\infty$. %%@
And if $s < t$, then $H^s_{\delta}(X) > \delta^{s - t} H^t_{\delta}(X)$. This implies that if %%@
$H^t(X)$ is positive, then $H^s(X)$ is infinite. So there is a unique value, %%@
${\rm{dim}}_H(X)$, the {\em Hausdorff dimension} of $X$, such that
\be
H^s(X) = \infty \; {\rm{if}} \; 0 \leq s \leq {\rm{dim}}_H(X) \;\; ; \; {\rm{and}} \;\;
H^s(X) = 0 \; {\rm{if}} \; {\rm{dim}}_H(X) \leq s \leq \infty \, .
\ee
A rich theory ensues: for its beginning, cf. Falconer (2003, Chapter 2).

{\em Box dimension}: The idea is to count the minimum number $N(\delta)$ of closed $n$-cubes %%@
of a given edge-length $\delta$ that cover the set in question, and to consider the limit as %%@
$\delta \raw 0$. In the now-familiar way suggested by the scaling dimension, $\log m / \log %%@
k$, of eq. \ref{eqscaldim}, the dimension is defined as 
\be
{\rm{lim}}_{\delta \raw 0} \; \frac{\log N(\delta)}{- \log \delta} \; .
\ee
The minus sign is needed to make the dimension positive, since $\log \delta \raw - \infty$ as %%@
$\delta \raw 0$. In fact, we can work, equivalently and conveniently, with the smallest %%@
number of closed balls of radius $\delta$. 

As to the conditions for the limit to exist, I here just recall that any sequence $\{ a_n \}$ %%@
of real numbers has a lim inf and a lim sup (which may equal $\pm \infty$), defined as %%@
follows: lim inf $a_n$ is the number $a$ such that: (i) for all $\varepsilon > 0$, $a_n$ is %%@
eventually forever greater than $a - \varepsilon$, i.e. $\forall \varepsilon > 0, \exists N, %%@
\forall M > N, a_M > a - \varepsilon$; and (ii) for all $\varepsilon > 0$, the sequence %%@
forever returns to being less than $a + \varepsilon$, i.e. $\forall \varepsilon > 0, \forall %%@
N, \exists M > N, a_M < a + \varepsilon$. The requirements (i) and (ii) imply that such a %%@
number $a$ is unique; and if there is no such real number, we set lim inf $a_n$ = $- \infty$. %%@
Similarly for lim sup. (One could summarize in topological jargon by saying that lim inf %%@
$a_n$ is the (possibly infinite) smallest of the sequence's accumulation points; and lim sup %%@
$a_n$ is the (possibly infinite) largest of its accumulation points.) So for any bounded set %%@
$X \subset \mathR^n$, we can define the {\em lower} and {\em upper box dimension} by
\be
{\rm{dim}}^L_B (X) \, := \, {\rm{lim \; inf}}_{\delta \raw 0} \, \frac{\log N(\delta, X)}{- %%@
\log \delta} \;\; ; \;\;
{\rm{dim}}^U_B (X) \, := \, {\rm{lim \; sup}}_{\delta \raw 0} \, \frac{\log N(\delta, X)}{- %%@
\log \delta} ;
\ee
and then we say that if these values are equal, that value is $X$'s {\em box dimension} $= : %%@
{\rm{dim}}_B (X)$.

Again, a rich theory ensues (Falconer 2003, Chapter 3; Barnsley 1988, Chapter 5). For %%@
example: (i) familiar ``regular'' sets like points, lines and planes have box dimension equal %%@
to their topological dimension; (ii) for any set $X$, ${\rm{dim}}_H (X) \leq {\rm{dim}}^L_B %%@
(X) \leq {\rm{dim}}^U_B (X)$; and (iii) for a wide class of sets, the box and Hausdorff %%@
dimension are equal---but the box dimension has the advantage that it is often easier to %%@
calculate.

\subsection{The claims illustrated by emergent dimensions}\label{emergedim}
I turn to describing how the non-integral dimensions of Section \ref{scaldim} count as %%@
emergent behaviour in my sense, and how they illustrate my claims (1:Deduce), (2:Before) and %%@
(3:Herring) (listed in Section \ref{prosp}). As I announced, the illustrations do not need %%@
all the details, especially of Section \ref{moddimtheory}. To keep things simple and brief, I %%@
specialize to sets like the Cantor set and Koch snowflake (Section \ref{Cantor}) that are %%@
defined by taking a limit of an iterated process of definition. Then the illustrations unfold %%@
immediately, once we stipulate that having a non-integral dimension is to be the emergent %%@
property or behaviour: i.e. novel (or surprising) and robust, relative to a comparison class.

Certainly, non-integer dimensions are novel (more so than Section \ref{maf}'s limiting %%@
probabilities). And they are `robust' in at least two senses. First, the scaling dimension of %%@
Section \ref{Cantor} obviously takes the same value for congruent sets of points, and for %%@
enlarged and reduced versions of a given set: this invariance is a kind of robustness. Second %%@
and more interesting: as discussed in Section \ref{moddimtheory}, there are various  novel %%@
notions of dimension which can take non-integer values, and which are ``cousins'' of each %%@
other in various ways. They share the ideas of dimension as an exponent, and of taking %%@
successively finer covers or grids; and for wide classes of sets, their values agree. In %%@
particular, the values of Section \ref{Cantor}'s scaling dimension are endorsed by Section %%@
\ref{moddimtheory}'s Hausdorff and box dimension. So indeed it is fair to talk of `emergent %%@
dimensions'.

\subsubsection{Emergence in the limit: with reduction---and without}\label{dimwithwithout}
As to (1:Deduce): we have `reduction as deduction'  in as strong a sense as you could %%@
demand---provided we take the limit. The general situation is that at stage $N = 0$, a %%@
``regular'' set is given. Here ``regular'' can mean various things depending on the context, %%@
but I will take it to always imply having a well-defined topological dimension. Another set %%@
is then defined, yielding stage $N = 1$, by a process that can be iterated to give sets at %%@
stages $N = 2, 3, ...$. At all finite stages, the defined sets are regular. And for a wide %%@
class of cases (including Section \ref{Cantor}'s Cantor set $C$ and and Koch snowflake $K$), %%@
the stages' dimensions are all equal---and is the integer you would expect. For example, at %%@
stage $N$ for the Cantor set $C$, the defined set, $C_N$, is a union of closed sub-intervals %%@
of the unit interval; and its topological dimension is 1, as you would expect. Similarly for %%@
the stages in defining $K$. But the ``irregular'' set is defined by taking the limit $N \raw %%@
\infty$. In general it has a different topological dimension: thus dim($C$) = 0 $\neq$ %%@
dim($C_N$) $\equiv$ 1. So topological dimension is not continuous in the limit; (footnote %%@
\ref{Wayne} notes how this shows discontinuous limits do not imply emergence). And more %%@
important for us: according to one or more of the novel notions of dimension (scaling, %%@
Hausdorff, box etc.), the set has a non-integral dimension. For example, $C$'s dimension %%@
(according to all three notions) is about 0.63.

Thus the non-integral dimension, the emergent behaviour, is indeed deduced (and so reduced!) %%@
in the limit. In terms of my mnemonic notations: (1:Deduce) is illustrated as follows. Take %%@
as $T_b$ the theory of scaling dimension, and-or one or more of its generalizations like the %%@
Hausdorff or box dimension; and if you wish, include, as a sub-theory, the topological theory %%@
of dimension. Take as $T_t$ the assignments of non-integral dimensions to sets like $C, K$; %%@
(and if $T_b$ includes the generalizations, to other sets that are not exactly self-similar). %%@
Then clearly, we have reduction: $T_b$ contains $T_t$! (Or in terms of Section %%@
\ref{dissolvemyst}'s quantity $f$ whose value, 1 or 0, records the presence or absence of the %%@
emergent property: $f_{\infty} = 1$.)

But there is ``the other side of the coin'': the emergent behaviour is not deducible if we do %%@
not take the limit. Notice that the situation is a bit different from that for the method of %%@
arbitrary functions (Section \ref{emergelimitMAF}). There, all one needed so as to deduce the %%@
emergent behaviour was consideration of the limit. Here, one needs ideas that go beyond the %%@
topological notion of dimension---discontinuous though it is, in the limits concerned. One %%@
needs the idea of dimension as an exponent, as developed in scaling dimension or its %%@
generalizations. But notwithstanding this difference, the main point is that (1:Deduce)'s %%@
second claim holds true again. Namely: if $T_b$ is just the traditional theory of dimension, %%@
there is no reduction; and because this weaker theory is salient, it is tempting to speak of %%@
irreducibility.

Finally, note another contrast with the method of arbitrary functions. Section %%@
\ref{emergelimitMAF} ended by noting that no roulette wheel has infinitely many arcs; nor is %%@
any wheel spun infinitely fast. In Section \ref{deflatelimits}'s notation: there was no %%@
infinite system $\sigma(\infty)$. But in the fractals example, there are such infinite %%@
systems---the sets $C, K \subset \mathR^n$ etc.---and the whole discussion focusses on them.

\subsubsection{Emergence before the limit}\label{dimbefore}
(2:Before) claims that before the limit, there is emergence in a weaker but still vivid %%@
sense. It is illustrated in a manner parallel to the method of arbitrary functions. Thus %%@
recall Section \ref{equiprobbefore}'s discussion of approximate equiprobability in, for %%@
example, a casino's roulette wheel. For fractals, the obvious analogue of the wheel is a %%@
computer running some software so as to produce a simulation of some fractal set, by %%@
iterating the steps of its definition some finite number $N$ of times. The most obvious case %%@
is computer graphics software, producing an approximate or coarse-grained image of a fractal %%@
set. Nowadays, such images are ubiquitous in films and games, superseding the static images %%@
in yesteryear's lavish books (e.g. Peitgen and Richter 1986).

It is easy to check that all of Section \ref{equiprobbefore}'s discussion---about how one can %%@
calculate, perhaps numerically, how closely a set-up approximates equiprobability, and how we %%@
philosophers can leave it to the casino-owners to worry about how close is close enough to be %%@
indiscernible by prospective gamblers---carries over to fractals, {\em mutatis mutandis}. I %%@
will save space by not spelling this out. In short: what was said there, about the practical %%@
purposes of a casino in making a wheel fair enough that a gambler cannot profit from %%@
assiduously observing its long-run statistics, carries over here to the practical purposes of %%@
a film studio in making a simulated image look fractal at spatial scales so small that even %%@
the most hawk-eyed cinema-goer cannot see that it is in fact {\em not} fractal.

But there are two other topics worth pausing over. One is obvious from the mention of %%@
computer graphics: the use of fractals to model naturally occurring objects like mountains, %%@
rocks, trees and leaves. This merits a separate discussion; cf. Section \ref{nature?}. 

The other topic is an analogue for fractals of the quantifier-shift that Section %%@
\ref{emergelimitMAF} discussed as underlying the ``two sides of the coin'' in (1:Deduce). %%@
(This topic is also connected to the robustness requirement in my notion of emergence; but I %%@
will not pursue this.)

Thus take a traditional geometrical variable magnitude: in philosophers' jargon, a %%@
determinable property of a geometrical figure $F$. For example, consider `contains a %%@
continuous arc of length greater than $L$' (variable $L$). And suppose we have an repeatable %%@
definitional process, that at its $M$th iteration defines a figure (subset of $\mathR^n$), %%@
$F_M$, and that introduces successively finer structure so that for each value $L$ of the %%@
variable, $F_M$ lacks the property for sufficiently large $M$. That is: the property is lost %%@
after sufficiently many iterations. Or to put it more positively: an approximate or %%@
coarse-grained version of a fractal-like property is {\em gained}. For example, the %%@
definitional process might imply: $\forall L > 0, \exists N, \; \forall M > N$: the figure %%@
$F_M$ lacks  arcs of length greater than $L$. But for smaller $L$, more iterations will be %%@
needed.

To make an analogy with Section \ref{emergelimitMAF}'s quantifier-shift, we now develop this %%@
idea so as to both:\\
 \indent (a) use an `resolution' $\varepsilon$, as is usual in definitions of convergence;\\
 \indent  (b) make a ``pointwise vs. uniform'' contrast, by quantifying over some set $\CG$ %%@
of geometrical properties, or sub-figures, of the figure $F_M$.

 Thus suppose that in the figure $F_M$ at stage $M$, the only, or the largest, example of a %%@
property or sub-figure $G \in \CG$ is of size (say, length) $L$. I will write this as: %%@
Size$(F_M, G) = \varepsilon$. Then the successive loss of properties $G \in \CG$---more %%@
exactly: the loss of visible, large-enough-to-be-seen, $G \in \CG$---by a sequence of figures %%@
$\{ F_M \}$ can happen: either pointwise across $\CG$, viz.\\
 \indent \indent $\forall \varepsilon, \forall G \in \CG, \exists N \; \forall M > N: %%@
\mbox{Size}(F_M,G) < \varepsilon  $; \\
 or uniformly across $\CG$, viz.\\
 \indent \indent $\forall \varepsilon, \exists N \; \forall G \in \CG,  \forall M > N: %%@
\mbox{Size}(F_M,G) < \varepsilon . $ \\
Besides, there are alternatives to using such a set $\CG$ so as to make the pointwise/uniform %%@
contrast. We could instead use different parts of the figures $F_M$. Thus one can imagine the %%@
stages of the definitional process to proceed at different ``rates'' in different regions: in %%@
different thirds, ninths,..., of the Cantor set; or sides, sub-sides, sub-sub-sides,..., of %%@
the Koch snowflake. If the rates vary in a suitably ever-slower way, across a denumerable %%@
sequence of sub-regions, one would get convergence to the fractal structure that is merely %%@
pointwise across the set.

\subsubsection{Supervenience is a red herring}\label{redherring2}
I shall be very brief about my third claim, (3:Herring): that although various supervenience %%@
theses are true, they yield little or no insight into emergence, or more generally, into %%@
``what is going on'' in the example. For the situation is again like that for the method of %%@
arbitrary functions (Section \ref{redherring1}): my claim holds true, essentially because %%@
supervenience makes no connection with the main ideas of the example---self-similarity and %%@
dimension as an exponent.

For any finite $N$, the property of interest, dimension, of the object concerned, i.e. of a %%@
subset $X \subset \mathR^n$, ``supervenes on how $X$ is constituted from points''---in at %%@
least two obvious senses of this phrase. Namely: (i) the trivially strong sense in which only %%@
$X$ itself contains those very points (cf. set-theory's axiom of extensionality); (ii) the %%@
marginally weaker sense in which as regards its constitution from points, $X$ matches any %%@
congruent or scaled copy of $X$. And since in this example, there are infinite systems %%@
$\sigma(\infty)$, i.e. the ``irregular'' sets $C, K \subset \mathR^n$ etc., the same goes for %%@
$N = \infty$. That is: the dimension of these sets, in any of the several senses of %%@
dimension, thus supervenes. 

But such supervenience theses are trivial and useless, for the two now-familiar reasons. (a): %%@
They provide no control on the infinity (infinite disjunction) they are concerned with, %%@
because no kind of limit is taken. (b): Their infinity makes no connection with the limit, $N %%@
\raw \infty$, that the example is concerned with. In particular, the supervenience thesis %%@
gives no hint that we can use the idea of dimension as an exponent so as to define %%@
non-integral dimensions.

\subsection{The fractal geometry of nature?}\label{nature?}
So far, the pure mathematics of dimension has dominated the discussion. But fractals have %%@
many empirical applications. As I discussed in Section \ref{moddimtheory}.A, countless %%@
empirical studies have found power law behaviour with a dimension as a non-integral exponent: %%@
recall the examples of the coastline and electrodeposited copper. And Section \ref{dimbefore} %%@
mentioned computer graphics' use of fractals to model objects like mountains, trees and %%@
leaves. This representational power of fractals is remarkable, indeed amazing.\footnote{And %%@
noticed by the wider culture: in Stoppard's play {\em Arcadia} (1993), the hero Valentine %%@
describes a stage-by-stage computer-simulation: `If you knew the algorithm and fed it back %%@
say ten thousand times, each time there'd be a dot somewhere on the screen. You'd never know %%@
where to expect the next dot. But gradually you'd start to see this shape, because every dot %%@
would be inside the shape of this leaf. It wouldn't be a leaf, it would be a mathematical %%@
object.' In another passage he is lyrical about fractals' representation of other %%@
`ordinary-sized stuff which is our lives, the things people write poetry about---clouds, %%@
daffodils, waterfalls'.} Thus fractals have been hailed as revealing the true geometry of %%@
nature, e.g. by Mandelbrot (1982). But this claim has been disputed (Shenker 1994, especially %%@
Sections 3-5; Smith 1998, pp. 31-38): hence this Subsection's title. 

I will argue that with my claims (2:Before) and (4:Unreal), we can put this controversy to %%@
rest. I will distinguish two senses of the phrase `geometry of nature', and propose that %%@
fractal geometry is {\em a} geometry of nature, in the second sense but not the first. It %%@
will be clear that (2:Before) corresponds to the second sense, while (4:Unreal) corresponds %%@
to the first. Finally, I will introduce an ``abstract'', rather than ``natural history'', %%@
sense of the phrase. In this last sense, fractal geometry is again a geometry of nature; and %%@
this again corresponds to (2:Before). 

Suppose first that `geometry of nature' means `the completely accurate description of the %%@
shapes and sizes of macroscopic objects'. Then it {\em sure looks like} fractal geometry is %%@
the geometry of nature---as many a film with computer-generated graphics attests. But authors %%@
such as Shenker have objected that a fractal has an infinite sequence of intricate but %%@
similar structure on ever smaller length scales; while a mountain, rock, tree, fern and leaf %%@
do not, thanks to their atomic constitution. This objection is of course correct: recall my %%@
claim (4:Unreal) of Section \ref{unreal}. So despite initial appearances, fractal geometry is %%@
{\em not} in this sense the geometry of nature.

Indeed, the objection can be sharpened, in two ways: one theoretical, one practical. (Neither %%@
seems to have been noticed in the literature.) I touched on the theoretical sharpening, %%@
already in footnote \ref{onlyeucsimilar}, when I noted that while Euclidean geometry admits %%@
the similarity of triangles and other figures, on which self-similarity and so fractals %%@
depend, non-Euclidean geometries do not. This means that if physical space is in fact %%@
slightly non-Euclidean on even the tiniest scales, as general relativity and cosmology %%@
nowadays say, then macroscopic objects could  not be {\em exactly} fractal---even if atomism %%@
was false and they were instead composed of continuous matter, even on arbitrarily small %%@
length scales. So here again, we meet my claim (4:Unreal).\footnote{I stress the phrases %%@
`nowadays say' and `macroscopic objects'. I of course agree that, for all we know, fractals %%@
may be involved as fundamental structures in the ultimate theory, at present unknown, of %%@
matter and-or space. But that is not our concern.}
 
The practical sharpening concerns the details of Section \ref{moddimtheory}.A's empirical %%@
studies of power laws with a quantity $f$ proportional to a non-integral power of a %%@
resolution $\delta$: $f = {\rm{constant}} \, \times \, \delta^{D}$. Suppose that faced with %%@
such a study, we ask: how many orders of magnitude of  $\delta$ does the data report---or %%@
does the analysis in fact probe? The answer can be: disappointingly few. A survey of %%@
ninety-six {\em Physical Review} articles (in the years 1990-1996) reporting fractal analysis %%@
of data found that among these articles: (i) the average spread of resolutions that were %%@
probed was 1.3 orders of magnitude; and (ii) at most three orders of magnitude were probed %%@
(Avnir et al. 1998). In terms of measuring the length of a coastline: an ``average paper'' in %%@
the set surveyed by Avnir et al. would describe the coastline or its length as `fractal', %%@
though the authors considered a spread of resolutions that went only from some length $L$ to %%@
about thirteen times $L$, $\approx 13 L$. And even the papers that were most stringent, or %%@
cautious, in describing their phenomenon as `fractal' probed their resolution only up to %%@
three orders of magnitude.\footnote{Thanks to Leo Kadanoff for commenting that, happily, the %%@
range probed can also be much larger. He mentions the work of Libchaber and co-authors on %%@
turbulence, and Nagel and co-authors on glassy behaviour. Indeed, the former have probed five %%@
orders of magnitude (e.g. Castaing et al. 1989), and the latter have probed thirteen (Dixon %%@
et al. 1990). I presume that the latter group's {\em Physical Review} papers have been %%@
omitted from Avnir et al's survey for the ironic reason that they meritoriously avoid using %%@
the word `fractal'.}

To sum up about this first sense of `geometry of nature':  if we ask the question,
\begin{quote}
Do fractals describe, with complete accuracy, the shapes and sizes of naturally occurring %%@
macroscopic objects?
\end{quote}
we have to answer `No'. 

But despite this answer `No', the representational power of fractals remains very striking.  %%@
Power laws with a non-integral exponent describe very many phenomena; and our understanding %%@
of the phenomenon is often enhanced, empirically as well as theoretically, by adding to the %%@
bare power law, the suggestive language and exact theorems of fractal geometry. Here again we %%@
see that in a suitably weak sense, emergence can occur before the relevant limit: (2:Before) %%@
again! 

Besides, this is consistent with (4:Unreal), since (2:Before) applies to values of the %%@
parameter $N$ which are typically much smaller than those making true (4:Unreal). That is: %%@
our `No' answer turned upon our question's requiring complete accuracy. If instead we ask, in %%@
the context of modelling some specific phenomenon involving naturally occurring macroscopic %%@
objects, `Do fractals describe, with sufficient accuracy, the shapes and sizes of these %%@
objects?', our answer would very often be `Yes'. In this weaker sense, fractal geometry %%@
undoubtedly is {\em a} geometry of nature.

There is another aspect to this resolution of the controversy; (which, like the foregoing, %%@
should not be controversial!). So far I have considered the shapes and sizes of macroscopic %%@
objects in physical space. But suppose we allow that `geometry' applies to objects or %%@
structures in {\em other} spaces: in physical theories' state-spaces, which indeed often %%@
include objects and structures the theory calls `geometric'. Some of these theories are %%@
strikingly successful, in the depth and accuracy of their theoretical descriptions and %%@
observational predictions. So some of these postulated geometric descriptions surely deserve %%@
to be called (a) `geometry of nature': i.e. a geometry of an object or structure in a %%@
physically real space, albeit a more abstract (and often less visualizable) space than %%@
physical space. So if we ask instead a third question,
\begin{quote}
Do some of our successful physical theories use fractals to describe certain subsets of their %%@
abstract spaces, in particular attributing a non-integer dimension to such objects?
\end{quote}
the answer is again: {\em Yes}. 

Here are just two examples, out of many one could cite. One can check that in each of them, %%@
the justification for using fractals, i.e. for the $N \raw \infty$ limit, is the %%@
Straightforward Justification of Section \ref{strtfwd}, with its two obvious reasons, %%@
mathematical convenience and empirical success.\\
\indent (1): In classical mechanics, there are physically important fractal subsets of the %%@
phase spaces of systems. In particular, the famous Lorenz and H\'{e}non attractors have %%@
fractal dimension. (The philosophical literature on chaos theory has discussed these, along %%@
the lines of the Straightforward Justification; e.g. Smith's (1998, pp. 41-43, 50-56) %%@
discussion of the Lorenz attractor.)\\
\indent (2): Statistical mechanics describes some aspects (viz. critical points) of some %%@
processes (phase transitions, like boiling and freezing) with scale-free (regimes of) %%@
theories, which involve power-law behaviour on all scales and self-similarity, and therefore %%@
fractals. Section \ref{phasetr} will give more details. In particular, Section %%@
\ref{crossover} will discuss the phenomenon of cross-over: in which, as the parameter $N$ %%@
grows, the system's behaviour crosses over from illustrating (2:Before), at intermediate %%@
values of $N$, to illustrating (4:Unreal), at larger values of $N$, to again illustrating %%@
(2:Before) at yet larger values of $N$. Thus cross-over will be a vivid illustration of my %%@
swings-and-roundabouts, (2:Before)-and-(4:Unreal), Yes-and-No, answer to the question `Is %%@
fractal geometry the geometry of nature?'

\subsection{The story so far: summing up fractals}\label{fractalsumup}
Let me sum up the fractals example as a list of six morals. It will be obvious, without %%@
making explicit my four claims, or Section \ref{strtfwd}'s Straightforward Justification, or %%@
the parallels with the method of arbitrary functions, that this list also sums up the whole %%@
story so far.

\indent (i): The large finite is often well-modelled by the infinite. \\
\indent (ii): Such models are often justified in a straightforward, even obvious, way, by %%@
mathematical convenience and empirical success.\\
\indent (iii): The infinite often brings new mathematical structure: in the fractals example, %%@
non-integer dimension.\\
\indent (iv): Nevertheless, there is often a reduction: the emergent non-integer dimensions %%@
are reducible to a sufficiently rich theory that takes the infinite limit.\\
\indent (v): On the other hand, one can often see emergent behaviour on the way to the limit. %%@
Thus in the fractals example: the larger i.e. worse the spatial resolution---the worse your %%@
eyesight---the sooner in the iterative definitional process you see (more precisely: think %%@
you see!) the fractal structure.\\
\indent (vi): Various supervenience theses hold---but they are trivial, or at least %%@
scientifically useless.

\section{Superselection}\label{superselec}
\subsection{Introduction}\label{qmintro}
I turn to the first of my two examples from physics proper: superselection in the $N \raw %%@
\infty$ limit of quantum mechanics. As in the previous two examples, I will first expound the %%@
technical details without reference to my claims (Section \ref{supersubsec}), and then %%@
illustrate the claims (Section \ref{emergessn}). But  to make those illustrations richer, I %%@
will give more details than previously. So the aim of this Subsection is to describe the lie %%@
of the land---and so indicate which details matter more. I will do this by describing (i) the %%@
general topic of the quantum-classical transition (Section \ref{soup}) and (ii) the basic %%@
idea of superselection in the limit (Section \ref{nutshell}). 

\subsubsection{Out of the quantum soup}\label{soup}
This example is an aspect of a much larger topic: the emergence of the classical world from %%@
the quantum world. This is often discussed in terms of a limit $\hbar \raw 0$. But the topic %%@
involves a lot else, such as proposals for the importance of specific states (e.g. coherent %%@
states), and-or of specific physical  processes (e.g. decoherence). My example is just one %%@
case of the general idea that classical physics should emerge for ``large'' quantum systems: %%@
so in this example of my theme that $N \raw \infty$, $N$ will be the number of degrees of %%@
freedom. (We will  see how this $N \raw \infty$ relates to $\hbar \raw 0$. But as I announced %%@
at the end of Section \ref{unreal}, I will not in this Section pursue my fourth claim %%@
(4:Unreal).)

By way of glimpsing the larger topic, I note that there are now many well-understood examples %%@
of classical physics emerging for large (i.e. $N \raw \infty$) quantum systems. Much of this %%@
work uses the algebraic approach to quantum theory, in which systems are primarily described %%@
by algebras of quantities, on which the states are linear expectation functionals; and I %%@
shall follow suit. For example, Sewell's recent monograph uses this approach to articulate a %%@
`rather general scheme for ... deriving the irreversible deterministic macroscopic dynamical %%@
laws of many-particle systems, such as those of hydrodynamics or heat conduction, from their %%@
underlying quantum dynamics' (2002, p. 87).  This scheme forms a girder across the rest of %%@
Sewell's book: it is realised in detail, in several examples.\footnote{The first is a %%@
toy-model reminiscent of quantum Brownian motion: a massive particle at
one end of a semi-infinite chain of much lighter particles, with harmonic nearest-neighbour %%@
interactions; p. 94-106. His Chapters 7, 10 and 11 describe much more advanced cases, e.g. %%@
lasers.} 

To illustrate my claims, I can treat superselection more simply than by following Sewell's %%@
scheme: the main simplification will lie in ignoring dynamics (especially, the deduction of %%@
classical equations of motion) and aiming only to deduce, as $N \raw \infty$, classical %%@
kinematics (more precisely, commutativity of quantities). But to give a general perspective, %%@
it is worth first quoting Sewell's scheme. (These details are not needed later.)

Sewell's macroscopic picture is given by a classical dynamical system ${\cal M} = ({\cal %%@
Y},T)$ where ${\cal Y}$ is a topological space, and $\{T(t) \mid t \in \mathR_+ \}$ is a %%@
one-parameter semigroup of transformations of ${\cal Y}$. ${\cal M}$ is to correspond to the %%@
dynamics of a one-parameter family $Y_{\Omega}$ of finite sets of quantities of a quantum %%@
system $\Sigma$; where $\Sigma$'s evolution will be given by a one-parameter group %%@
$\alpha(\mathR)$ of automorphisms of $\Sigma$'s algebra of quantities. So we write %%@
$Y_{\Omega} = \{ Y^{(1)}_{\Omega},...,Y^{(k)}_{\Omega} \}$.\\
\indent Here $\Omega$ is a ``large'', dimensionless, positive parameter whose magnitude %%@
provides a measure of the quantities' macroscopicality. We then require that there is a set %%@
$\Delta$ of states of $\Sigma$ such that:\\
\indent \indent (a): For each state $\phi \in \Delta$, the means and dispersions of %%@
$Y_{\Omega}$ converge to limits, $Y(\phi) \equiv Y$ and 0, respectively, as $\Omega %%@
\rightarrow \infty$.\\
\indent \indent (b): As $\phi$ runs through $\Delta$, the resultant range of the limiting %%@
values $Y \equiv Y(\phi)$ is just the classical phase space ${\cal Y}$. (So ${\cal Y}$ is a %%@
subset of $\mathR^k$.)\\
\indent \indent (c) The classical dynamical semigroup $T(\mathR_+)$ of ${\cal M}$ is induced %%@
by the evolution of $\Sigma$ from states in $\Delta$, on a ``macroscopic'' time scale %%@
$\Omega^{\gamma}$, with $\gamma > 0$. To be precise: we require that the mean and dispersion, %%@
for a state $\phi \in \Delta$, of the $k$ macroscopically time-evolved quantities %%@
$\alpha(\Omega^{\gamma}t)Y_{\Omega}$ should converge to $T(t)Y$ and 0, respectively, as %%@
$\Omega \rightarrow \infty$.

Accordingly, Section \ref{supersubsec} will give analogues of Sewell's kinematic (a) and (b), %%@
though not of his dynamical (c). But in two other respects, I will go beyond the above %%@
scheme. First, I will emphasise that there are {\em two} different infinite limits to be %%@
considered:\\
\indent (i) the classical limit of ever-larger (increasing $N$) quantum systems, which will %%@
have {\em all} quantities commuting, and which corresponds to Sewell's (a) and (b); and \\
\indent (ii) the quantum limit of ever-larger quantum systems, which will exhibit %%@
superselection, i.e. some quantities apart from multiples of the identity operator commuting %%@
with all quantities: in algebraic terms, the centre of the algebra of quantities being %%@
non-trivial.

Second, I will discuss these limits in terms of {\em deformation quantization}: which has the %%@
merits not only of generality and precision, but also of showing how (both of) these limits %%@
are {\em continuous}---{\em pace} the frequent emphasis on the singularity of $\hbar \raw 0$, %%@
and philosophers' frequent emphasis on discontinuous limits as a signature of emergence. %%@
(Recall the discussions in Sections \ref{peace} and \ref{2genl}.) In both these respects, my %%@
treatment follows (but simplifies) some material in Landsman's masterly account of the %%@
quantum-classical transition (2006).

\subsubsection{The idea of superselection in the limit}\label{nutshell}
Setting aside all details of both mathematics and physics, I will state in a nutshell the %%@
idea of superselection emerging in the limit $N \raw \infty$, by using the idea that a %%@
sequence of numbers, each less than one, has in general an infinite product equal to zero. %%@
Thinking of each number as an inner product of two vectors allows us to think of this zero as %%@
orthogonality: a hallmark of superselection.

Imagine that we assign two real vectors $\in \mathR^3$, each of unit-length, with angle %%@
$\theta$ between them, the ``score'' $\cos \theta$.  So if they are parallel, the score is %%@
$\cos 0 = 1$; but if they are not parallel, it is less than 1.
And suppose we assign two sequences of unit vectors, each with $N$ members, a score that is %%@
the product of  the cosines of the angles between corresponding members:
\begin{eqnarray}\label{score}
{\rm{score}} ( <v_1,v_2,...,v_N> \; , \; <u_1,u_2,...,u_N > ) := \\ \nonumber
 \cos \theta_{v_1 u_1} \cos \theta_{v_2 u_2} ... \cos \theta_{v_N u_N}
\end{eqnarray}
We now let $N$ tend to infinity, and consider the limiting values of the score we have %%@
defined. We note two sorts of case:---\\
\indent (1): A pair of infinite sequences $<v_i>, <u_i>$ in which the vectors at %%@
corresponding positions $i$  are {\em not} parallel, only for finitely many $i$. Then only %%@
finitely many factors in the score will be {\em different} from 1; infinitely many factors %%@
will be 1. So the total infinite product of numbers is a product of finitely many cosines %%@
each less than 1. This is some real number of modulus less than 1. It {\em might} be zero: %%@
namely if at least one pair of vector $v_i, u_i$ are perpendicular.\\
\indent (2): A pair of infinite sequences $<v_i>, <u_i>$ in which the vectors at %%@
corresponding  $i$  are {\em not} parallel, for infinitely many $i$. So the total infinite %%@
product of numbers is a product including infinitely many numbers that are each less than 1. %%@
In general, this infinite
product is zero (and even if there are {\em also} infinitely many factors each equal to 1).

These elementary considerations underly the emergence of superselection in the $N \raw %%@
\infty$ limit of quantum mechanics. (1) corresponds to two quantum states (two infinite %%@
sequences of unit-vectors) being in the same superselection sector. And (2) corresponds to %%@
two quantum states  being in different  superselection sectors. 

Now I enter in to details. As in my previous two examples, I first expound the technicalities %%@
without reference to my claims (Section \ref{supersubsec}), and then illustrate the claims %%@
(Section \ref{emergessn}).

\subsection{Superselection in the $N \raw \infty$ limit of quantum %%@
mechanics}\label{supersubsec}
In Section \ref{chains}, I describe our prototype systems: finite and infinite spin chains. %%@
This will already exhibit the idea just expounded, of a ``score'' of sequences of vectors %%@
that converges to zero. In Section \ref{deform}, I introduce the ideas of deformation %%@
quantization (especially
continuous fields of algebras of quantities), in terms of which our $N \raw \infty$ limits %%@
will be continuous. In Section \ref{macro}, I treat the classical infinite-$N$ limit of %%@
quantum systems: for spin chains, this means identifying macroscopic quantities with averages %%@
of quantum observables, with the average being taken over greater and greater segments of the %%@
chain. In Section \ref{quasilocal}, I treat the quantum infinite-$N$ limit: here the limiting %%@
quantities are local in the sense that they ``ignore'' all but a finite part of the system. %%@
In Section \ref{compareFinetti}, I discuss classical states of the quantum-infinite, and thus %%@
connect with superselection. NB: Each Section begins with a brief statement and then, after %%@
the announcement `{\em In Detail}', gives details---which the reader in a hurry can {\em %%@
skip}. 

For all this material, and much more, I recommend Landsman (2006, especially Sections 4.3, %%@
6.1-6.4), who also gives many references (and whose notation I adopt). Compared with %%@
Landsman, I have chosen to down-play infinite tensor products and the representation of %%@
algebras, and to emphasise deformation quantization. This avoids repeating material which is %%@
well-known in the philosophy of physics literature; and more important, gives what we need %%@
for Section \ref{emergessn}'s illustrations of my claims. 

\subsubsection{Spin chains}\label{chains}
I begin with three claims about spin chains, which serve as ideal infinite models of %%@
ferromagnets. I shall take a doubly infinite spin-half chain, with sites labelled by the %%@
integers $\bf Z$. But it will be clear that similar claims would hold for higher spin, and %%@
for a one-dimensional half-infinite chain (sites labelled by $\bf N$) or for a two- or %%@
three-dimensional spin lattice (${\bf Z}^2$ or ${\bf Z}^3$). In these claims, we eschew %%@
infinite tensor products and non-separable Hilbert spaces. Rather we define a continuous %%@
family of separable Hilbert spaces, each of which will later turn out to be a superselection %%@
sector. 

 The overall physical idea is that:\\
\indent \indent (a) The vacuum i.e. ground state has all the spins aligned in one spatial %%@
direction, with  other (higher-energy) states built up by flipping a {\em finite} number of %%@
spins from the preferred direction (and superposing); yielding a separable Hilbert space %%@
representing the spin-algebra.   \\
\indent \indent (b) But nature does not prefer one such direction. So for any direction, %%@
there is a vacuum state with the spins thus aligned; and the higher-energy states built up %%@
from this vacuum yield an associated Hilbert space.\\
\indent \indent (c) These Hilbert space representations differ in a global/macroscopic %%@
quantity (spin density).\\
\indent \indent (d) The representations are thereby unitarily inequivalent---even though %%@
intuitively, a global rotation (an element of SO(3)) of course rotates one vacuum to another.

\indent {\em In Detail}: One shows four claims, e.g. for the one-dimensional doubly infinite %%@
spin chain; (cf. e.g. Sewell 1986: 16-18, or 2002: 15-18).

\indent\indent  (a): For each direction (where positive and negative $z$-directions count as %%@
two directions), there is an irreducible representation of the infinite spin algebra %%@
generated by denumerably many pairwise-commuting copies of the trio of the Pauli matrices, %%@
i.e. generated by  matrices: $\{{\bf \sigma_n} := (\sigma_{n, x}, \sigma_{n, y},\sigma_{n, %%@
z}) \mid n = 0,\pm 1, \pm 2, \dots \}$.

Let $S$ be the set of doubly infinite $(+1,-1)$-sequences, $s = (s_n)_{n \in \Z}, s_n = \pm %%@
1$. We think of this as the set of ``classical'' configurations of the eigenvalues $\pm 1$ of %%@
$\sigma_z$ at each site. Let $S^{(+)} \subset S$ be those configurations $s$ with all but %%@
finitely many $s_n = +1$. We think of these as local modifications of the $z$-up (classical) %%@
vacuum $s^{(+)} = (.., 1, ...)$. Let $\H^{(+)}$ be the square-summable functions on %%@
$S^{(+)}$:
\begin{equation}
\H^{(+)} : = \{ \phi: S^{(+)} \raw \mathbb{C} \; \mid \; \Sigma_{s \in S^{(+)}} |\phi(s)|^2 < \infty %%@
\} \; ,
\end{equation}
with inner product
\begin{equation}
\langle \phi, \psi \rangle^{(+)} := \Sigma_{s \in S^{(+)}} {\ovl {\phi}}(s)\psi(s)  \; .
\end{equation}
The vectors $\phi^{(+)}_s, s \in S^{(+)}$ defined by $\phi^{(+)}_s(s') = \delta_{s,s'}$, $s, %%@
s' \in S^{(+)}$, are in one-one correspondence with the configurations $s$. They form an %%@
orthonormal basis of $H^{(+)}$.

We can now define operators whose action on the basis vectors $\phi^{(+)}_s$ is the analogue %%@
of the action of the three  Pauli matrices on a single spin. 
That is, we define $\{ \sigma^{(+)}_{n, u} \; | \; n \in \Z, \; u = x,y,z \}$ on $\H^{(+)}$ %%@
in the obvious way, so as to 
build a representation on $\H^{(+)}$ of the abstract spin-algebra:
\begin{equation}
[\sigma_{n, x}, \sigma_{n, y}] = 2 i \sigma_{n, z} \; {\mbox{etc.}} \; ; \;
 [\sigma_{m, x}, \sigma_{n, y}] = 0 \; \mbox{for} \; m \neq n  \; \mbox{etc.}
    \label{absspin}
\end{equation}
The representation is irreducible since we can pass from any  basis vector $\phi^{(+)}_s$ to %%@
any other $\phi^{(+)}_{s'}$ by a sequence of operators $\sigma^{(+)}_{n, u}$.

\indent\indent Note that because we fixed on the denumerable $S^{(+)} \subset S$, not the %%@
continuously large $S$, we got a separable Hilbert space $\H^{(+)}$. So this representation %%@
does not requires us to make sense of a denumerable tensor product (of copies of $\mathC^2$).

(b): We can ``play the same game'', starting with $z$-down. That is: let $S^{(-)} \subset S$ %%@
be those configurations $s$ with all but finitely many $s_n = -1$. We think of these as local %%@
modifications of the $z$-down (classical) vacuum. We get a representation on $\H^{(-)}$.

 Of course, the choice of $z$ was arbitrary. So we can build a representation for any %%@
direction. And each such representation: (a) takes the ground state to have all spins aligned %%@
along the direction; and (b) builds elements of the representation as all linear combinations %%@
of product states obtained from the ground state by a finite number of unitary %%@
transformations (spin-flips or rotations) at individual sites. 

 (c):  For each such representation, every state in it (indeed every density matrix on it) %%@
has a common value of a {\em global quantity}, (aka: {\em classical} or {\em macroscopic %%@
quantity}): namely, the (vector) {\em spin density} defined as the limit as $N \rightarrow %%@
\infty$ of the average of the spin matrices at the sites  $-N, -N + 1, \dots, N-1, N$, i.e.
$
{\rm lim}_{N \rightarrow \infty} \frac{1}{2N+1} \Sigma^{+N}_{n = - N}
{\bf \sigma_n}
\label{definespindensity}
$. 

In more detail: On our first space $\H^{(+)}$, define
\begin{equation}
{\bf m}^{(+)}_N := \frac{1}{2N+1} \Sigma^{+N}_{n = -N}
{\bf \sigma}^{(+)}_{\bf n} \; ;
\label{defNm}
\end{equation}
so that the expectation value on basis states is
\begin{equation}
\langle \phi^{(+)}_s, {\bf m}^{(+)}_N \phi^{(+)}_s \rangle = \left(0, 0, \frac{1}{2N+1} %%@
\Sigma^{+N}_{n = -N}  s_n \right) \; .
\label{mfiniteexpect}
\end{equation}
Since all but a finite number of $s_n$ are $+1$, \er{mfiniteexpect} implies, with $\bf k$ the %%@
unit vector along $0z$:
\begin{equation}
\lni \langle \phi^{(+)}_s, {\bf m}^{(+)}_N \phi^{(+)}_s \rangle = {\bf k} \; , \forall s \in %%@
S^{(+)} \; .
\label{mfiniteexpect1}
\end{equation}
Similarly, one has:
\begin{equation}
\lni \langle \phi^{(+)}_s, {\bf m}^{(+)}_N \phi^{(+)}_{s'} \rangle = 0 \; , \mbox{\; for \;} %%@
s \neq s' \; :
\label{limselection}
\end{equation}
for in the $z$-component, ${\bf m}^{(+)}_N$ gives no spin-flip or rotation, so that the %%@
orthogonality, $\langle \phi^{(+)}_s, \phi^{(+)}_{s'} \rangle = 0$, yields 0; while in the %%@
$x$ and $y$ components, the eventual agreement of $s$ and $s'$ in having thereafter always %%@
the value $+1$ means that ${\bf m}^{(+)}_N$'s spin-flips and rotations give inner product %%@
zero at those sites, while the increasing $2N + 1$ denominator kills any initial non-zero %%@
contribution got from ${\bf m}^{(+)}_N$'s action. Thus from \er{mfiniteexpect} and %%@
\er{limselection}, it follows that for any unit vector $\Psi^{(+)} \in \H^{(+)}$,
\begin{equation}
\lni \langle \Psi^{(+)}, {\bf m}^{(+)}_N \Psi^{(+)} \rangle = {\bf k} \; .
\label{mfiniteexpect2}
\end{equation}
So this limiting spin density is a global property of the representation. And the %%@
representations built from other vacua will have different unit vectors in $\mathR^3$ as %%@
their limiting spin densities.\\
\indent As we will discuss in more detail: these are the system's superselection sectors. %%@
There are continuously many sectors, each of denumerable dimension.

 (d): Each such representation is unitarily inequivalent to every other. For example: if the %%@
representations on $\H^{(+)}$ and $\H^{(-)}$ were unitarily equivalent, there would be a %%@
unitary $U: \H^{(+)} \raw H^{(-)}$ with $U {\sigma}^{(+)}_{\bf n} U^{-1} = %%@
\sigma^{(-)}_{\bf n}$. This would imply that $U {\bf m}^{(+)}_N U^{-1} = {\bf m}^{(-)}_N$. %%@
This in turn would imply that, for any unit vectors $\Psi^{({\pm})} \in \H^{(\pm)}$ with  %%@
$\Psi^{(+)} = U^{-1} \Psi^{(-)}$
\begin{equation}
\langle \Psi^{(+)}, {\bf m}^{(+)}_N \Psi^{(+)} \rangle =
\langle \Psi^{(-)}, {\bf m}^{(-)}_N \Psi^{(-)} \rangle  \; :
\end{equation}
but this must be false, since the two sides have different limits, viz. $\pm {\bf k}$, as $N %%@
\raw \infty$.

\subsubsection{Continuous fields of algebras and deformation quantization}\label{deform}
The main idea we need from the modern theory of deformation quantization (in the %%@
$C^*$-algebraic approach) is the idea of a continuous field of algebras of quantities. For %%@
Sections \ref{macro} and \ref{quasilocal} will define such fields in such a way that their $N %%@
\raw \infty$ limits are continuous. But when I enter details, I will also define two other %%@
central ideas of the theory.
 
The idea is that a continuous field of algebras of quantities is like a bundle in %%@
differential geometry. The base space is a set $I \ni \hbar$ of real numbers, and the fibre %%@
above each point $\hbar \in I$ is a $C^*$-algebra $\CA_{\hbar}$ representing the system's %%@
quantities for that value of $\hbar$. (Of course, this is not meant to suggest that the value %%@
of $\hbar$ really varies. As usual, this is shorthand, essentially for the ratio of $\hbar$ %%@
to the typical values of the system's, or problem's, actions.) The topology of the bundle is %%@
defined indirectly by specifying what are its continuous sections. As we will see: the value %%@
$\hbar = 0$ can correspond to a $N \raw \infty$ limit; and we can choose the algebras (and %%@
the definitions of continuous sections) so as to make this either a classical limit or a %%@
quantum limit.

More formally: a  continuous field of $C^*$-algebras over $I \ni \hbar$   consists of a
 \ca\ $\CA$, and a collection of \ca s $\{\CA_{\hbar}\}_{{\hbar}\in I}$, subject to certain %%@
conditions which imply that:\\
\indent (i): the  family $(\CA_{\hbar})_{\hbar\in I}$ of \ca s is glued together by %%@
specifying the space of continuous sections of the bundle 
$\coprod_{\hbar\in I}\CA_{\hbar}$ (where $\coprod$ indicates disjoint union);\\
\indent (ii): the \ca\ $\CA$ can be identified with this space of sections.

{\em In Detail}: A {\em continuous field} of $C^*$-algebras over $I \ni \hbar$   consists of %%@
a
 \ca\ $\CA$, a collection of \ca s $\{\CA_{\hbar}\}_{{\hbar}\in I}$, and a surjective %%@
morphism $\phv_{\hbar}:\CA\raw\CA_{\hbar}$ for each $\hbar\in I$ such that:\\
\indent (i): the function ${\hbar}\mapsto  \|\phv_{\hbar}(A)\|_{\hbar}$ is in $C_0(I)$ for %%@
all $A\in\CA$;\footnote{Here $\| \; \|_{\hbar}$ is the norm in $\CA_{\hbar}$, and $C_0(I)$ is %%@
the continuous complex functions on $I$ that vanish at infinity in the usual sense that for %%@
any $\varepsilon > 0$, there is a compact set beyond which the function is less than %%@
$\varepsilon$. For any locally compact Hausdorff space $X$, $C_0(X)$ is a $C^*$-algebra when %%@
equipped with the supremum norm, $\left\| f \right\|_{\infty}$ := sup$_{x \in X} | f(x) |$. %%@
In any case, for our examples $I$ is itself compact.}\\
\indent (ii): the norm of any $A\in\CA$ is $\| A\|=\sup_{{\hbar}\in I}\|\phv_{\hbar}(A)\|$;\\
\indent (iii): for any $f\in C_0(I)$ and $A\in\CA$ there is an element $fA\in\CA$ for
which $\phv_{\hbar}(fA)=f({\hbar})\phv_{\hbar}(A)$ for all ${\hbar}\in I$.\\
The idea is that the \ca s $(\CA_{\hbar})_{\hbar\in I}$ are glued together by a topology on %%@
the disjoint union
$\coprod_{\hbar\in [0,1]}\CA_{\hbar}$. But this topology is  defined  indirectly, by %%@
specifying the space of continuous sections of the ``bundle''.
 Namely, a {\it continuous section}
of the field is defined to be  a map $\hbar\mapsto A_{\hbar}$
where $A_{\hbar}\in \CA_{\hbar}$
 for which there is an $A\in \CA$ such that $A_{\hbar}=\phv_{\hbar}(A)$ for all ${\hbar}\in %%@
I$. So the \ca\ $\CA$ may actually be identified with the space of continuous sections of the %%@
field: if we do so, the morphism $\phv_{\hbar}$ is just the evaluation
 map at $\hbar$.

With this idea of a continuous field of algebras, we can define a deformation quantization. %%@
We begin with a classical phase space $M$, on which the continuous complex functions $f$ %%@
represent quantities (through their real and imaginary parts). We want to define, for each %%@
value of $\hbar$, a quantization map $\qh$ mapping such functions $f$ to elements of %%@
$\CA_{\hbar}$, subject to various conditions---in particular, the ``Dirac'' condition that %%@
the Poisson bracket on $M$ is mapped to the quantum mechanical commutator (times %%@
$\frac{i}{\hbar}$). Two technical comments: (i) $M$ can be a Poisson manifold---a mild %%@
generalization of the usual symplectic manifold; (ii) on the other hand, we take $f$ to be %%@
smooth with compact support---the space $\cci(M)$ of these functions is a norm-dense %%@
sub-algebra of $C_0(M)$.

Thus we define: a {\em deformation quantization} of a phase space $M$ is a continuous field %%@
of \ca s $(\CA_{\hbar})_{\hbar\in [0,1]}$ (with $\CA_0=C_0(M)$), along with a family of   %%@
linear maps $\qh:\cci(M)\raw\CA_{\hbar}$, $\hbar\in(0,1]$, that are self-adjoint (i.e. %%@
$\qh(\ovl{f})=\qh(f)^*$), and such that:\\
\indent (i) for each $f\in\cci(M)$ the map defined by $0\mapsto f$ and
$\hbar\mapsto\qh(f)$ ($\hbar\neq 0$) is a continuous section of the given continuous field of %%@
algebras;\\
\indent (ii) for all $f,g\in \cci(M)$ one has the ``Dirac'' condition
\begin{equation}
\lim_{\hbar\rightarrow 0}
\left\|\frac{i}{\hbar}[\CQ_{\hbar}(f),\CQ_{\hbar}(g)]-\CQ_{\hbar}(\{f,g\})\right\|_{\hbar} %%@
=0. \label{Dirac}
\end{equation}
This definition turns out to imply other natural ``meshing'' conditions, such as:
\begin{equation}
\lim_{\hbar\rightarrow 0}
\left\|\CQ_{\hbar}(f)\CQ_{\hbar}(g)-\CQ_{\hbar}(fg)\right\|_{\hbar} = 0 \; \; ; \;\;
\lim_{\hbar\rightarrow 0}
\left\|\CQ_{\hbar}(f)\right\|_{\hbar} = \left\| f \right\|_{\infty}
 \label{Diracsconsequences}
\end{equation}
where $\left\| f \right\|_{\infty}$ is the supremum norm sup$_{z \in M} | f(z) |$ on %%@
$\CA_0=C_0(M)$.

By way of indicating the power of deformation quantization, let me sketch how a definition %%@
which it enables, viz. of a continuous field of states (treated as linear functionals on %%@
quantities), gives a natural generalization of the notion of coherent states---which are the %%@
focus of many discussions of the $\hbar \raw 0$ limit of quantum mechanics. (The details of %%@
Sections \ref{compareFinetti} and \ref{chainsagain} will also need this definition.) 

Given a state $\om_{\hbar}$ on $\CA_{\hbar}$
for each $\hbar\in[0,1]$: we define
 the family of states to be a {\it continuous field} (relative to a given deformation %%@
quantization), whenever the function $\hbar\mapsto \om_{\hbar}(\qh(f))$ is continuous on %%@
$[0,1]$ for each $f\in\cci(M)$. (In fact, this notion is  intrinsically defined by the %%@
continuous field of \ca s, and is therefore independent of the quantization maps $\qh$.)  In %%@
particular, one has
\begin{equation}
\lho \om_{\hbar}(\qh(f))=\om_0(f).
\label{CFS0}
\end{equation}

Now recall the idea of a sequence of coherent states $(\Psi^{\hbar}_z)_{\hbar\in[0,1]}$ that %%@
tend, as $\hbar \raw 0$, to be ever more peaked about the classical phase space state $z \in %%@
M$, so that in the limit, the quantum expectation value tends to the value of the classical %%@
quantity at the state $z$: 
\begin{equation}
\lho (\Psi^{\hbar}_z , \qh(f) \Psi^{\hbar}_z)  = f(z).
\label{coherentidea}
\end{equation}
(This of course exemplifies (a) and (b) in Sewell's scheme, reported in Section \ref{soup}: %%@
the coherent states are Sewell's set $\Delta$ of quantum states, $\hbar$ is the reciprocal of %%@
his $\Omega$, and $M$ is his phase space $\cal Y$.) 
Clearly, eq. \ref{CFS0} generalizes eq. \ref{coherentidea}. Indeed, one can show that %%@
coherent states are examples of continuous fields of (pure) states; (for details, cf. %%@
Landsman (2006, Section 4.2, 5.1).

\subsubsection{The classical infinite: macroscopic quantities from symmetric %%@
sequences}\label{macro}
I turn to the classical $N \raw \infty$ limit of $N$ copies of a quantum system that has an %%@
algebra of quantities $\CA_1$. The idea is to identify a classical, macroscopic quantity with %%@
a limiting average of a microscopic quantum quantity. Thus the microscopic quantity is %%@
defined over say $M$ copies of the system; (so it is often called an ``$M$-particle %%@
observable''). But this quantity is averaged over $N$ copies ($N > M$), and then we take the %%@
limit $N \raw \infty$. For spin chains (Section \ref{chains}), this means the average is %%@
taken over greater and greater segments of the chain. 

This is made precise in terms of a continuous field of \ca s $\CA^{\mathrm(c)}$ over %%@
different values of $N$. The four main points are:---\\
\indent (i): To conform to Section \ref{deform}'s notation for $\hbar \raw 0$, we in fact use %%@
the reciprocal of $N$, $1/N$, rather than $N$. Thus  $\CA_{1/N}^\mathrm(c)$ will be the %%@
usual algebra of quantities for $N$ copies of the basic quantum system, viz. $\hat{\ot}^N %%@
\CA_1$, i.e. the $N$-fold tensor power of the basic algebra $\CA_1$. \\
\indent (ii): We associate a macroscopic quantity with a sequence $A=(A_1,A_2,\cdots)$ of %%@
algebra-elements, with $A_N\in \hat{\ot}^N \CA_1 =: \CA_1^N$, that is {\em symmetric} in the %%@
sense that its tail consists of some finite-particle quantity averaged over an ever-larger %%@
number of systems or sites. That is: a sequence is symmetric if each of its elements $A_P$ %%@
for all $P$ greater than some fixed $N$ consists of some quantity on  $M$ (with $M < N$) %%@
copies of the system (an ``$M$-particle observable'') averaged over $P$ copies. \\
\indent (iii): In the limit $N \raw \infty$, symmetric sequences commute; and this will mean %%@
that the macroscopic quantities form a commutative \ca . In fact this algebra is isomorphic %%@
to $C({\cal S}(\CA_1))$, i.e. the continuous complex functions on the quantum state-space for %%@
(the density matrices on) the basic algebra $\CA_1$.  \\
\indent (iv): The important features (both here and in subsequent Sections) are present in %%@
the construction for the simplest basic algebra  $\CA_1 := M_2(\mathC)$, i.e. a spin chain %%@
with a spin-half system at each site. So we can throughout the discussion keep this system in %%@
mind. For example, $\CS(M_2(\mathC))$ is the Bloch sphere $B^3$ in $\R^3$, with pure states %%@
on the boundary $\partial B^3 = S^2$. So according to (iii), the macroscopic quantities for a %%@
chain of spin-halves are given by the continuous complex functions on the Bloch sphere.

{\em In Detail}: From  $\CA_1$ = say $M_2(\mathC)$, we will construct a continuous field of %%@
\ca s $\CA^{\mathrm (c)}$ over the discrete set
\begin{equation} I=0\cup 1/\N=\{0,\ldots, 1/N,\ldots, \third,\half,1\}\subset [0,1], %%@
\label{interval}\end{equation}
by putting
\begin{eqnarray}
\CA_0^{\mathrm (c)} :=  C(\CS(\CA_1));\nonumber \\
\CA_{1/N}^{\mathrm (c)} :=  \CA_1^N := \hat{\ot}^N \CA_1. \label{fibers}
\end{eqnarray}
Thus if $\CA_1 = M_2(\mathC)$, then $\CS(M_2(\mathC))$ is the Bloch sphere $B^3$ in $\R^3$, %%@
and $\CA_0^{\mathrm (c)}$ is the set of continuous functions on $B^3$.

To define symmetric sequences, we first say that
the {\em symmetrization operator} $S_N: \CA^N_1\raw \CA^N_1$ is given by (linear and %%@
continuous) extension of
\begin{equation}
S_N(B_1\ot\cdots \ot B_N) := \frac{1}{N!}\sum_{\sg\in \GS_N} B_{\sg(1)}\ot\cdots\ot %%@
B_{\sg(N)}, \label{landc}
\end{equation}
where $\GS_N$ is the permutation group on $N$ elements and $B_i\in\CA_1$ for all %%@
$i=1,\ldots,N$. Then we define {\it Symmetrization maps} $j_{NM}:  \CA^M_1\raw  \CA^N_1$  by
\begin{equation}
j_{NM}(A_M)=S_N(A_M\ot 1\ot\cdots \ot 1); \label{symmaps}
\end{equation}
with $N-M$ copies of  $1\in\CA_1$ so as to get an element of $\CA_1^N$. For example, %%@
$j_{N1}:\CA_1\raw\CA_1^N$ is given by
\begin{equation}
j_{N1}(B)=
\ovl{B}^{(N)}=\frac{1}{N}\sum_{k=1}^N 1\ot\cdots\ot B_{(k)}\ot 1\cdots \ot1,
\end{equation}
where $B_{(k)}$ is $B$  as an element of the $k$'th copy of $\CA_1$ within $\CA_1^N$. As  %%@
$\ovl{B}^{(N)}$ indicates, this is  the `average' of $B$ over all copies of $\CA_1$. More %%@
generally, in forming $j_{NM}(A_M)$ an
operator $A_M\in\CA_1^M$ that involves $M$ sites is averaged over $N\geq M$ sites. When %%@
$N\raw\infty$ this means that one forms a macroscopic average of an $M$-particle operator.

A sequence $A=(A_1,A_2,\cdots)$ of algebra elements with $A_N\in\CA_1^N$
is {\it symmetric} when
\begin{equation} A_N=j_{NM}(A_M) \label{ass}
\end{equation}  for some fixed $M$ and all $N\geq M$.  So the tail of a symmetric sequence  %%@
consists of `averaged' quantities, which become macroscopic in the limit $N\raw\infty$. The %%@
important point is that symmetric sequences commute in this limit; more precisely
\begin{equation}
\lni \| A_NA_N'-A_N'A_N\|=0. \label{aprc}
\end{equation}
The averaging of 1-particle spin operators provides a clear example, and illustrates how $N %%@
\raw \infty$ corresponds to $\hbar \raw 0$. Thus let $A_N := j_{N1}(B)$ and $A_N' := %%@
j_{N1}(C)$ with $B,C\in\CA_1$. Then the fact that $[B_{(k)},C_{(l)}]=0$ for $k\neq l$
 implies that
\begin{equation} %%@
\left[\ovl{B}^{(N)},\ovl{C}^{(N)}\right]=\frac{1}{N}\ovl{\left[B,C\right]}^{(N)}.
\label{av0}
\end{equation}
For example, if $\CA_1=M_2(\mathC)$ and if for $B$ and $C$ one takes the spin-$\half$ %%@
operators
$S_j=\frac{\hbar}{2}\sg_j$ for $j=1,2,3$ ($\sg_j$ the Pauli matrices), then
\begin{equation}
\left[\ovl{S}_j^{(N)},\ovl{S}_k^{(N)}\right]=i\frac{\hbar}{N}\epsilon_{jkl} \ovl{S}_l^{(N)}.
\label{av0spin}
\end{equation}
Thus we get commutation relations formally like those of the one-particle operators, except %%@
that Planck's constant $\hbar$ is replaced by $\hbar/N$.

We are now ready to define our continuous field of algebras. A section of the field with %%@
fibers \er{fibers} is a sequence $A=(A_0,A_1,A_2,\cdots)$, with $A_0\in\CA_0^{\mathrm (c)}$ %%@
and $A_N\in\CA_1^N$. We say that
a sequence $A$ defines a  {\it continuous} section of the field iff:
\begin{itemize}
\item $(A_1,A_2,\cdots)$  is {\it approximately symmetric}; i.e. for any $\varep>0$ there is %%@
an $N_{\varep}$ and a symmetric sequence $A'$ such that
$\|A_N-A_N'\|< \varep$ for all $N\geq N_{\varep}$;
\item $A_0(\om)=\lni\om^N(A_N)$, where $\om\in\CS(\CA_1)$ and $\om^N\in \CS(\CA_1^N)$ is the %%@
tensor product of $N$ copies of $\om$, i.e.
\begin{equation} \om^N(B_1\ot\cdots\ot B_N)=\om(B_1)\cdots\om(B_N).
\label{omN}
\end{equation}
\end{itemize}
This choice of continuous sections defines a continuous field of \ca s  over $I=0\cup 1/\N$ %%@
with fibers \er{fibers}. In fact it follows that
\begin{equation}
\lni \| A_N\|=\|A_0\| \; . \label{normeq}
\end{equation}
\indent To sum up: the main point  is that, in accordance with \er{aprc},  the macroscopic %%@
quantities are organized in the limit $N\raw\infty$ in to a commutative \ca\ isomorphic to %%@
$C(\CS(\CA_1))$.

\subsubsection{The quantum infinite: quasi-local sequences}\label{quasilocal}
I now treat the quantum $N \raw \infty$ limit of $N$ copies of the algebra $\CA_1$. The key %%@
idea will be that of a quasilocal sequence $(A_1,A_2,\cdots)$ of algebra elements, with %%@
$A_N\in\CA_1^N$. In constructing the continuous field of  \ca s, this notion will play an %%@
analogous role to that played in Section \ref{macro} by symmetric sequences.  The idea is %%@
that the tail of a quasilocal sequence becomes arbitrarily close (in norm) to an element $A_M %%@
\ot 1 \ot 1 \ot \cdots$, so that the sequence ``ignores, in the limit'' all but a finite part %%@
of the system. The intuition behind this restriction is that human limitation means we can %%@
observe only a finite part of the system. In any case, the restriction allows us to make %%@
precise the heuristic idea of the algebra for an infinite quantum system, which we might %%@
write heuristically as  $\CA_1^{\infty}$ = say $M_2(\mathC)^{\infty}$.  

Formally, the infinite quantum system is the {\it inductive limit} \ca\ 
\begin{equation} \ovl{\cup_{N\in\N} \CA_1^N}\label{ILCAruf}
\end{equation}
of the  family of \ca s  $(\CA_1^N)$. Eq. \ref{ILCAruf} consists of all equivalence classes %%@
$[A] \equiv A_0$ of quasilocal sequences $A = (A_1,A_2,\cdots)$, under the equivalence %%@
relation $A\sim B$ iff $\lni\|A_N-B_N\|=0$. As the notation suggests, each $\CA_1^N$ is %%@
contained in  $\ovl{\cup_{N\in\N} \CA_1^N}$ as a $C^*$-subalgebra by identifying $A_N\in %%@
\CA_1^N$ with
a quasilocal sequence that after the $N$th term just tensors with the identity, viz. the %%@
sequence $A=(0,\cdots,0,A_N, A_N\ot 1,A_N\ot 1\ot1, \cdots)$, and forming its equivalence %%@
class $[A] \equiv A_0$ in $\ovl{\cup_{N\in\N} \CA_1^N}$.

{\em In Detail}:  A sequence $A=(A_1,A_2,\cdots)$ ($A_N\in\CA_1^N$)  is  {\it local} if for %%@
some fixed $M$ and all $N\geq M$: $A_N=A_M\ot 1\ot\cdots\ot 1$
 (with $N-M$ copies of the unit $1\in\CA_1$). A sequence is  {\it quasilocal} when  for any %%@
$\varep>0$ there is an $N_{\varep}$
and a local sequence $A'$ such that
$\|A_N-A_N'\|< \varep$ for all $N\geq N_{\varep}$. 

We now define
the {\it inductive limit} \ca\
\begin{equation} \ovl{\cup_{N\in\N} \CA_1^N}\label{ILCA}
\end{equation}
of the  family of \ca s  $(\CA_1^N)$ with respect to the inclusion maps %%@
$\CA_1^N\hookrightarrow\CA_1^{N+1}$ given by $A_N\mapsto A_N\ot 1$. As a set,  \er{ILCA} %%@
consists of all equivalence classes $[A]\equiv A_0$ of quasilocal sequences $A$ under the %%@
equivalence relation $A\sim B$ when $\lni\|A_N-B_N\|=0$.  The norm on $\ovl{\cup_{N\in\N} %%@
\CA_1^N}$ is
\begin{equation} \| A_0\|=\lni \|A_N\|, \label{normput}\end{equation}
and other \ca ic structure is inherited from the quasilocal sequences
in the obvious way (e.g., $A_0^*=[A^*]$ with $A^*=(A_1^*,A_2^*,\cdots)$, etc.). Thus each %%@
$\CA_1^N$ is contained in  $\ovl{\cup_{N\in\N} \CA_1^N}$ as a $C^*$-subalgebra by identifying %%@
$A_N\in \CA_1^N$ with
the local sequence $A=(0,\cdots,0,A_N, A_N\ot 1,A_N\ot 1\ot1, \cdots)$, and forming its %%@
equivalence class $A_0$ in $\ovl{\cup_{N\in\N} \CA_1^N}$.

So we define a second continuous field of \ca s $\CA^{\mathrm (q)}$ over $0\cup 1/\N$, with %%@
fibers
 \begin{eqnarray}
\CA^{\mathrm (q)}_0&=& \ovl{\cup_{N\in\N} \CA_1^N};\nonumber \\
\CA^{\mathrm (q)}_{1/N}&=& \CA_1^N. \label{fibers2}
\end{eqnarray}
 by declaring that the continuous sections are of the form $(A_0,A_1,A_2,\cdots)$
 where $(A_1,A_2,\cdots)$ is quasilocal and $A_0$ is defined to be the equivalence class of  %%@
this quasilocal sequence, as just explained.

For $N<\infty$ this field has the same fibers
\begin{equation} \CA^{\mathrm (q)}_{1/N}=\CA_{1/N}^{\mathrm (c)}= \CA_1^N\end{equation}
 as Section \ref{macro}'s continuous field $\CA^{\mathrm (c)}$ , but the fiber $\CA^{\mathrm %%@
(q)}_0$ is completely different from $\CA_0^{\mathrm (c)}$. For if $\CA_1$ is noncommutative %%@
then so is $\CA^{\mathrm (q)}_0$, since it contains all $\CA_1^N$.

\subsubsection{Comparing the classical and quantum limits: classical states and the de %%@
Finetti theorem}\label{compareFinetti}
One natural way to study the relations between the fields $\CA^{\mathrm (c)}$ and  %%@
$\CA^{\mathrm (q)}$ is to consider those  families of abstract states %%@
$(\om_1,\om_{1/2},\cdots,\om_{1/N},\cdots)$
($\om_{1/N}$ is a state on $\CA_1^N$) that have appropriate limit states on both %%@
$\CA^{\mathrm (c)}$ and  $\CA^{\mathrm (q)}$. (Here `appropriate' is made precise in terms of %%@
Section \ref{deform}'s notion of a continuous field of states.) In this Section, I introduce %%@
the most important such family, the {\em permutation-invariant} states. We will see in %%@
Section \ref{chainsagain} how they yield superselection sectors, especially in Section %%@
\ref{chains}'s prototype system, spin chains.

Of course, any state $\om_0^{\mathrm (q)}$ on $\CA^{\mathrm (q)}_0$ defines  a state
$\om_{0|1/N}^{\mathrm (q)}$ on each $\CA_1^N$ by restriction; and the sequence of these %%@
states have the given $\om_0^{\mathrm (q)}$ as their appropriate limit state on $\CA^{\mathrm %%@
(q)}_0$. If this sequence of states {\em also} converges with respect to the other continuous %%@
field of algebras $\CA^{\mathrm (c)}$, i.e. converges to some state $\om_0^{\mathrm (c)}$ on %%@
$\CA^{\mathrm (c)}_0$, then the given state $\om_0^{\mathrm (q)}$ is called {\em classical}.

We specialize to an important class of classical states, viz. those that are ``indifferent %%@
to'' the label $N$ of the component systems (e.g. of the sites in the spin chain). Formally, %%@
we say that a state $\om_0^{\mathrm (q)}$ on $\CA^{\mathrm (q)}_0$ is {\em %%@
permutation-invariant} if its restrictions to each of the $\CA_1^N$ is invariant under the %%@
natural action of the symmetric group $S_N$ on $\CA_1^N$.

For our purposes, the important point about these states is that they give a close quantum %%@
analogue of the de Finetti representation theorem. Roughly speaking, this theorem  says that %%@
any classical probability measure on an infinite Cartesian power probability space %%@
$X^{\infty} := X \times X \times \cdots$ that is permutation-invariant (under permutations %%@
between copies of the factor space $X$) is a unique mixture of (i.e. has a unique integral %%@
decomposition in terms of) infinite product probability measures $p^{\infty}$ given by %%@
$p^{\infty}(Y_1 \times Y_2 \times \cdots) := p(Y_1).p(Y_2). ...$ (with $Y_i$ in the %%@
sigma-algebra of the $i$th copy of $X$) for some probability measure $p$ on $X$.
Section \ref{chainsagain} will describe how the quantum analogue of this theorem gives a %%@
precise yet general framework for the emergence of superselection.

{\em In Detail}: Consider those families of states %%@
$(\om_1,\om_{1/2},\cdots,\om_{1/N},\cdots)$
(where $\om_{1/N}$ is a state on $\CA_1^N$) that have limit states {\it both} $\om_0^{\mathrm %%@
(c)}$  on $\CA_0^{\mathrm (c)}$ and {\it and} $\om_0^{\mathrm (q)}$ on $\CA_0^{\mathrm (q)}$, %%@
such that the ensuing families $(\om_0^{\mathrm (c)},\om_1,\om_{1/2},\cdots)$ and %%@
$(\om_0^{\mathrm (q)},\om_1,\om_{1/2},\cdots)$ are  {\it continuous} fields of states on  %%@
$\CA^{\mathrm (c)}$ and on $\CA^{\mathrm (q)}$ (in the sense of Section \ref{deform}).

Any state $\om_0^{\mathrm (q)}$ on $\CA^{\mathrm (q)}_0$ defines  a state
$\om_{0|1/N}^{\mathrm (q)}$ on $\CA_1^N$ by restriction, and the ensuing field of states on %%@
$\CA^{\mathrm (q)}$ is clearly continuous. (Conversely, any continuous field
$(\om_0^{\mathrm (q)},\om_1,\om_{1/2},\ldots, \om_{1/N},\ldots)$ of states on $\CA^{\mathrm %%@
(q)}$
becomes arbitrarily close to a field of the above type for $N$ large.) But the restrictions %%@
$\om_{0|1/N}^{\mathrm (q)}$ of a given state $\om_0^{\mathrm (q)}$ on $\CA^{\mathrm (q)}_0$ %%@
to $\CA_1^N$ may well {\em not} converge to a state $\om_0^{\mathrm (c)}$ on $\CA_0^{\mathrm %%@
(c)}$ for $N\raw\infty$.
States $\om_{0}^{\mathrm (q)}$ on
$\ovl{\cup_{N\in\N} \CA_1^N}$ whose restrictions $\om_{0|1/N}^{\mathrm (q)}$ {\em do} %%@
converge to a state $\om_0^{\mathrm (c)}$ on $\CA_0^{\mathrm (c)}$ are called {\it %%@
classical}.

In other words (cf. the definition of $\CA_0^{\mathrm (c)}$ especially eq. \ref{omN}): %%@
$\om_{0}^{\mathrm (q)}$ is classical when there exists a probability measure $\mu_0$ on %%@
$\CS(\CA_1)$ such that
\begin{equation}
\lim_{N\raw\infty} \int_{\CS(\CA_1)} d\mu_0(\rh)\, (\rh^N(A_N)-\om_{0|1/N}^{\mathrm %%@
(q)}(A_N))=0 \label{PMOQ}\end{equation}
for each approximately symmetric sequence $(A_1, A_2,\ldots)$. In other words: a classical %%@
state $\om_0^{\mathrm (q)}$ with limit state $\om_0^{\mathrm (c)}$ on
$C(\CS(\CA_1))$ defines a probability measure $\mu_0$ on $\CS(\CA_1)$ by
\begin{equation}
 \om_0^{\mathrm (c)}(f)=\int_{\CS(\CA_1)} d\mu_0\, f, \label{probm} \end{equation}
which describes the probability distribution of the macroscopic quantities in that state.

We now make this more concrete by specializing to an important class of classical states.
We say that a state $\om$ on $\ovl{\cup_{N\in\N} \CA_1^N}$  is {\it permutation-invariant} %%@
when each of its restrictions  to $\CA_1^N$ is invariant under the natural action of  $\GS_N$ %%@
on  $\CA_1^N$ (i.e.\ $\sg\in\GS_N$ maps an elementary tensor $A_N=B_1\ot\cdots\ot %%@
B_N\in\CA_1^N$ to $B_{\sg(1)}\ot\cdots\ot B_{\sg(N)}$, cf.\ \er{landc}).

We can now state the quantum analogue of the de Finetti representation theorem: a %%@
permutation-invariant state on $\CA^{\mathrm (q)}_0 = \ovl{\cup_{N\in\N} \CA_1^N}$ is a %%@
unique mixture of (i.e. has a unique integral decomposition in terms of) infinite product %%@
states $\rh^{\infty}$, that are defined (with $\rh\in\CS(\CA_1)$) by saying that  if $A_0\in %%@
\CA^{\mathrm (q)}_0$ is an equivalence class $[A_1,A_2,\cdots]$, then (cf.\ \er{omN})
\begin{equation} \rh^{\infty}(A_0)=\lni \rh^N(A_N) \; .
\label{rhoinfinite}
\end{equation}
An equivalent definition is to say that the restriction of $\rh^{\infty}$ to any %%@
$\CA_1^N\subset \CA^{\mathrm (q)}_0$ is given by $\ot^N \rh$.

In other words, the theorem says: any permutation-invariant state $\om^{\mathrm (q)}_0$ has a %%@
unique decomposition
\begin{equation}
\om^{\mathrm (q)}_0(A_0)=\int_{\CS(\CA_1)} d\mu(\rh)\, \rh^{\infty}(A_0), \label{Unn}
\end{equation}
where $\mu$ is a probability measure on $\CS(\CA_1)$ and $A_0\in \CA^{\mathrm (q)}_0$. We can %%@
also state this in more geometric language, as follows. The set $\CS^{\GS}$ of all %%@
permutation-invariant states in $\CS(\CA^{\mathrm (q)}_0)$ is a compact convex set, and is %%@
the (weak$\mbox{}^*$-closed)  convex hull of  its (extreme) boundary $\partial_e  \CS^{\GS}$. %%@
So the claim of the theorem is that this boundary consists of the infinite product states %%@
(and so is isomorphic to $\CS(\CA_1)$  in the obvious way).

To sum up this Section, especially  \er{PMOQ}, \er{probm} and \er{Unn}:---
 If $\om^{\mathrm (q)}_0$ is permutation-invariant, then it is classical.
The  associated limit state $\om_0^{\mathrm (c)}$
on $\CA_0^{\mathrm (c)}$ is characterized by the fact that
 the measure $\mu_0$ in \er{probm} coincides with the measure $\mu$ in \er{Unn}.

\subsection{The claims illustrated by emergent superselection}\label{emergessn}
The material in Section \ref{supersubsec} gives a precise and powerful framework for %%@
understanding the emergence of superselection in the $N \raw \infty$ limit. It will be %%@
clearest to first summarize this  (Section \ref{chainsagain}; again following Landsman %%@
(2006)), and then spell out the illustrations of my claims (Section \ref{ssnwithwithout} et %%@
seq.).

\subsubsection{Superselection from permutation-invariant states, in spin %%@
chains}\label{chainsagain}
I begin with generalities, and then return to spin-chains. In the algebraic approach to %%@
quantum theory, a superselection sector is taken to be an appropriate  equivalence class %%@
(under unitary isomorphism) of \irrep s of the system's abstract the algebra of quantities %%@
$\CA$. The word `appropriate' reflects the fact that most \irrep s of a typical \ca\ $\CA$ %%@
used in physics are physically irrelevant, and so need to be excluded; (jargon: one needs a %%@
{\it selection criterion}).  Here, we take the algebra of quantities to be $\CA^{\mathrm %%@
(q)}_0$; and take an (equivalence class of) \irrep s of $\CA^{\mathrm (q)}_0$ to be a %%@
superselection sector iff it corresponds to a permutation-invariant pure state on  %%@
$\CA^{\mathrm (q)}_0$. (Here `correspond' is made precise using the GNS theorem, viz. as %%@
equivalence to the GNS-representation of the permutation-invariant state.) With this %%@
selection criterion, the results in Section \ref{supersubsec}, especially at \er %%@
{rhoinfinite} onwards, trivially imply that there is a bijective correspondence between pure %%@
states on $\CA_1$ and superselection sectors of $\CA_0^{\mathrm (q)}$. 

 The results are vividly illustrated by spin chains. In Section \ref{chains}, I did not give %%@
details about how to build the infinite tensor product state space $\H_1^{\infty}$, with say %%@
$\H_1=\mathC^2$. But as one would hope, we have: If $(e_i)$ is some basis of $\mathC^2$, an %%@
orthonormal basis of $\H_1^{\infty}$ consists of all different infinite strings $e_{i_1}\ot %%@
\cdots e_{i_n}\ot \cdots$, where $e_{i_n}$ is $e_i$ regarded as a vector in $\mathC^2$. (And %%@
similarly, when we choose the ``building-block'' algebra $\CA_1 \neq \mathC^2$.)  We denote %%@
the multi-index $(i_1,\ldots, i_n,\ldots)$ simply by $I$,
and the corresponding basis vector by $e_{I}$. This  \Hs\ $\H^{\infty}_1$  carries a natural %%@
faithful \rep\ $\pi$ of $\CA^{\mathrm (q)}_0$:  if $A_0\in \CA^{\mathrm (q)}_0$ is an %%@
equivalence class $[A_1,A_2,\cdots]$, then $\pi(A_0)e_I=\lni A_Ne_i$, where $A_N$ acts on the %%@
first $N$ components of $e_I$ and leaves the remainder unchanged.

 Now the important point is that
although each $\CA_1^N$ acts irreducibly on $\H^N_1$, the \rep\ $\pi(\CA^{\mathrm (q)}_0)$
on $\H^{\infty}_1$ thus constructed is highly reducible. The reason for this is that by %%@
definition (quasi-) local elements of $\CA^{\mathrm (q)}_0$ leave the infinite tail of a %%@
vector in $\H^{\infty}_1$ (almost) unaffected, so that vectors with different tails lie in %%@
different superselection sectors.
Without the quasi-locality condition on the elements of $\CA^{\mathrm (q)}_0$, no %%@
superselection rules would arise.

 For example, in terms of the usual basis
$\left\{\up=\left( \begin{array}{c} 1 \\ 0 \end{array} \right), \down=\left( \begin{array}{c} %%@
0 \\ 1 \end{array} \right)\right\}$
 of $\mathC^2$, the vectors  $\Ps_{\up}=\up\ot\up\cdots \up\cdots$ (i.e.\ an infinite product %%@
of `up' vectors) and $\Ps_{\down}=\down\ot\down\cdots \down\cdots$ (i.e.\ an infinite product %%@
of `down' vectors) lie in different sectors. (Cf. items (a) and (b) in Section \ref{chains}, %%@
and (iv) in Section \ref{macro}.)  The reason why the inner product
$(\Ps_{\up}, \pi(A)\Ps_{\down})$ vanishes  for any $A\in\CA^{\mathrm (q)}_0$ is that for %%@
local quantities $A$ one has
$\pi(A)=A_M\ot 1\ot\cdots 1\cdots$ for some  $A_M\in\CB(\H_M)$;
the inner product in question therefore involves infinitely many factors $(\up, %%@
1\down)=(\up,\down)=0$. For quasilocal $A$ the operator $\pi(A)$ might have a small %%@
nontrivial tail, but again the inner product  vanishes by an approximation argument.

 More generally,
elementary analysis shows that $(\Ps_u, \pi(A)\Ps_v)=0$ whenever $\Ps_u=\ot^{\infty}u$ and
$\Ps_v=\ot^{\infty}v$ for unit vectors $u,v\in\mathC^2$ with $u\neq v$. The corresponding %%@
vector states $\ps_u$ and $\ps_v$ on $\CA^{\mathrm (q)}_0$ (i.e.\ $\ps_u(A)=(\Ps_u, %%@
\pi(A)\Ps_u)$ etc.) are obviously permutation-invariant and hence classical. Identifying
$\CS(M_2(\mathC))$ with $B^3 \subset \R^3$, the corresponding limit state
$(\ps_u)_0$ on $\CA_0^{\mathrm (c)}$ defined by $\ps_u$ is  given by (evaluation at) the %%@
point $\til{u}=(x,y,z)$ of
$\partial  B^3=S^2$ (i.e.\ the two-sphere) for which the corresponding density matrix %%@
$\rh(\til{u})$ is the projection operator onto $u$.
(It follows that  $\ps_u$ and $\ps_v$ are unitarily inequivalent.)

We conclude that each unit vector $u\in\mathC^2$ determines a superselection sector $\pi_u$, %%@
namely the GNS-\rep\ of the corresponding state $\ps_u$, and that each such sector is %%@
realized as a subspace $\H_u$ of $\H^{\infty}_1$ (viz.\ $\H_u=\ovl{\pi(\CA^{\mathrm %%@
(q)}_0)\Ps_u}$). Moreover, since a permutation-invariant state on $\CA^{\mathrm (q)}_0$
is pure iff it is of the form $\ps_u$, these are all the superselection sectors.
Thus we have the subspace
(of $\H^{\infty}_1$) and sub\rep\ (of $\pi$)
\begin{eqnarray}
\H_{\GS}&:=&\oplus_{\til{u}\in S^2} \H_u; \nonumber \\
\pi_{\GS}(\CA^{\mathrm (q)}_0)&:=& \oplus_{\til{u}\in S^2} \pi_u(\CA^{\mathrm %%@
(q)}_0),\label{repsubrep}
\end{eqnarray}
where $\pi_u$ is simply the restriction of $\pi$ to $\H_u\subset \H^{\infty}_1$.

In the presence of superselection, there are operators that distinguish different sectors %%@
whilst being a multiple of the unit in each sector; cf. items (c) and (d) of Section %%@
\ref{chains}. In the framework developed from Section \ref{deform} onwards, these operators %%@
are the macroscopic quantities of Section \ref{macro}. In fact, one can show for any %%@
approximately symmetric sequence $(A_1,A_2,\cdots)$
the limit
\begin{equation} \ovl{A}=\lni\pi_{\GS}(A_N) \label{slim}\end{equation}
exists in the strong operator topology on $\CB(\H_{\GS})$. Moreover, let $A_0\in %%@
\CA_0^{\mathrm (c)}=C(\CS(\CA_1))$ be the function defined by the given sequence (recall that %%@
$A_0(\om)=\lni \om^N(A_N)$, cf. eq. \ref{omN}). Then the map $A_0\mapsto \ovl{A}$ defines a %%@
faithful \rep\ of $\CA_0^{\mathrm (c)}$ on $\H_{\GS}$, which we again call $\pi_{\GS}$.  A  %%@
calculation shows that $\pi_{\GS}(A_0)\Ps= A_0(\til{u})\Ps$ for $\Ps\in \H_u$, or, in other %%@
words,
\begin{equation}
\pi_{\GS}(A_0)=\oplus_{\til{u}\in S^2} A_0(\til{u})1_{\H_u}. 
\label{macrojob}
\end{equation}
Thus the $\pi_{\GS}(A_0)$ are indeed the promised operators.

\subsubsection{Emergence in the limit: with reduction---and without}\label{ssnwithwithout}
I turn at last to how the example of superselection illustrates my claims. It is clear from %%@
the wealth of details in the preceding Sections that my positive claims, (1:Deduce) and %%@
(2:Before), can be richly illustrated, in terms of both quantities and states. So I will %%@
confine myself to mentioning the obvious illustrations, and referring to the previous %%@
discussions and equations. 

As to (1:Deduce): we again have `reduction as deduction'  in as strong a sense as you could %%@
demand---provided we take the limit. There are several illustrations, some concerning %%@
quantities or algebras of them, some concerning states. I begin with two brief cases, %%@
referring back to equations in Subsections of Section \ref{supersubsec}. 

An obvious case, which concerns quantities, is that symmetric sequences $(A_N)$ of %%@
quantities, $A_N \in \CA_1^N$, commute in the limit $N \raw \infty$. Recall eq. \ref{aprc} in %%@
Section \ref{macro}. Another obvious case, which equally concerns states, is spin density in %%@
spin chains, as in part (c) of Section \ref{chains}. Thus recall the limiting behaviour of %%@
the 
average, over $2N+1$ sites, of the spin matrices, i.e. the limiting behaviour of ${\bf %%@
m}_N^{(+)}$ defined by  eq. \ref{defNm}. The discussion from eq. \ref{mfiniteexpect} to eq. %%@
\ref{mfiniteexpect2} deduced that the limiting spin density has the same value, $\bf k$, for %%@
all states in the representation ${\cal H}^{(+)}$; and so it is a classical quantity, or %%@
superselection operator. 

 Here are three more substantial cases of such a deduction, using my mnemonic notations, %%@
$T_b$ and $T_t$: that is, cases where $T_b$ implies $T_t$. The first two concern quantities, %%@
the third concerns states. Here, I shall mostly refer back to the summary in Section %%@
\ref{chainsagain}.

\indent (A): Take as $T_b$ the continuous field of \ca s $\CA^{\mathrm (q)}$, together with %%@
its representation theory: or more modestly, together with that part of its representation %%@
theory that deals with permutation-invariant states. Take as $T_t$ a statement of %%@
superselection, e.g. that $\CA_0^{\mathrm (q)}$ acts (highly!) reducibly on the %%@
infinite-system state space $\H^{\infty}_1$. Then we have just seen in Section %%@
\ref{chainsagain} that $T_b$ implies $T_t$. For the requirement for sequences of operators to %%@
be quasi-local makes the inner products between states in different sectors of %%@
$\H^{\infty}_1$ vanish. Compare the discussion leading up to eq. \ref{repsubrep}.\\
\indent
(B): Let $T_b$ be the continuous field of \ca s $\CA^{\mathrm (c)}$, together with %%@
$\pi_{\GS}$, the 
faithful \rep\ of $\CA_0^{\mathrm (c)}$ on $\H_{\GS}$  defined by eq. \ref{slim}.
Take $T_t$ to state that there are superselection operators (classical quantities) that %%@
restrict to the identity on each sector $\H_{\GS}$.  Then we have just seen at the end of %%@
Section \ref{chainsagain} that $T_b$ implies $T_t$. For compare the discussion leading up to %%@
eq. \ref{macrojob}: any approximately symmetric sequence $(A_N)$ (and so any continuous %%@
section of the field) defines such a classical quantity, viz. $\pi_{\GS}(A_0)$. \\
\indent
(C): We can focus instead on states.  Take $T_b$ to encompass both continuous fields of \ca %%@
s, $\CA^{\mathrm (q)}$ and $\CA^{\mathrm (c)}$, and to consider families of states on them.   %%@
Take $T_t$ to be the theory of classical states, especially permutation-invariant, states %%@
$\om^{\mathrm (q)}_0$; or more modestly, to be just the quantum de Finetti representation %%@
theorem, eq. \ref{Unn}. Then Section \ref{compareFinetti} shows that $T_b$ implies $T_t$. 

I turn to the ``the other side of the coin'' of (1:Deduce): the failure of the reduction at %%@
finite $N$. For (A) and (B), the point is familiar from elementary quantum theory. Finite $N$ %%@
means there is no algebra $\CA_0^{\mathrm (c)}$ or $\CA_0^{\mathrm (q)}$ to consider: there %%@
are only the algebras $\CA_N^{\mathrm (c)} \equiv \CA_N^{\mathrm (q)} := \CA_1^N$. But in %%@
general, if $\CA_1$ acts irreducibly on $\H_1$, then so does $\CA_1^N$ on $\H_1^N$. So (A) %%@
fails. And then Schur's Lemma---that if an algebra acts irreducibly, its commutant is trivial %%@
---immmediately implies that (B) fails, i.e. there are only trivial superselection operators.

As to (C), the point is less familiar. But it is again straightforward, especially if we %%@
limit ourselves by choosing $T_t$ to be the quantum de Finetti representation theorem, not %%@
the whole theory of classical states; (of course choosing a weaker $T_t$ adds to the %%@
dialectical force of the point that it is not implied). Here the point is that for finite %%@
$N$, the theorem fails. The reason is essentially the same as for the corresponding failure %%@
of the classical theorem for finite $N$. In short, finite $N$ corresponds to drawing from an %%@
urn without replacement, rather than with replacement. In the classical case, this means that %%@
a permutation-invariant probability measure is a unique mixture, not of product measures, but %%@
of hypergeometric measures (cf. Diaconis 1977, Diaconis and Freedman 1980, Jeffrey 1988, pp. %%@
240-245). Similar remarks apply in the quantum case: (C) of Section \ref{ssnbefore} will give %%@
a few more details.

Finally, I note two respects in which this example's illustration of (1:Deduce) is similar to %%@
that given by fractals, and dissimilar to that given by the method of arbitrary functions. I %%@
noted these respects at the end of Section \ref{dimwithwithout}. First: the emergent %%@
behaviour involves mathematical structures which are new at the limit (such as unitarily %%@
inequivalent representations, and so superselection). This is like fractals, where the %%@
emergent behaviour involved non-integer dimension at the limit; but unlike the method of %%@
arbitrary functions, where the emergent behaviour  involved, more simply, deducing the %%@
limiting behaviour of functions given at finite $N$. 

The second respect is related to the first. Namely, this example considers an infinite %%@
quantum system: in Section \ref{deflatelimits}'s notation, an infinite system %%@
$\sigma(\infty)$. Again, this is like fractals, where the Cantor set, Koch snowflake etc. %%@
count as infinite systems; but unlike the method of arbitrary functions, since, for example, %%@
no roulette wheel has infinitely many arcs.

\subsubsection{Emergence before the limit}\label{ssnbefore}
(2:Before) claims that before the limit, there is emergence in a weaker but still vivid %%@
sense. This claim is illustrated in a manner parallel to my previous examples: the method of %%@
arbitrary functions, and fractals. In short, one just has to interpret (sensibly!) the %%@
example's formalism describing finite $N$. And as in those examples, the discussion can be %%@
made vivid by referring to practical purposes. There, I cited the practical purposes of a %%@
casino in making a wheel that is fair enough; and the purposes of a film studio in making an %%@
image look fractal at small enough spatial scales. Similarly here: imagine an experimental %%@
physicist making, or a theoretical physicist describing, a sample or device comprising $N$ %%@
atoms or sites (perhaps for a nanotechnology project), with $N$ large enough for behaviour %%@
characteristic of superselection to occur.   

Again, one can pick several illustrations. I shall give three. The first concerns states in %%@
spin chains as in Section \ref{chains}; the second concerns symmetrized quantities as in %%@
Section \ref{macro}; the third returns us to the quantum de Finetti representation theorem, %%@
as in Section \ref{ssnwithwithout}. All three have the merit (absent from my discussion of my %%@
previous examples!) of being {\em quantitative} about the ``rate of emergence'': about how %%@
large an $N$ is needed for emergent behaviour.

\indent (A):
The discussion in part (c) of Section \ref{chains} concentrated on the limiting behaviour of %%@
the expectation value of ${\bf m}_N^{(+)}$ on states in ${\cal H}^{(+)}$. We deduced (cf. eq. %%@
\ref{mfiniteexpect} to eq. \ref{mfiniteexpect2})  that the limiting  value is $\bf k$, the %%@
unit vector in the $z$-direction. But that discussion can of course be adapted to entail %%@
statements about the situation for finite $N$. To give one example: consider the $x$ or $y$ %%@
component of the matrix element of  ${\bf m}_N^{(+)}$ between two elements $\phi_s^{(+)}, %%@
\phi_{s'}^{(+)}$ of the basis of ${\cal H}^{(+)}$ that correspond to two doubly infinite %%@
(+1,-1)-sequences, $s = (s_n)_{n \in Z}$ and $s' = (s'_n)_{n \in Z}$, both with all but %%@
finitely many elements ($s_n$ or $s'_n$) equal to +1. In part (c) of Section \ref{chains}, we %%@
argued that these matrix elements tend to 0 as $N \raw \infty$. But as to finite $N$: if we %%@
know that $s_n = s'_n$ for all $n$ with $| n | > M$, we can readily calculate how large $N$ %%@
needs to be for the $2N + 1$ denominator to make the $x$ or $y$ component of the matrix %%@
element have modulus less than any given $\varepsilon$. 

\indent (B): The discussion surrounding eq. \ref{av0} and \ref{av0spin} in Section %%@
\ref{chains} concentrated on the limiting behaviour, as $N \raw \infty$, of the commutator of %%@
averages over $N$ particles of 1-particle spin operators. But as in (A), that discussion  of %%@
course entails statements about the situation for finite $N$. Indeed, eq. \ref{av0} and %%@
\ref{av0spin} say explicitly that these commutators are proportional to $1/N$. 

(C): Section \ref{ssnwithwithout} reported that the quantum de Finetti representation %%@
theorem, eq. \ref{Unn}, failed for finite $N$: a permutation-invariant state on $N$ quantum %%@
systems need not be mixture of product states. But in recent years, various finite-$N$  %%@
analogues have been proven, with the following flavour: any state of $M$ systems, with $M < %%@
N$, that is obtained from a permutation-invariant state on $N$ systems, by tracing out %%@
(partial tracing over) $N - M$ systems, can be approximated by a mixture of product states on %%@
the $M$ systems (i.e. states $\omega^{\otimes M}$), with an error that goes to zero with the %%@
ratio $M/N$, e.g. an error like $O(M/N)$; (Koenig and Renner 2005, Renner 2007, Koenig and %%@
Mitchison 2007). 

Note finally that as with my previous two examples, in these three illustrations of %%@
(2:Before) we again see the Straightforward Justification of Section \ref{strtfwd} in action. %%@
I will not labour the point: I will just quote Landsman's own statement of the idea of that %%@
Section's fifth paragraph, as applied to the averaging process used (in Section \ref{macro}) %%@
in defining macroscopic quantities. Thus Landsman writes: `the limit $N \raw \infty$ is valid %%@
whenever {\em averaging} over $N = 10^{23}$ particles is well approximated by averaging over %%@
an arbitrarily larger number $N$ (which, then, one might as well let go to infinity)' %%@
(Landsman 2006, preamble to Section 6; p. 493).  

\subsubsection{Supervenience is a red herring}\label{redherring3}
As in two previous examples, I shall be brief about my third claim, (3:Herring): that
although various supervenience theses are true, they give little or no insight into the %%@
emergent behaviour, or more generally into ``what is going on'' in the example. The reason is %%@
as in the previous examples: there is no connection between supervenience's idea of a variety %%@
of ways to have a higher-level property $P$ (in particular the example's emergent property) %%@
and the limit processes on which the example turns. 

But to save space, I will not formulate any such supervenience theses. I leave it as an %%@
exercise (!) to formulate how a system's having an emergent property, such as a specific %%@
value for a specific classical quantity, or a superselection operator,  supervenes on the %%@
system's microstate, i.e. on the sequence of states assigned to the algebras ${\cal A}_1^N$ %%@
for successively larger $N$-particle (or site) sub-systems. Suffice it to say here that the %%@
important points will be that:\\
\indent (i) the value (indeed, even just the well-definedness) of the quantity or operator, %%@
in the strong sense of (1:Deduce) but not the weak sense of (2:Before), needs one to take the %%@
limit $N \raw \infty$, in the classical (Section \ref{macro}) and-or quantum (Section %%@
\ref{quasilocal}) version; cf. the end of Section \ref{chainsagain};\\
\indent (ii) the idea of supervenience on the microstate makes no connection with taking this  %%@
limit.

\subsection{Summing up superselection}\label{ssnsumup}
To emphasise the parallel between this long example and the previous two, let me sum up with %%@
a list of six morals which are parallel to those in Section \ref{fractalsumup}. As in that %%@
Section, it would be flogging a dead horse to again make explicit my four claims, or Section %%@
\ref{strtfwd}'s Straightforward Justification, or the parallels with previous examples.

\indent (i): The large finite is often well-modelled by the infinite. \\
\indent (ii): Such models are often justified in a straightforward, even obvious, way, by %%@
mathematical convenience and empirical success.\\
\indent (iii): The infinite often brings new mathematical structure: in this example, %%@
superselection, and associated notions like unitarily inequivalent representations.\\
\indent (iv): Nevertheless, there is often a reduction: the emergent superselection %%@
properties, and the associated behaviour, are reducible to a sufficiently rich theory that %%@
takes the infinite limit.\\
\indent (v): On the other hand, one can often see emergent behaviour on the way to the limit. %%@
Indeed, the larger your error bar, e.g. for detecting experimental statistics characteristic %%@
of some commuting operators, the lower the number of particles (or lattice sites) $N$ for %%@
which your experiments will confirm (more precisely: you think your experiments confirm!) the %%@
properties and behaviour that are characteristic of superselection.\\
\indent (vi): Various supervenience theses hold---but they are trivial, or at least %%@
scientifically useless.

\section{Phase transitions}\label{phasetr}
Lack of space means I must deal much more briefly with my fourth example. This is %%@
unfortunate, for two reasons. Scientifically, this example represents a much larger and more %%@
controversial topic than my previous ones. And philosophically, it has been a prominent %%@
example in recent controversy about whether the $N = \infty$ limit is ``physical real'', and %%@
whether a ``singular'' limit is necessary for emergence (e.g. Callender (2001, Section 5, pp. %%@
547-551), Liu (2001, Sections 2-3, pp. S326-S341), Batterman (2005, Section 4, pp. 233-237); %%@
and more recently, Mainwood (2006, Chapters 3,4; 2006a), Bangu (2009, Section 5, pp. %%@
496-502), Menon and Callender (2011, especially Sections 3, 4)). But sufficient unto the day %%@
is the work thereof! I have already declared my general position in these controversies %%@
(especially in Sections \ref{Intr} and \ref{ssqv}). So in this Section, it will be enough to %%@
sketch: (i)  how statistical mechanics treats phase transitions by taking a limit, in which %%@
the number of constituent particles (or sites in a lattice) $N$ goes to infinity (Section %%@
\ref{phtrexpound}); and (ii) how this treatment illustrates my claims (Section %%@
\ref{emergephtrn}).

For more information, I especially recommend: (i)  accounts by masters of the subject, such %%@
as Emch and Liu (2002, Chapters 11-14) and Kadanoff (2009, 2010, 2010a), which treat the %%@
technicalities and history, as well as the conceptual foundations, of the subject; and (ii) %%@
Mainwood (2006, Chapters 3,4; 2006a), to which I am much indebted, especially in Section %%@
\ref{phtrnwithwithoutpaul}'s treatment of phase transitions in finite $N$ systems, and %%@
Section \ref{crossover}'s discussion of cross-over.     

\subsection{Phase transitions and thermodynamics}\label{phtrexpound}
 
\subsubsection{Separating issues and limiting scope}\label{seplimiting}
This example is an aspect of a very large topic, the ``emergence'' of thermodynamics from %%@
statistical mechanics---around which debates about the reducibility of one to the other %%@
continue. This topic is very large, for various reasons. Three obvious ones are: (i) both %%@
thermodynamics and statistical mechanics are entire sciences; (ii) statistical mechanics, and %%@
so this topic, can be developed in either classical or quantum terms; (iii) there is no %%@
single agreed formalism for statistical mechanics (unlike e.g. quantum mechanics).

Phase transitions are themselves a large topic: there are several classification schemes for %%@
them, and various approaches to understanding them---some of which come in both classical and %%@
quantum versions. I will specialize to just one aspect; which will however be enough to %%@
illustrate my claims. Namely: the fact (on most approaches!) that for statistical mechanical %%@
systems, getting a (theoretical description of) a phase transition requires that one take a %%@
limit (often called `the thermodynamic limit'), in which the number of constituent particles %%@
(or sites in a lattice) $N$ goes to infinity. In brief, this means something like: both the %%@
number $N$  of constituent particles (or sites), and the volume $V$ of the system  tend to %%@
infinity, while the density
$\rho = N/V$ remains fixed. More details in Section \ref{needTD}.

Even for this one aspect, I will have to restrict myself severely. Three main limitations %%@
are:\\
\indent 1): As regards philosophy, I impose a self-denying ordinance about the controversies %%@
mentioned in this Section's preamble: apart from what I have already said in Sections %%@
\ref{Intr} and \ref{ssqv}---and one remark in Section \ref{phtrnwithwithoutpaul}! \\
\indent 2): In physics, how to understand phase transitions is an ongoing research area. For %%@
our purposes, the main limiting (i.e. embarrassing!) fact is that most systems do {\em not} %%@
have a well-defined thermodynamic limit---so that all that follows is of limited scope.\\
\indent 3): Various detailed justifications can be given for phase transitions requiring us %%@
to take the thermodynamic limit; and Section \ref{needTD} will only sketch a general %%@
argument, and mention two examples. Much needs to be (and has been!) said by way of assessing %%@
these justifications---but I will not enter into this here. But by way of emphasizing how %%@
open all these issues are, I note that some physicists have developed frameworks for %%@
understanding phase transitions {\em without} taking the thermodynamic limit (Gross (2001)). 

\subsubsection{The thermodynamic limit}\label{needTD}
I will first give a broad description of the need for the thermodynamic limit; and then a %%@
classical and a quantum example to show how the limit secures new mathematical structure %%@
appropriate for describing phase transitions; (Section \ref{needTD}.A). Then in Section %%@
\ref{needTD}.B, I will give a classical example of approaching the limit. This example will %%@
be the phase transition of a ferromagnet at sub-critical temperatures. It has the merit of %%@
being very simple, and of developing Section \ref{dissolvemyst}'s dissolution of the %%@
``mystery'' of describing a finite-$N$ system with a model using infinite $N$.

{\em 7.1.2.A: The need for the limit}:--- For classical physics, the brutal summary of why we %%@
need this limit is as follows. Statistical mechanics follows thermodynamics in representing %%@
phase transitions by non-analyticities of the free energy $F$. But a non-analyticity cannot %%@
occur for the free energy of a system with finitely many constituent particles (or %%@
analogously: lattice sites). So statistical mechanics considers a system with infinitely many %%@
particles or sites, $N = \infty$. One gets some control over this idea by subjecting the %%@
limiting process, $N \raw \infty$, to physically-motivated conditions like keeping the %%@
density constant, i.e. letting the volume $V$ of the system also go to infinity, while $N/V$ %%@
is constant. As we would expect---especially given my previous examples!---this infinite %%@
limit gives new mathematical structures: which happily turn out to describe phase %%@
transitions---in many cases, in remarkable quantitative detail.

But to make my ferromagnet example comprehensible, I need to spell out this line of argument %%@
in a bit more detail. In Gibbsian statistical mechanics, we postulate that the probability of %%@
a state $s$ is proportional to $\exp(-H(s)/kT) \equiv \exp(-\beta H(s))$, where $\beta := %%@
\frac{1}{kT}$ is the inverse temperature and $k$ is Boltzmann's constant. That is:
\be
{\rm{prob}}(s) = \exp(-\beta H(s)) / Z 
\label{classlGibbs}
\ee
where the normalization factor $Z$, the partition function, is the sum (or integral) over all %%@
states, and defines the free energy $F$ as:
\be
Z \; = \; \Sigma_s \exp(-\beta H(s)) \; =: \; \exp(-  \beta F) \; .
\label{defZF}
\ee  
Thus $F$ is essentially, the logarithm of the partition function; which is itself the sum (or %%@
integral) over all states of the exponential of the Hamiltonian. It turns out that $Z$ and %%@
$F$ encode, in their functional forms, a great deal of information about the system: various %%@
quantities, in particular the system's thermodynamic quantities, can be obtained from them, %%@
especially by taking their derivatives. For example, in a ferromagnet, the magnetization is %%@
the first derivative of the free energy with respect to the applied magnetic field, and the %%@
magnetic susceptibility is the second derivative. 

Now, broadly speaking: phase transitions involve abrupt changes, in time and-or space, in %%@
thermodynamic quantities: for example, think of the change of particle density in a %%@
solid-liquid, or liquid-gas, transition. Thermodynamics describes these changes as %%@
discontinuities in thermodynamic quantities (or their derivatives), and statistical mechanics %%@
follows  suit. This means that the statistical mechanical description of phase transitions %%@
requires non-analyticities of the free energy $F$. But under widely applicable assumptions, %%@
the free energy of a system with finitely many constituent particles (or analogously: sites) %%@
is an analytic function of the thermodynamic quantities within it. For example, we will soon %%@
see that in the Ising model with $N$ sites, the Hamiltonian $H$ is a quadratic polynomial in %%@
spin variables (cf. eq. \ref{IsiHamn}). This means that the partition function $Z$, which by %%@
eq. \ref{defZF} is a sum of exponentials of $- \beta H$, is analytic; and so also is its %%@
logarithm, and the free energy. (This and similar arguments about more general forms of the %%@
Hamiltonian (or partition function or free energy), are widespread: e.g. Ruelle (1969, p. %%@
108f.), Thompson (1972, p. 79), Le Bellac (1991, p. 9), Lavis and Bell (1999, pp. 72-3)) .

I of course admit that---as my phrases `broadly speaking' and `under widely applicable %%@
assumptions' indicate---this argument why phase transitions need the thermodynamic limit is %%@
not a rigorous theorem. Hence the effort mentioned in Section \ref{seplimiting} to develop a %%@
theory of phase transitions in finite systems; and the philosophical debate among Callender %%@
et al. mentioned in this Section's preamble. Hence also the historical struggle to recognize %%@
the need for infinite systems: both Emch and Liu (2002, p. 394) and Kadanoff (2009, p. 782; %%@
2010, Section 4.4) cite the famous incident of Kramers putting the matter to a vote at a %%@
meeting in memory of Van der Waals in 1937.\footnote{Of course, since I have not precisely %%@
defined `thermodynamic limit'---let alone `phase transition'!---the argument could hardly be %%@
a rigorous theorem. Ruelle (1969, Sections 2.3-4, 3.3-5) rigorously discusses conditions  for %%@
the thermodynamic limit; (cf. also Emch (1972, p. 299; 2006, p. 1159) Lavis and Bell (1999a, %%@
pp. 116, 260)). Such discussions bring out how in some models, the  limit is not just the %%@
idea that, keeping the density constant, the number $N$ of molecules or sites tends to %%@
infinity: there are also conditions on the limiting behaviour of short-range forces. This %%@
means that the models eventually run up against both aspects of my fourth claim, (4:Unreal), %%@
i.e. against what Section \ref{unreal} called `atomism' and `cosmology'.} 

 But this argument, although not a rigorous theorem, is very ``robust''---and recognized as %%@
such by the literature. For example, Kadanoff makes it one of the main themes of his recent %%@
discussions, and even  dubs it the `extended singularity theorem' (2010, Sections 2.2, 6.7.1; %%@
2010a, Section 4.1). He also makes it a playful variation on Anderson's slogan that `more is %%@
different' (as I mentioned in footnote \ref{leocredit}). Namely, he summarizes it in Section %%@
titles like `more is the same; infinitely more is different' (2009, Section 1.5; 2010, %%@
Section 3). In any case, for the rest of this paper, I accept the argument.  

Taking the thermodynamic limit introduces new mathematical structures. But (as one might %%@
expect from my previous examples) the variety of formalisms in statistical mechanics (and %%@
indeed, the variety of justifications for taking the limit) means that there is a concomitant %%@
variety of new structures that in the limit get revealed. I mention one classical, and one %%@
quantum, example.

In Yang-Lee theory (initiated by Yang and Lee 1952), one uses complex generalizations of the %%@
partition function and free energy, and then argues that for any $z \in \mathC$, there can be %%@
a phase transition (i.e. a non-analyticity of $F$ or $Z$) at $z$ only if there are zeroes of %%@
$Z$ arbitrarily close to $z$. For finite $N$, $Z$ has finitely many zeroes, so that there can %%@
be a phase transition only at the zeroes themselves: but all of them lie off the real line, %%@
and so are not physical. Taking the limit $N \raw \infty$ ``breaks'' this last argument: %%@
there can be a curve of zeroes that intersects the real axis. Indeed, in Yang-Lee theory one %%@
goes on to classify phase transitions in terms of the behaviour of the density of zeroes in %%@
$\mathC$: (cf. Thompson 1972, pp. 85-88; Ruelle 1969, pp. 108-112; Lavis and Bell 1999a, pp. %%@
114, 125-134).

My quantum example concerns Gibbs states and KMS states. This  follows on from Section %%@
\ref{superselec}'s discussion of superselection: especially Section \ref{chains} on unitarily %%@
inequivalent representations of an algebra of quantities, and these representations differing %%@
in the value of a global/macroscopic quantity. (For more details, cf. Emch (1972, pp. %%@
213-223; 2006, Section 5.6-7, pp. 1144-1154); Sewell (1986, pp. 73-80; 2002, pp. 113-123); %%@
Emch and Liu (2002, pp. 346-357),  Liu and Emch (2005, pp. 142-145, 157-161).) 

Thus we recall that the Gibbs state of a finite quantum system with Hamiltonian $H$ at %%@
inverse temperature $\beta = \frac{1}{kT}$ is given by the density matrix
\begin{equation}
\rho = \exp(- \beta H) / {\rm tr}(\exp(- \beta H))
\end{equation}
and represents the (Gibbsian) equilibrium state of the system. (Note the beautiful analogy %%@
with eq.s \ref{classlGibbs} and \ref{defZF}!) It is unique (for given $\beta$): thereby %%@
precluding the representation of two phases of the system at a common temperature---as one %%@
would want for a phase transition.\footnote{This uniqueness also precludes spontaneous %%@
symmetry breaking, understood (as usual) as the allowance of distinct equilibria that differ %%@
by a dynamical symmetry---e.g. in Section \ref{chains}'s scenario, ground states each with %%@
all spins aligned, but in different spatial directions. Spontaneous symmetry breaking is (yet %%@
another!) important aspect of phase transitions which I cannot here pursue: a fine recent %%@
philosophical discussion is Liu and Emch (2005).}

So how can we give a quantum description of phase transitions? The algebraic approach to %%@
quantum statistical mechanics proposes some states, viz. KMS states, which are defined on %%@
infinite quantum systems  and which generalize the notion of a Gibbs state in a way that is %%@
(a) compelling mathematically, and (b) well-suited to describing phase transitions. A word %%@
about each of (a) and (b):---\\
\indent (a): A mathematical property of Gibbs states (the `KMS condition') is made into a %%@
definition of an equilibrium state that is applicable to both infinite and finite systems: %%@
(for the latter it coincides with the Gibbs state at the given temperature). KMS states can %%@
be shown to have various stability or robustness properties that makes them very well suited %%@
to describe (stable) physical equilibria. (Emch (2006, Section 5.4, pp. 1128-1142) is a %%@
excellent survey of these properties. Such a survey brings out how KMS states could %%@
themselves form an example of emergent behaviour, in my sense of novel and robust %%@
properties!)   \\
\indent (b): The set $K_{\beta}$ of KMS states at a given inverse temperature $\beta$ is in %%@
general not a singleton set. Rather, it is convex, with: (i) every element having a unique %%@
expression as a mixture of its extremal points; and (ii) its extremal points being %%@
well-suited to describe pure thermodynamical phases (mathematically, they are factor states). %%@
Taken together, (i) and (ii) suggest that a compelling representative of the state of a %%@
system undergoing a phase transition at inverse temperature $\beta$ is a non-extremal $\om %%@
\in K_{\beta}$. 

So much by way of indicating justifications for taking ``the'' thermodynamic limit. I turn to %%@
discussing {\em the approach} to the limit.

{\em 7.1.2.B: Approaching the limit}:---
I will give a classical example of approaching the limit $N \raw \infty$ of infinitely many %%@
sites in a lattice: namely, the phase transition (change of magnetization) of a ferromagnet %%@
at sub-critical temperatures, as described by the Ising model with $N$ sites (in two or more %%@
spatial dimensions). As I mentioned at the start of this Section, this will develop Section %%@
\ref{dissolvemyst}'s dissolving of the ``mystery'' of describing a finite-$N$ system with a %%@
model using infinite $N$.

The Ising model postulates that at each of $N$ sites, a classical ``spin'' variable $\sigma$ %%@
(which we think of as defined with respect to some spatial direction) takes the values $\pm %%@
1$. To do Gibbsian statistical mechanics, i.e. to apply eq.s \ref{classlGibbs} and %%@
\ref{defZF}, we need to define a Hamiltonian and then sum over configurations. The %%@
Hamiltonian is chosen to give a simple representation of the ideas that (i) neighbouring %%@
spins interact and tend to be aligned (i.e. their having equal values has lower energy) and %%@
(ii) the spins are coupled to an external magnetic field which points along the given spatial %%@
direction. Thus the Hamiltonian is
\be
H \; \; = \; \; J \; \Sigma_{nn} \sigma_p \sigma_q \;\; +\; \; J' \; \Sigma \sigma_p \; . 
\label{IsiHamn}
\ee  
where: the first sum is over all pairs of nearest-neighbour sites, the second sum is over all %%@
sites, $J$ (with the dimension energy) is negative to represent that the neighbouring spins %%@
``like'' to be aligned, and $J'$ is given by the magnetic moment times the external magnetic %%@
field.

The simplest possible case is the case of $N = 1$! With only one site, the Hamiltonian  %%@
becomes
\be
H = J' \sigma  \; ;
\ee
so that if we define a dimensionless coupling $h := -J'/kT$, then eq.s \ref{classlGibbs} and %%@
\ref{defZF} give
\be
{\rm{prob}}(+1) = e^h / z \;\; {\rm{and}} \;\; {\rm{prob}}(-1) = e^{-h} / z \;\; , \; %%@
{\rm{with}} \;\;
z = e^{h} + e^{-h} = 2 \cosh h \; .
\label{Isi1}
\ee
This implies that the magnetization, i.e. the average value of the spin, is
\be
\langle \sigma \rangle = e^{h}/z - e^{-h}/z = \tanh h \; .
\ee
This is as we would hope: the statistical mechanical treatment of a single spin predicts the %%@
magnetization increases smoothly from -1, through zero, to +1 as the applied field along the %%@
given axis increases from minus infinity through zero to plus infinity.

What about larger $N$? The analytical problem becomes much more complicated (though the %%@
magnetization is still a smooth function of the applied field). But the effect is what we %%@
would expect: a larger $N$ acts as a brake on the ferromagnet's response to the applied field %%@
increasing from negative to positive values (along the given axis).  That is: the increased %%@
number of nearest neighbours means that the ferromagnet ``lingers longer'', has ``more %%@
inertia'', before the rising value of the applied field succeeds in flipping the %%@
magnetization from -1 to +1. More precisely: as $N$ increases, most of the change in the %%@
magnetization occurs more and more steeply, i.e. occurs in a smaller and smaller interval %%@
around the applied field being zero.  Thus the magnetic susceptibility, defined as the %%@
derivative of magnetization with respect to magnetic field, is, in the neighbourhood of 0, %%@
larger for larger $N$, and tends to infinity as $N \rightarrow \infty$. As Kadanoff says: `at %%@
a very large number of sites ... the transition will become so steep that the causal observer %%@
might say that it has occurred suddenly. The astute observer will look more closely, see that %%@
there is a very steep rise, and perhaps conclude that the discontinuous jump occurs only in %%@
the infinite system' (2009, p. 783; and Figure 4; cf. also 2010, p. 20, Figure 5).

Clearly, this example corresponds closely to that in Section \ref{dissolvemyst}'s dissolving %%@
of the ``mystery''. And this general picture of the approach to the $N \raw \infty$ limits %%@
applies much more widely. In particular, very similar remarks apply to liquid-gas phase %%@
transition, i.e. boiling. There the quantity which becomes infinite in the $N \rightarrow %%@
\infty$ limit, i.e. the analogue of the magnetic susceptibility, is the compressibility, %%@
defined as the derivative of the density with respect to the pressure. And I am happy to give %%@
a hostage to fortune: as I declared in Section \ref{dissolvemyst}'s footnote \ref{hunch}, I %%@
believe this example is a good prototype for dissolving the corresponding alleged mystery in %%@
physics' other `singular' limits. Though I cannot argue for that here, I note that the views %%@
of some philosophers discussing phase transitions seem to mesh with it: e.g. Mainwood's %%@
proposal in Section \ref{phtrnwithwithoutpaul} below; Bangu's appeal (2009, p. 497f.) to %%@
Bogen and Woodward's (1988) distinction between data and phenomena; and Menon and Callender %%@
(2011, Sections 3, 4).

\subsection{The claims illustrated by emergent phase transitions}\label{emergephtrn}
I will not now devote a Subsection to each of my three main claims, (1:Deduce), (2:Before) %%@
and (3:Herring), as I did for my first three examples. It would take too much space---and %%@
much more detail that Section \ref{phtrexpound} has supplied---to do so. Thus (1:Deduce) %%@
would require me to properly define: (a) a handful of novel and robust behaviours shown in %%@
phase transitions (a handful of $T_t$s), and (b) a corresponding handful of statistical %%@
mechanical theories $T_b$ in which the behaviours are rigorously deducible if one takes an %%@
appropriate  version of ``the'' thermodynamic limit, $N \raw \infty$---but not otherwise. %%@
Similarly, for (2:Before) and (3:Herring). Doing all that properly would require a Section as %%@
long as Section \ref{superselec} ... and as to (3:Herring), I would anyway prefer not to flog %%@
horses that by now should be dead! 

Instead, I will just summarize how phase transitions illustrate the three claims, and endorse %%@
a  proposal of Mainwood's about emergence before the limit: i.e. about how to think of phase %%@
transitions in finite-$N$ systems (Section \ref{phtrnwithwithoutpaul}). Then I will briefly %%@
report a remarkable class of phenomena associated with phase transitions, viz. cross-over %%@
phenomena; (Section \ref{crossover}). These phenomena make emergence before the limit even %%@
more vivid than it was in my previous examples; for they show how an emergent phenomenon can %%@
be first gained, then {\em lost}, as we approach a phase transition. And this will illustrate %%@
(4:Unreal) as well as (2:Before). (For (ii), as for (i), I learnt what follows from Paul %%@
Mainwood (2006, especially Chapters 3 and 4). So all this Subsection owes a great deal to %%@
him.)

\subsubsection{Emergence in the limit, and before it: Mainwood's %%@
proposal}\label{phtrnwithwithoutpaul}
My main claims are my two reconciling claims, (1:Deduce) and (2:Before). Applied to phase %%@
transitions, they would  say, roughly speaking:\\
\indent (1:Deduce): Some of the emergent behaviours shown in phase
transitions are, when understood (as in thermodynamics) in terms of non-analyticities, %%@
rigorously deducible within a statistical mechanical theory that takes an appropriate version %%@
of the $N \raw \infty$ limit. But they are not deducible in a theory that sticks to finite %%@
$N$; so that if one concentrates on finite $N$, one will claim irreducibility.\\
\indent (2:Before): But these behaviours can also be understood more weakly; (no doubt, this %%@
is in part a matter of understanding them phenomenologically). And thus understood, they %%@
occur before the limit, i.e. in finite-$N$ systems.

Here I admit that the phrases `some of the emergent', `appropriate version' and `can be %%@
understood more weakly' are vague in ways which, as I said in the preamble, I have not the %%@
space to make  precise: hence my saying `roughly speaking'. But I still submit that the %%@
claims are true, for a wide class of emergent behaviours; and that Section \ref{needTD} gives %%@
good evidence for this. More precisely, Section \ref{needTD}.A supports (1:Deduce), and %%@
Section \ref{needTD}.B supports (2:Before).\footnote{Besides, a theory of phase transitions %%@
in finite systems of the kind argued for by Gross (mentioned in Section \ref{seplimiting}) %%@
would surely illustrate (2:Before), rather than refute (1:Deduce)'s negative claim of %%@
non-deducibility at finite $N$. For of course, a theory like Gross' cannot overturn pure %%@
mathematical arguments, for example about the analyticity of certain forms of partition %%@
function.} By way of summing up: one can check that the six previous morals, (i) to (vi), %%@
that were used in Sections \ref{fractalsumup} and \ref{ssnsumup} to sum up the fractals and %%@
superselection examples, apply again.

Finally, I would like to briefly report and endorse a proposal of Mainwood's (2006, Section %%@
4.4.1, p. 238; 2006a, Section 4.1) which fits well with the swings-and-roundabouts flavour of %%@
my combining (1:Deduce), especially its negative claim of non-deducibility at finite $N$, %%@
with (2:Before). Mainwood's topic is, not emergence in general, but the recent philosophical %%@
debate about phase transitions in finite systems, especially as focussed by Callender's %%@
(2001, p. 549) presentation of four jointly contradictory propositions about phase %%@
transitions. Mainwood first gives a very judicious survey of the pros and cons of denying %%@
each of the four propositions, and then uses its conclusions to argue for a proposal that %%@
evidently reconciles: (a) statistical mechanics' use of the thermodynamic limit to describe %%@
phase transitions in terms of non-analyticities; and  (b) our saying that phase transitions %%@
occur in the finite system. 
That is, to take a stock example: Mainwood's proposal secures that a kettle of water, though %%@
a finite system, can boil!

Mainwood's proposal is attractively simple. It is that for a system with $N$ degrees of %%@
freedom, with a free energy $F_N$ that has a well-defined thermodynamic limit, $F_N \raw %%@
F_{\infty}$, we should just say:\\
\indent\indent  phase transitions occur in the finite system iff $F_{\infty}$ has %%@
non-analyticities.\\
And if we wish, we can add requirements that avoid our having to say that small systems (e.g. %%@
a lattice of four Ising spins laid out in a square) undergo phase transitions. Namely: we  %%@
can add to the above right-hand side conditions along the following lines: {\em and} if $N$ %%@
is large enough, or the gradient of $F_N$ is steep enough etc. (Of course, `large enough' %%@
etc. are vague words. But Mainwood thinks that the consequent vagueness about whether a phase %%@
transition occurs is acceptable; and I agree---after all, `boil' etc. are vague.)

\subsubsection{Cross-over: gaining and losing emergence at finite $N$}\label{crossover}
I end by describing {\em cross-over phenomena}. I again follow Mainwood, who uses it (2006, %%@
Sections 4.4.2-3, pp. 242-247; 2006a, Section 4.2) to illustrate and defend his proposal for %%@
phase transitions in finite systems. I concur with that use of it. But my own aims are rather %%@
different. The main idea (mentioned at the start of Section \ref{unreal} and the end of %%@
Section \ref{nature?}) will be that cross-over phenomena yield striking illustrations of %%@
``oscillations'' between (2:Before) and (4:Unreal). That is: a system can be:\\
\indent (i) first,  manipulated so as  to illustrate (2:Before), i.e. an emergent behaviour %%@
at finite $N$; and \\
\indent (ii) then manipulated so as  to lose this behaviour, i.e. to illustrate (4:Unreal); %%@
by the manipulation corresponding to higher, and unrealistic, values of $N$; and  \\
\indent  (iii) then manipulated so as  to either (a) enter a regime illustrating some other %%@
emergent behaviour, or (b) revert to the first emergent behaviour; so that either (a) or (b) %%@
illustrate (2:Before) again.\\
In short: cross-over will illustrate my swings-and-roundabouts combination of (2:Before) and %%@
(4:Unreal). 

Besides, cross-over will illustrate a simpler point about (2:Before), which we already saw %%@
for my first two examples: viz., how the emergent behaviour that one ``sees'' at large but %%@
finite $N$, can be ``lost'' if one alters certain features of the situation. We saw this:\\
  \indent (a) for the method of arbitrary functions: by raising one's standard of how close %%@
to exact equiprobability was ``close enough'', or by widening the class of density functions %%@
with respect to each of which one required approximate equiprobability; (Sections %%@
\ref{emergelimitMAF}, \ref{equiprobbefore}) and \\
  \indent (b) for fractals: by ``getting better eyesight'', i.e. by reducing the length-scale %%@
on which one could resolve spatial structure---and so see that at the given finite $N$, there %%@
was not yet the infinitely descending tower of structure characteristic of a fractal; %%@
(Section \ref{dimbefore}).

Cross-over occurs near a {\em critical} phase transition. This is one where a quantity called  %%@
the {\em correlation length} $\xi$, which summarizes the average length-scale on which %%@
microscopic quantities' values are correlated, diverges (in the modest sense of growing %%@
without bound, reviewed in Section \ref{deflatelimits}). Understanding many such transitions %%@
(quantitatively as well as qualitatively), and understanding cross-over in particular, is one %%@
of the great successes of the renormalization group (RG) techniques that have been developed %%@
over the last fifty years. 

Beware: some recent philosophical literature, waxing enthusiastic about the RG, suggests that %%@
every phase transition has a ``singular'' thermodynamic limit and-or infinite correlation %%@
length. Not so: witness the fact that until now, this Section has not had to mention the RG! %%@
More generally, I reiterate my point (e.g. in the preamble to Section \ref{ssqv} and in %%@
Section \ref{deform}) that there can be emergent limiting behaviour, with nothing %%@
``singular'' about the limit. But enough admonition! I turn to describing cross-over; (for %%@
details, cf. e.g. Yeomans (1992, p. 112), Cardy (1997, pp. 61, 69-72), Chaikin and Lubensky %%@
(2000, pp. 216, 270-3); Hadzibabic et al. (2006) is a recent example of experiments).

As substances approach a critical phase transition, they typically show behaviour %%@
characteristic of one of a small number of {\em universality classes}. Cross-over happens %%@
when a substance appears to show behaviour characteristic of one universality class, but then %%@
suddenly changes to another as it is brought even closer to its critical point. To explain %%@
this, we first recall the basic idea of the RG, as follows.\\
\indent (1): We define a space $X$ coordinatized by the parameter values that define the %%@
microscopic Hamiltonian, e.g. interaction strengths between particles, and the strength of %%@
the coupling to an external field; (and typically, also temperature).\\
\indent (2): We define a transformation $T$ on $X$ designed to preserve the large-scale %%@
physics of the system. Typically, $T$ is a coarse-graining, defined by local collective %%@
variables that take some sort of majority vote about the local quantities' values, followed %%@
by a rescaling, so that the resulting system can be assigned to a point in %%@
$X$.\footnote{Hence another playful variation by Kadanoff on Anderson's slogan (cf. footnote %%@
\ref{leocredit}); Section 6.4 of Kadanoff (2010) is called `Less is the same'.} \\
\indent (3): This assignment of the resulting system to a point within X enables one to %%@
consider {\em iterating} $T$, so that we get a flow on $X$. Critical points where $\xi$ %%@
diverges will be among the fixed points of this flow. For the fact that $\xi$ diverges means %%@
that the system ``looks the same'' at all length-scales, so that $T$ fixes (makes no change %%@
in) the description of the system. (Besides, this scale-invariance can involve power-law %%@
behaviour on all scales and self-similarity, and so lead to the use of fractals, e.g. to %%@
describe the distribution of sizes of the ``islands'' of aligned spins in a two- or %%@
three-dimensional Ising model; cf. the end of Section \ref{nature?}) 
 
I can now describe cross-over. I choose a kind called {\em finite-size cross-over}. This %%@
occurs when the ratio of $\xi$ to the system's size determines the fixed point towards which %%@
the RG flow sends the system. When $\xi$ is small compared to the size of the system, though %%@
very large on a microscopic length-scale, the system flows towards a certain fixed point %%@
representing a phase transition; and so exemplifies a certain universality class. Or to put %%@
it more prosaically: coarser and coarser (and suitably rescaled) descriptions of the system %%@
are more and more like descriptions of a phase transition. So in the jargon of my claims: the %%@
system illustrates (2:Before). But as $\xi$ grows even larger, and becomes comparable with %%@
the system size, the flow crosses over and moves away---in general, eventually, towards a %%@
different fixed point. In my jargon: the system runs up against (4:Unreal), and goes over to %%@
another universality class---eventually to another behaviour illustrating (2:Before).

 Of course, the correlation length will only approach a system's physical size when the %%@
system has been brought very close indeed to the phase transition, well within the usual %%@
experimental error. That is: until we enter the cross-over regime, experimental data about %%@
quantities such as the gradient of the free energy will strongly suggest non-analyticities, %%@
such as a sharp corner or an infinite peak. Or in other words: until we enter this regime, %%@
the behaviour will be as if the system is infinite in extent. But once we enter this regime, %%@
and the cross-over occurs, the appearance of non-analyticities goes away: peaks become tall %%@
and narrow---but finitely high. Again, in my jargon, we have: (2:Before) followed by %%@
(4:Unreal).

Finally, I remark that a similar discussion would apply to other kinds of cross-over, such as %%@
{\em dimensional cross-over}. For example, this occurs when the behaviour of a thin film %%@
crosses over from a universality class typical of three-dimensional systems to one for %%@
two-dimensional systems, as $\xi$ becomes comparable with the film's thickness.

\subsection{Envoi}\label{envoi}
I believe that my claims, in particular my two main ones, (1:Deduce) and (2:Before), are %%@
illustrated by many examples beyond the four I have chosen. For instance, to stick to the %%@
area of my main physics example, viz. the $N \raw \infty$ limit of quantum theory: there are %%@
Sewell's own examples of his scheme in Section \ref{soup}, and KMS states' description of %%@
thermodynamic phases (Section \ref{needTD}.A). Showing my claims in many such examples would %%@
indeed be strong testimony to the reconciliation of emergence and reduction. Work for another %%@
day! \\ \\

{\em Acknowledgements}:--\\
I dedicate this paper to the memory of Peter Lipton, a superb philosopher and a wonderful %%@
person.  
I am indebted to many audiences and colleagues. For invitations to lecture, I am very %%@
grateful to: the Royal Society, Bristol University, the Italian Society for Philosophy of %%@
Science, Lee Gohlicke and the Seven Pines Symposium, the Lorentz Centre at the University of %%@
Leiden, the University of Pittsburgh and Sandra Mitchell, Princeton University, and the %%@
organizers of the conferences Frontiers of Fundamental Physics 11, and Emergence in Physics. %%@
I am also grateful: to my hosts and audiences at these occasions, and at seminars in %%@
Cambridge, Chicago (twice), Maryland, Minnesota, Oxford and Sydney; to A. Caulton, R. Bishop, %%@
S. Bangu, E. Curiel, C. Callender, P. Humphreys, O. Shenker, E. Sober and A. Wayne, and %%@
especially to Paul Mainwood, Leo Kadanoff and two referees, for comments on previous versions %%@
and discussions; and to the editors---not least for their patience.

\section{References}

Avnir, D., Biham O., et al (1998), `Is the geometry of nature fractal?' {\em Science} {\bf %%@
279} , 39-40.

Anderson, P. (1972), `More is different', {\em Science} {\bf 177}, pp. 393-396; reprinted in %%@
Bedau and Humphreys (2008).

Bangu, S. (2009), `Understanding thermodynamic singularities: phase transitions, data and %%@
phenomena', {\em Philosophy of Science} {\bf 76}, pp. 488-505.

Barnsley, M. (1988), {\em Fractals Everywhere}, Academic Press, Boston, 1988.

Barrenblatt, G. (1996), {\em Scaling, Self-similarity and Intermediate Asymptotics}, %%@
Cambridge University Press.

Batterman, R. (1992), `Explanatory instability', {\em Nous} {\bf 26}, pp. 325-348.

Batterman, R. (2002), {\em The Devil in the Details}, Oxford University Press.

Batterman, R. (2005), `Critical phenomena and breaking drops: infinite idealizations in %%@
physics', {\em Studies in History and Philosophy of Modern Physics} {\bf 36B}, pp. 225-244.

Batterman, R. (2006), `Hydrodynamic vs. molecular dynamics: intertheory relations in %%@
condensed matters physics', {\em Philosophy of Science} {\bf 73}, pp. 888-904.

Batterman, R. (2009), `Emergence, singularities, and symmetry breaking', Pittsburgh arXive.

Batterman, R. (2010), `On the explanatory role of mathematics in empirical science', {\em %%@
British Journal for the Philosophy of Science} {\bf 61}, pp. 1-25.

Bedau, M. and Humphreys, P. (eds.) (2008), {\em Emergence: contemporary readings in %%@
philosophy and science}, MIT Press: Bradford Books. 

Belot, G. (2005), `Whose Devil? Which Details?, {\em Philosophy of Science} {\bf 72}, pp. %%@
128-153; a fuller version is available at: http://philsci-archive.pitt.edu/archive/00001515/

Berry, M. (1994), `Asymptotics, singularities and the reduction of theories', in D. Prawitz, %%@
B. Skyrms and D. Westerdahl eds. {\em Logic, Methodology and Philosophy of Science IX: %%@
Proceedings of the Ninth International Congress of Logic, Methodology and Philosophy of %%@
Science, Uppsala, Sweden 1991}, Elsevier Science, pp. 597-607.

Bogen, J. and Woodward, J. (1988), `Saving the phenomena', {\em Philosophical Review} {\bf %%@
97}, pp. 303-352.

Bouatta, N. and Butterfield, J. (2011), `Emergence and reduction combined in phase %%@
transitions', forthcoming in J. Kouneiher ed. {\em Frontiers of Fundamental Physics 11}, %%@
(Proceedings of the Conference), American Institute of Physics.

Brady, R. and Ball, R. (1984), `Fractal growth of copper electrodeposits', {\em Nature} {\bf %%@
309}, Issue 5965, pp. 225-229. 

Butterfield, J. (2010), `Emergence, reduction and supervenience: a varied landscape', in this %%@
issue of {\em Foundations of Physics}.

Callender, C. (2001), `Taking thermodynamics too seriously', {\em Studies in History and %%@
Philosophy of Modern Physics} {\bf 32B}, pp. 539-554. 

Cardy, J. (1997), {\em Scaling and Renormalization in Statistical Physics}, Cambridge Lecture %%@
Notes in Physics, volume 5; Cambridge University Press.

Castaing, B., Gunaratne, G. et al. (1989), `Scaling of hard thermal turbulence in %%@
Rayleigh-Benard convection', {\em Journal of Fluid Mechanics} {\bf 204}, pp. 1-30.

Cat, J., (1998), `The physicists' debates on unification in physics at the end of the %%@
twentieth century', {\em Historical Studies in the Physical and Biological Sciences}, {\bf %%@
28} pp. 253-300.

Chaikin, P. and Lubensky, T. (2000), {\em Principles of Condensed Matter Physics}, Cambridge %%@
University Press.

Colyvan, M. (2005), `Probability and ecological complexity': a review of Strevens (2003), %%@
{\em Biology and Philosophy} {\bf 20}, pp. 869-879. 

Diaconis, P. (1977), `Finite forms of de Finetti's theorem on exchangeability'  {\em %%@
Synthese} {\bf 36}, pp. 271-281. 

Diaconis, P. and Freedman D. (1980), `De Finetti's generalizations of exchangeability', in R. %%@
Jeffrey ed. {\em Studies in Inductive Logic and Probability}, volume 2, University of %%@
California Press, pp. 233-249. 

Diaconis, P. Holmes, S. and Montgomery R.(2007), `Dynamical bias in the coin toss', {\em SIAM %%@
Review} {\bf 49} pp. 211-235. 

Dixon, P., Wu, L. et al. (1990), `Scaling in the relaxation of supercooled liquids', {\em %%@
Physical Review Letters} {\bf 65}, pp. 1108-1111.

Emch, G. (1972), {\em Algebraic Methods in Statistical Mechanics and Quantum Field Theory}, %%@
John Wiley.

Emch, G. (2006), `Quantum Statistical Physics', in {\em Philosophy of Physics, Part B}, a %%@
volume of {\em The Handbook of the Philosophy of Science}, ed. J. Butterfield and J. Earman, %%@
North Holland, pp. 1075-1182. 

Emch, G., and Liu, C. (2002), {\em The Logic of Thermo-statistical Physics}, Springer-Verlag 

Engel, E. (1992), {\em A Road to Randomness in Physical Systems}, Springer. 
 
Falconer, K. (2003), {\em Fractal Geometry}, John Wiley. 

Frigg, R. and Hoefer, C. (2010), `Determinism and chance from a Humean perspective', %%@
forthcoming in {\em The Present Situation in the Philosophy of Science}, edited by D. Dieks, %%@
W. Gonzalez, S. Hartmann, F. Stadler, T.Uebel and M. Weber, Springer. 

Goldenfeld, N., Martin O. and Oono, Y. (1989), `Intermediate asymptotics and renormalization %%@
group theory', {\em Journal of Scientific Computing} {\bf 4}, pp. 355-372.  

Gross, D. (2001), {\em Microcanonocal Thermodynamics; phase transitions in small systems}, %%@
World Scientific.

Hadzibabic, Z. et al. (2006), `Berezinskii-Kosterlitz-Thouless crossover in a trapped atomic %%@
gas', {\em Nature} {\bf 441}, 29 June 2006, pp. 1118-1121.

Hastings, H. and Sugihara, G. (1993), {\em Fractals: a User's Guide for the Natural %%@
Sciences}, Oxford University Press.

Hooker, C. (2004), `Asymptotics, reduction and emergence', {\em British Journal for the %%@
Philosophy of Science} {\bf 55}, pp. 435-479.

Jeffrey, R. (1988), `Conditioning, Kinenatics and Exchangeability', in B. Skyrms and W. %%@
Harper ed.s, {\em Causation, Chance and Credence} volume 1, pp. 221-255, Kluwer.

Kadanoff, L.  (2009), `More is the same: phase transitions and mean field theories', {\em %%@
Journal of Statistical Physics} {\bf 137}, pp. 777-797; available at %%@
http://arxiv.org/abs/0906.0653

Kadanoff, L. (2010), `Theories of matter: infinities and renormalization', forthcoming in R. %%@
Batterman ed., {\em The Oxford Handbook of the Philosophy of Physics}, Oxford University %%@
Press; available at http://arxiv.org/abs/1002.2985; and at \\ %%@
http://jfi.uchicago.edu/$\sim$leop/AboutPapers/Trans2.pdf.

Kadanoff, L. (2010a), `Relating theories via renormalization', available at \\ %%@
http://jfi.uchicago.edu/$\sim$leop/AboutPapers/RenormalizationV4.0.pdf 
   
Keller, J. (1986), `The probability of heads', {\em American Mathematical Monthly} {\bf 93}, %%@
pp. 191-197.

Koenig, R. and Renner, R. (2005), `A de Finetti representation for finite symmetric quantum %%@
states', {\em Journal of Mathematical Physics} {\bf 46}, 012105; available at %%@
arXiv:quant-ph/0410229v1 

Koenig, R. and Mitchison, G. (2007), `A most compendious and facile quantum de Finetti %%@
theorem', {\em Journal of Mathematical Physics} {\bf 50}, 012105; available at %%@
arXiv:quant-ph/0703210

Kritzer, P. (2003), {\em Sensitivity and Randomness: the Development of the Theory of %%@
Arbitrary Functions}, Diplom thesis, University of Salzburg. 

Landsman, N. (2006), `Between Classical and Quantum', in {\em Philosophy of Physics, Part A}, %%@
a volume of {\em The Handbook of the Philosophy of Science}, ed. J. Butterfield and J. %%@
Earman, North Holland, pp. 417-554; available at arxiv:quant-ph/0506082 and at %%@
http://philsci-archive.pitt.edu/archive/00002328. 

Lavis, D. and Bell, G. (1999), {\em Statistical Mechanics of Lattice Systems 1; Closed Forms %%@
and Exact Solutions}, Springer: second and enlarged edition.

Lavis, D. and Bell, G. (1999a), {\em Statistical Mechanics of Lattice Systems 2; Exact, %%@
Series and Renormalization Group Methods}, Springer.

Le Bellac, M. (1991), {\em Quantum and Statistical Field Theory}, (translated by G. Barton) %%@
Oxford University Press.

Liu, C. (2001), `Infinite systems in SM explanation: thermodynamic limit, renormalization %%@
(semi)-groups and ireversibility' {\em Philosophy of Science} {\bf 68} (Proceedings), pp. %%@
S325-S344.

Liu, C. and Emch, G. (2005), `Explaining quantum spontaneous symmetry breaking', {\em Studies %%@
in History and Philosophy of Modern Physics} {\bf 36}, pp. 137-164.  

Mainwood, P. (2006), {\em Is More Different? Emergent Properties in Physics}, D.Phil. %%@
dissertation, Oxford University. At: http://philsci-archive.pitt.edu/8339/

Mainwood, P. (2006a), `Phase transitions in finite systems', unpublished MS; (corresponds to %%@
Chapter 4 of Mainwood (2006). At: http://philsci-archive.pitt.edu/8340/

Mandelbrot, B. (1982), {\em The Fractal Geometry of Nature}. San Francisco: Freeman.

Menon, T. and Callender C. (2011), `Going through a phase: philosophical questions raised by %%@
phase transitions', forthcoming in R. Batterman ed. {\em The Oxford Handbook of Philosophy of %%@
Physics}, Oxford University Press.    

Myrvold, W. (2011), `Deterministic laws and epistemic chances', unpublished manuscript. .
   
Nagel, E. (1961), {\em The Structure of Science: problems in the Logic of Scientific %%@
Explanation}, Harcourt. 

Nelson, E. (1987), {\em Radicaly Elementary Probability Theory}, Princeton University Press, %%@
Annals of Mathematics Studies number 117.

Peitgen, H.-O. and Richter P. (1986), {\em The Beauty of Fractals}, Springer-Verlag, %%@
Heidelberg. 

Poincar\'{e}, H. (1896), {\em Calcul de Probabilities}, Gauthier-Villars, Paris; page %%@
reference to 1912 edition.

Richardson, L. (1961), `The problem of contiguity: an appendix of statistics of deadly %%@
quarrels', {\em General Systems Yearbook} {\bf 6}, pp. 139-187. 

Renner, R. (2007), `Symmetry implies independence', {\em Nature Physics} {\bf 3}, pp. 645 - %%@
649.

Rueger, A. (2000), `Physical emergence, diachronic and synchronic', {\em Synthese} {\bf 124}, %%@
pp. 297-322. 

Rueger, A. (2006), `Functional reduction and emergence in the physical sciences', {\em %%@
Synthese} {\bf 151}, pp. 335-346. 

Ruelle, D. (1969), {\em Statistical Mechanics: Rigourous Results}, W.A. Benjamin. 

Sewell, G. (1986), {\em Quantum Theory of Collective Phenomena}, Oxford University Press.

Sewell, G. (2002), {\em Quantum Mechanics and its Emergent Microphysics}, Princeton %%@
University Press.

Shenker, O. (1994), `Fractal Geometry is not the Geometry of Nature', {\em Studies in the %%@
History and Philosophy of Science} {\bf 25}, pp. 967-982.

Simon, H. (1996), `Alternative views of complexity', Chapter 7 of his {\em The Sciences of %%@
the Artificial}, third edition, MIT Press; the Chapter is reprinted in Bedau and Humphreys %%@
(2008); page reference to the reprint. 

Smith, P. (1998), {\em Explaining Chaos}, Cambridge University Press.
 
Sober, E. (2010), `Evolutionary Theory and the Reality of Macro-probabilities', in E. Eells %%@
and J. Fetzer ed.s, {\em The Place of Probability in Science} (Boston Studies in the %%@
Philosophy of Science volume 284), Springer, pp. 133-161.

Stoppard, T. (1993), {\em Arcadia}. London: Faber and Faber. 
  
Strevens, M. (2003), {\em Bigger than Chaos: Understanding Complexity through Probability}, %%@
Harvard University Press.

Thompson, C. (1972), {\em Mathematical Statistical Mechanics}, Princeton University Press.

von Plato, J. (1983), `The method of arbitrary functions', {\em British Journal for the %%@
Philosophy of Science} {\bf 34}, pp. 37-47.
 
von Plato, J. (1994), {\em Creating Modern Probability}, Cambridge University Press.

Wayne, A. (2009), `Emergence and singular limits', forthcoming in {\em Synthese}; available %%@
at http://philsci-archive.pitt.edu/archive/00004962/.

Weinberg, S. (1987), `Newtonianism, reductionism and the art of congressional testimony', %%@
{\em Nature} {\bf 330}, pp. 433-437; reprinted in Bedau and Humphreys (2008); page reference %%@
to the reprint; also reprinted in Weinberg, S. {\em Facing Up: Science and its Cultural %%@
Adversaries}, Harvard University Press, pp. 8-25.

Werndl, C. (2010), Review of Strevens (2003), forthcoming in {\em British Journal for the %%@
Philosophy of Science}. 

Wimsatt, W. (1997), `Aggregativity: reductive heuristics for finding emergence', {\em %%@
Philosophy of Science} {\bf 64}, pp. S372-S384; reprinted in Bedau and Humphreys (2008); page %%@
reference to the reprint.

Yeomans, J. (1992), {\em Statistical Mechanics of Phase Transitions}, Oxford University %%@
Press.

\end{document}